\documentclass[notitlepage, superscriptaddress
]{revtex4-1}

\usepackage[ruled, vlined]{algorithm2e}
\AtBeginDocument{
    \SetInd{0.5em}{1em}
}
\usepackage{nicematrix}
\usepackage[utf8]{inputenc}
\usepackage{bm}

\usepackage{bbold}
\usepackage{color}
\usepackage{hyperref}
\hypersetup{colorlinks=true}
\usepackage{tikz}
\usetikzlibrary{topaths,calc}
\usepackage{float}
\usepackage{amsmath}
\usepackage[normalem]{ulem}

\usepackage{braket}
\usepackage{svg}
\usepackage{graphicx}
\usepackage{subcaption}

\usepackage{mathtools}

\usepackage[english]{babel}
\usepackage{amsthm}
\usepackage{amssymb}

\usepackage{physics}
\usepackage{dsfont}

\usepackage{qcircuit}
\usepackage{enumitem}
\usepackage{quiver}

\definecolor{graphblue}{RGB}{0,0,232}
\definecolor{graphgreen}{RGB}{23,153,11}
\definecolor{graphorange}{RGB}{248,136,1}

\newcommand{\BaseConstraint}[6]{%
    \begin{NiceArray}{c|c|c}[cell-space-limits=3pt]
    \cline{2-3}
    1 & #1 & \multicolumn{1}{c|}{#2}  \\
    \cline{2-3}
    2 & #3 & \\
    \cline{2-2}
    3 & #4 & \\
    \cline{2-2}
    4 & #5 & \\
    \cline{2-2}
    5 & #6 & \\
    \cline{2-2}
    \end{NiceArray}%
}

\newcommand{\SequentialConstraint}[9]{%
    \begin{NiceArray}{c|c|c|c|}[cell-space-limits=3pt]
    \cline{2-4}
    1 & #1 & #2 & #3 \\
    \cline{2-4}
    2 & #4 & #5 & #6 \\
    \cline{2-4}
    3 & #7 & #8 & #9 \\
    \cline{2-4}
    \end{NiceArray}%
}

\newcommand{\UnstructuredConstraintOne}[9]{
\begin{NiceArray}{c|c|cc}
\cline{2-3}
1 & #1 & \multicolumn{1}{c|}{#2} & \\
\cline{2-3}
2 & #3  & & \\
\cline{2-4}
 3 & #4 & \multicolumn{1}{c|}{#5} & \multicolumn{1}{c|}{#6} \\
\cline{2-4}
4 & #7 & \\
\cline{2-2}
5 & #8 &  \\
\cline{2-3}
6 & #9 & \multicolumn{1}{c|}{1} \\
\cline{2-3}
\end{NiceArray}
}

\newcommand{\UnstructuredConstraintZero}[9]{
\begin{NiceArray}{c|c|cc}
\cline{2-3}
1 & #1 & \multicolumn{1}{c|}{#2} & \\
\cline{2-3}
2 & #3  & & \\
\cline{2-4}
 3 & #4 & \multicolumn{1}{c|}{#5} & \multicolumn{1}{c|}{#6} \\
\cline{2-4}
4 & #7 & \\
\cline{2-2}
5 & #8 &  \\
\cline{2-3}
6 & #9 & \multicolumn{1}{c|}{0} \\
\cline{2-3}
\end{NiceArray}
}

\newtheorem{theorem}{Theorem}[section]

\newtheorem{lemma}[theorem]{Lemma}
\newtheorem{definition}[theorem]{Definition}

\def\arrvline{\hfil\kern\arraycolsep\vline\kern-\arraycolsep\hfilneg}
\begin{document}
\title{Convergence guarantee for linearly-constrained combinatorial optimization with a quantum alternating operator ansatz}

\author{Brayden Goldstein-Gelb}
\affiliation{Brown University, Providence, RI 02912, USA}

\author{Phillip C. Lotshaw}
\affiliation{Quantum Information Science Section, Oak Ridge National Laboratory, Oak Ridge, TN 37381, USA}
\email[]{Lotshawpc@ornl.gov}
\thanks{This manuscript has been authored by UT-Battelle, LLC, under Contract No. DE-AC0500OR22725 with the U.S. Department of Energy. The United States Government retains and the publisher, by accepting the article for publication, acknowledges that the United States Government retains a non-exclusive, paid-up, irrevocable, world-wide license to publish or reproduce the published form of this manuscript, or allow others to do so, for the United States Government purposes. The Department of Energy will provide public access to these results of federally sponsored research in accordance with the DOE Public Access Plan.}

\begin{abstract}

We present a quantum alternating operator ansatz (QAOA$^+$) that solves a class of linearly constrained optimization problems by evolving a quantum state within a Hilbert subspace of feasible problem solutions. Our main focus is on a class of problems with a linear constraint containing sequential integer coefficients. For problems in this class, we devise QAOA$^+$ circuits that provably converge to the optimal solution as the number of circuit layers increases, generalizing previous guarantees for solving unconstrained problems or problems with symmetric constraints. Our approach includes asymmetric ``mixing" Hamiltonians that drive transitions between feasible states, as well as a method to incorporate an arbitrary known feasible solution as the initial state, each of which can be applied beyond the specific linear constraints considered here. This analysis extends QAOA$^+$ performance guarantees to a more general set of linearly-constrained problems and provides tools for future generalizations.
\end{abstract}
\date{\today}

\maketitle

\section{Introduction} 

Variational quantum algorithms have gained popularity as potential candidates for near-term quantum computational advantages and for exploring the capabilities of near-term quantum devices \cite{cerezo2021variational}.   Here we focus on the quantum approximate optimization algorithm (QAOA) \cite{QAOA} and its generalization in the quantum alternating operator ansatz (QAOA$^+$) \cite{hadfield2019quantum} as variational approaches to solving combinatorial optimization problems. There have been a wide variety of studies to assess performance \cite{lotshaw2021empirical,lotshaw2023simulations,crooks2018performance,zhou2020quantum}, scaling \cite{lotshaw2023approximate,lotshaw2022scaling,akshay2022circuit} and theoretical behaviors of these algorithms \cite{shaydulin2023parameter,farhi2020quantum,wurtz2022counterdiabaticity,shaydulin2021classical,akshay2020reachability,brady2021optimal,brady2021behavior}, as well as hardware demonstrations \cite{harrigan2021quantum,ebadi2022quantum,shaydulin2023qaoawith,pelofske2023quantum,pelofske2024short}, development of advanced compilation approaches \cite{herrman2021globally,moondra2024promise} and proposed algorithmic improvements \cite{morris2024performant,herrman2022multi,ponce2023graph,tate2023warm,chandarana2022digitized,zhu2022adaptive,bravyi2020obstacles,brady2023iterative}.  Advantages for QAOA over certain conventional algorithms have been predicted analytically in several cases, such as large size limits for the Sherrington-Kirkpatrick spin-glass model relative to conventional semi-definite programming \cite{farhi2022quantum} and for ``large girth" MaxCut instances relative to all algorithms known by the authors of Ref.~\cite{basso2021quantum} (though a classical algorithm with similar scaling is reported in \cite{hastings2021classical}).  An empirical scaling advantage for QAOA has also been reported for the low-autocorrelation binary sequences problem, based on scaling observed in numerical calculations with up to forty qubits and relative to state-of-the-art conventional algorithms \cite{shaydulin2024evidence}.   These important results encourage further study into characterizing and developing advanced implementations of QAOA and QAOA$^+$.

Despite significant progress on solving unconstrained problems with QAOA, there has been much less work on extensions to constrained combinatorial problems, which are important for a wide variety of practical applications. QAOA was originally designed to solve unconstrained combinatorial problems, building on previous work in quantum adiabatic algorithms \cite{farhi2000quantum,farhi2001quantum,albash2018adiabatic}, and hence can only handle constrains indirectly, in a Lagrangian approach. Limitations of these approaches were addressed theoretically in the context of quantum adiabatic evolution, using modified interactions that restrict the quantum evolution to a subspace of feasible problem solutions, consistent with the constraints \cite{hen2016quantum,hen2016driver,leipold2021constructing}. These approaches can require many-body interactions that may not be easy to implement in the adiabatic evolution of a physical system. QAOA$^+$ adapts similar ideas to a quantum circuit setting, where sequences of two-qubit gates can realize the many-body constraint-preserving interactions.

One shortcoming of the QAOA$^+$ approaches is that it is not always clear how to determine a constraint-preserving circuit. Previous research has considered  Hamming weight constraints corresponding to a linear constraint with equal coefficients, with applications in graph coloring \cite{wang2020xy}, graph isomorphism \cite{hen2016driver}, extractive text summarization \cite{niroula2022constrained}, and portfolio optimization \cite{he2023alignment}.   Explicit circuit constructions for nonlinear constraints were also considered in Ref.~\cite{hadfield2019quantum}, as well as previous work on maximum independent set \cite{saleem2020max} and various satisfiability problems \cite{hen2016quantum,hen2016driver}. It is desirable to generalize the previous approaches to address more general constraints, but finding suitable circuits is at least NP-hard in general \cite{leipold2021constructing}. Determining which classes of constraints can be addressed with explicit circuit constructions is an important open question.

Another shortcoming is that QAOA$^+$ approaches have not always focused on maintaining an adiabatic limit that guarantees it is possible to prepare an optimal solution state with sufficient circuit depth \cite{hadfield2019quantum}.  A large-depth performance guarantee of this type was a key founding principle of QAOA \cite{QAOA}, and without it, we are not aware of any theoretical reason to expect optimal or near-optimal solutions to be obtained from QAOA$^+$.  Cases where the performance guarantee does not hold have also been found to yield poor empirical performance, as in QAOA$^+$ simulations with a ``misaligned" mixing operator and initial state \cite{he2023alignment}. Recently Binkowski {\it et al.}~\cite{ConvergenceProof} have derived technical conditions under which QAOA$^+$ obtains a performance guarantee similar to QAOA, which will prove useful here. It is also worth noting that new alternatives to QAOA$^+$ have also been recently been developed, including quantum walks \cite{marsh2019quantum}, nonstandard adiabatic rotations \cite{perlin2024q}, and quantum Zeno dynamics with midcircuit measurements \cite{herman2023constrained}, but here we shall focus on developing the more standard QAOA$^+$-based approaches. 

We address the shortcomings identified in the previous paragraphs by developing a theoretical approach for solving a class of linearly constrained optimization problems with QAOA$^+$.  We generalize previous ``$XY$"-mixer \cite{wang2020xy} approaches for problems with a symmetric Hamming-weight constraint to solve a more general class of problems with a nonsymmetric linear constraint that contains sequential integer coefficients, and we demonstrate a large-depth performance guarantee is obtained for our constructions.  The approach also includes a ``warm-start" from an arbitrary known feasible solution, inspired by  Refs.~\cite{tate2023warm,wurtz2021classically,saleem2023approaches,maciejewski2024improving}.  This initial state overcomes a significant technical issue with preparing adiabatic initial states for nonsymmetric QAOA$^+$ ``mixing" Hamiltonians, as noted previously in Refs.~\cite{hen2016quantum,hen2016driver,leipold2021constructing}. It proves necessary here for obtaining a convergence guarantee while maintaining a simple circuit, and is broadly applicable to QAOA$^+$ implementations beyond the specific linearly-constrained problems we consider. 

%%%%%%%%%%%%%%%%%%%%%%%%%%%%%%%%%%%%%%%%%%%%%%%%%%%%%%%%%%%%%%%%%%
%%%%%%%%%%%%%%%%%%%%%%%%%%%%%%%%%%%%%%%%%%%%%%%%%%%%%%%%%%%%%%%%%%
%%%%%%%%%%%%%%%%%%%%%%%%%%%%%%%%%%%%%%%%%%%%%%%%%%%%%%%%%%%%%%%%%%
\section{Quantum combinatorial optimization with linear constraints}\label{quantum opt section}

This section reviews background on formulations of combinatorial optimization problems for conventional and quantum computers, the quantum algorithms QAOA and QAOA$^+$, and the convergence guarantees for these algorithms.  We shall also define a graphical notation for visualizing operators in Sec.~\ref{background}, which will be used throughout this work. Readers familiar with the background may wish to skip to our new results in Sec.~\ref{linear constraints theory section}. 

\subsection{Combinatorial optimization with linear constraints}
Combinatorial optimization problems seek to minimize an objective function $\mathcal{C}(\bm z)$ with a bitstring argument $\bm z = (z_1,z_2,\ldots,z_N)$ on $N$ bits $z_i \in \{0,1\}$. In many important cases the objective $\mathcal{C}(\bm z)$ can be expressed in a quadratic form \cite{lucas2014ising}%
\begin{equation} \mathcal{C}(\bm z) = \sum_{i<j} J_{ij} z_i z_j + \sum_i h_i z_i \end{equation} 
with real coefficients $J_{ij}$ and $h_i$ that specify a problem instance.  The solution bitstring may also be required to satisfy certain constraints, which limit the set of potential solutions from the set of all $2^N$ possible bitstrings $\bm z$ down to a smaller feasible subset $\mathcal{S}$. Here we will focus on linear constraints because they are relatively simple and appear in a variety of contexts such as NP-hard binary integer linear programming (BILP) problems \cite{lucas2014ising}.  Optimization problems with linear equality constraints and integer coefficients have the form 
\begin{align} \label{BILP general} \bm z_\mathrm{opt} =\ & \underset{\bm z}{\mathrm{arg\ min}}\  \mathcal{C}(\bm z) \nonumber\\
& \text{s.t.} \ \sum_{i} s_{\alpha,i} z_i  = b_\alpha \ \ \forall \alpha \in \{1,2,\ldots,t\} \nonumber\\
& z_i \in \{0,1\} \ \forall i 
\end{align}
where the matrix $\bm s$ with elements $s_{\alpha, i} \in \mathbb{N}$ and vector $\bm b$ with components $b_\alpha \in \mathbb{N}$ together encode $t$ total constraints. As mentioned previously, it is generally NP-hard to determine a solution satisfying the constraints \cite{leipold2021constructing}. To make progress analytically, we therefore focus on a restricted class of instances with a single constraint containing sequential integer coefficients

\begin{align} \label{sequential linear BILP single index} \bm z_\mathrm{opt} =\ & \underset{\bm z}{\mathrm{arg\ min}}\  \mathcal{C}(\bm z) \nonumber\\
& \text{s.t.} \ \sum_{i} s_{i} z_i  = b \nonumber\\
& z_i \in \{0,1\} \ \forall i 
\end{align}
where $s_i, b \in \mathbb{N}$. For our convergence proofs, we require sequential constraint coefficients $s_i \in \{1, 2, \ldots, k\}$ such that each unique value of $2 \leq s_i \leq k$ appears at least once in the constraint equation, while $s_i=1$ appears at least twice. For example, a constraint $z_1 + z_2 + 2z_3=b$ follows the assumptions in our proof because it contains at least two coefficients $s_i=1$ ($i=1,2)$ and at least one of each larger coefficient up to the maximum value $k=2$, in this case just the single coefficient $s_3=2$. A constraint $z_1 + z_2 + 3z_3=b$ does not follow our assumptions because it contains a constraint coefficient $s_3=3$ but does not contain at least one coefficient $s_i=2$. The set of allowed constraints generalizes the Hamming weight constraints with constant coefficients $s_i = 1$ that have been considered in previous QAOA$^+$ implementations with the $XY$-mixer \cite{wang2020xy,niroula2022constrained,cook2020quantum,he2023alignment,bartschi2020grover}. In Sec.~\ref{linear constraints theory section} we will provide explicit circuits to solve these problems using polynomially-many gates per circuit layer, as well as a convergence guarantee that the circuits can approach the optimal solution with probability arbitrarily close to one as the number of layers increases. These  circuits add to the body of known constraint preserving mixers \cite{hen2016driver,hadfield2019quantum,fuchs2022constraint,leipold2021constructing,saleem2020max}.

The set of feasible solutions $\mathcal{S}^{(\bm b)}_\text{classical}$ that satisfy the constraints for a given $\bm b$ is denoted
\begin{equation} \mathcal{S}^{(\bm b)}_\text{classical} = \{ \bm z\ |\ \bm s \cdot \bm z = \bm b\}, \end{equation} 
where the shorthand $\bm s \cdot \bm z = \bm b$ denotes the constraint. One way to solve these problems is to search for solutions only within the feasible set $\mathcal{S}^{(\bm b)}_\text{classical}$.  An alternative which may be simpler to implement is to search through the full $2^N$-dimensional space and minimize the Lagrangian
\begin{equation} \label{dual obj} \tilde{\mathcal{C}}(\bm z) = \mathcal{C}(\bm z) - \sum_\alpha \lambda_\alpha \left(\sum_i s_{\alpha,i}z_i - b_\alpha\right)^2 \end{equation}
where $\lambda_\alpha \in \mathbb{R}$ are Lagrange multipliers.  Choosing each $\lambda_\alpha$ to be sufficiently large guarantees that the optimal solution $\bm z_\mathrm{opt}$ is the same for $\mathcal{C}(\bm z)$ and $\tilde{\mathcal{C}}(\bm z)$, and in this sense $\tilde{\mathcal{C}}(\bm z)$ is an equivalent formulation. 

In the following section we review quantum algorithms for solving these problems, beginning with the quantum annealing and QAOA algorithms which can be formulated to solve the Lagrangian objective (\ref{dual obj}), and ending with QAOA$^+$ approaches which search only in the relevant feasible subspace to more directly solve an instance of (\ref{BILP general}). 

\subsection{Quantum algorithms for combinatorial optimization} \label{background}

To solve combinatorial optimization problems with quantum algorithms, we encode an optimization problem into the eigenspectrum of a quantum operator $C$, with 
\begin{equation} C\ket{\bm z} = \mathcal{C}(\bm z) \ket{\bm z},\end{equation} 
where $\ket{\bm z} = \ket{z_1, z_2, \ldots, z_N}$ is a computational basis state with $z_i \in \{0,1\}$ (a similar relation holds when using the Lagrangian formulation with $\tilde{\mathcal{C}}(\bm z)$ in place of $\mathcal{C}(\bm z)$). Then a quantum algorithm is used to generate and identify an exact or approximate ground state of $C$, providing an exact or approximate solution to the classical problem. For constrained optimization problems we define a feasible solution Hilbert subspace as 
\begin{equation} \mathcal{S}^{(b)} = \text{span}(\{ \ket{\bm z} \ | \ \bm s \cdot \bm z = \bm b\}) \end{equation}%
while the total Hilbert space is denoted $\mathcal{H} = \text{span}(\{\ket{\bm z}\})$.

In addition to $C$, the quantum algorithm must also employ at least one operator that does not commute with $C$, to drive transitions between basis states and identify a final solution.  We will find it useful to think of these operators in terms of two types of graphs, which describe their associated interactions among qubits and basis states.  Specifically, for an operator $O = \sum_i B_i,$ we define the following graphs associated with $\{B_i\}$

\begin{definition}\label{def:qubit interaction graph} Qubit interaction graph

    Let $G_\mathrm{qubit}(\{B_i\}) = (V_\mathrm{qubit},E_\mathrm{qubit})$ be the ``qubit interaction graph" for a set of operators $\{B_i\}$ acting on subspace $\mathcal{S}$ of a qubit Hilbert space $\mathcal{H}$.  The vertex set $V_\mathrm{qubit}$ contains a vertex for each qubit $i$ with basis states $\ket{z_i}$.  The edge set contains (hyper-)edges between each set of qubits that are acted on by the individual $B_i$.
\end{definition}

\begin{definition}\label{def:basis interaction graph} Basis state interaction graph

    Let $G_{\mathcal{S}}(\{B_i\}) = (V_{\mathrm{basis}}, E_{\mathrm{basis}})$ be the ``basis state interaction graph" for a set of operators $\{B_i\}$ acting on subspace $\mathcal{S}$ of a qubit Hilbert space $\mathcal{H}$.  The vertex set $V_{\mathrm{basis}}$ corresponds to a set of computational basis states $\ket{\bm z} = \bigotimes_{i=1}^{N} \ket{z_{i}} \in \mathcal{S}$.  The edge set $E_{\mathrm{basis}}$ contains edges between all $\ket{\bm z}, \ket{\bm z'} \in V_{\mathrm{basis}}$ such that for at least one $B_i$, $\bra{\bm z } B_i \ket{\bm z'} \neq 0$. 
\end{definition}

A simple example of each of these graphs is given in Fig.~\ref{fig:transition basis graphs}(a)-(b) for a transverse field operator 
\begin{equation} \label{transverse field} B = -\sum_i X_i \end{equation}
The qubit-interaction graph $G_{\text{qubit}}(\{X_i\})$ for the operator $B$ is shown in Fig.~\ref{fig:transition basis graphs}(a); it contains a set of self-loops on each qubit, since the operators $X_i$ each act on a single qubit (examples with operators acting on multiple qubits will be given later). Figure \ref{fig:transition basis graphs}(b) shows the basis state interaction graph $G_{\mathcal{H}}(\{X_i\})$ of transitions driven by $B$, which forms a hypercube on the set of computational-basis-state vertices in the total Hilbert space $\mathcal{H}$.

We shall use this graphical notation in the presentation and proofs that follow. 
We now review three quantum algorithms for exactly or approximately solving combinatorial problems, which will lead to our new approach to linearly constrained optimization in Sec.~\ref{linear constraints theory section}.

\subsubsection{Adiabatic evolution} \label{adiabatic section}

\begin{figure}
    \centering
    \includegraphics[width=\textwidth,height=15cm,keepaspectratio]{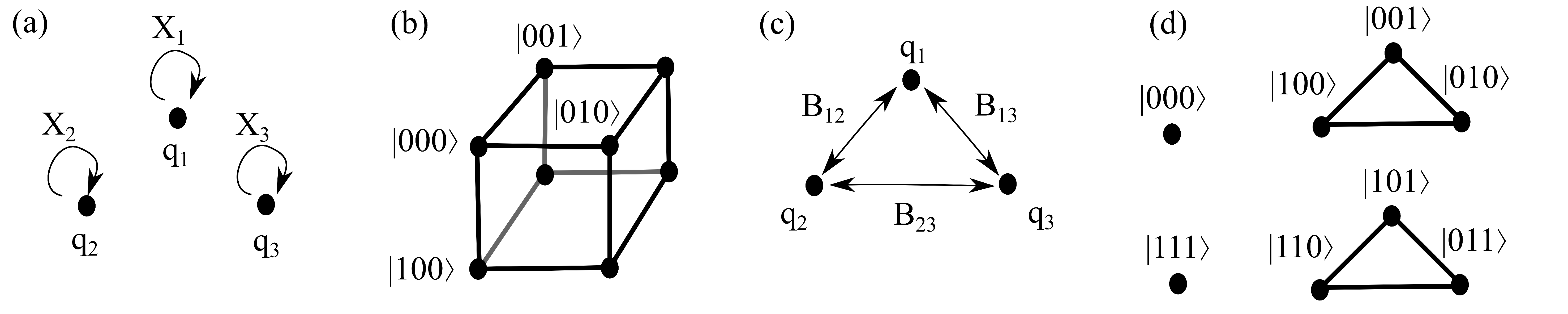}
    \caption{(a,c) Example qubit-interaction graphs on three qubits $q_1, q_2,$ and $q_3,$, along with (b,d) the respective basis-state-transition graphs on the computational Hilbert space $\mathcal{H}= \mathrm{span}(\{\ket{z_1,z_2,z_3}\})$. (a) The qubit-interaction graph for the transverse field mixer $B$, with single-qubit operators depicted as self-loops. (b) The transverse-field basis-state-transition graph is a hypercube, where edges denote transitions between basis states generated by $B$. A subset of vertices are labeled to show changes $0 \leftrightarrow 1$ along each dimension. (c) Qubit-interaction graph for the $XY$ mixer, with edges (shown as arrows) depicting two-qubit operations $B_{ij} = -\frac{1}{2}(X_i X_j + Y_i Y_j)$. (d) The $XY$-mixer basis-state-transition graph separates into components based on Hamming weight, each corresponding to a different feasible subspace $\mathcal{S}$. }
    \label{fig:transition basis graphs}
\end{figure}

Adiabatic evolution provides a systematic way to prepare an exact or approximate ground state of $C$.  We consider evolution beginning with an easy-to-prepare ground state $\ket{\psi_0}=\ket{+}^{\otimes n}$ of the simple Hamiltonian $B$ in (\ref{transverse field}). The operator $B$ drives transitions between all basis states, as shown in Fig.~\ref{fig:transition basis graphs}(b), and a suitable adiabatic protocol can use these transitions to drive the initial uniform superposition into the ground state of $C$.  For this the initial ground state $\ket{+}^{\otimes n}$ is evolved under the time-dependent Hamiltonian
\begin{equation} \label{Ht} H(s(t)) = (1-s(t))B + s(t)C,\end{equation}
with a schedule $s(t)$ following $s(0)=0$ and $s(T)=1$, where $T$ is the total evolution time. When the changes in $H(s(t))$ are sufficiently slow, then an initial ground state $\ket{E_0(0)}$ of $H(0)=B$ will evolve to the ground state $\ket{E_0(1)}$ of $H(1)=C$ at time $T$. To prepare the ground state, the runtime $T$ must be much greater than the adiabatic timescale $T_A = \max_s T_A(s)$, where  \cite{Albash2018RMP}
\begin{equation} \label{TA(s)} T_A(s) = \max_{j>0} \frac{|\bra{E_0(s)}\partial_s H |E_j(s)\rangle|}{|E_0(s) - E_j(s)|^2}. \end{equation}
where the sum runs over all other instantaneous eigenstates $\ket{E_j(s)}$ with eigenenergies $E_j(s)$. Hence a nonzero energy gap $|E_0(s) - E_j(s)|> 0$ between the ground and excited states during the annealing process yields a finite runtime $T_A$ to prepare ground states of $H(s)$  \cite{QAOA,ConvergenceProof}.  As $s$ approaches one these states approach the ground state of $C$, hence the quantum algorithm can solve the problem of identifying this state. The Perron-Frobenius theorem provides a sufficient, but not necessary, condition for a finite energy gap at $s<1$, which holds for $B$ in (\ref{transverse field}) and any $C$ that is diagonal in the computational basis \cite{ConvergenceProof}; it is also possible to guarantee a finite gap for alternative choices of $B$ that do not satisfy the conditions of the Perron-Frobenius theorem, and which have been considered in the context of ``warm-started" QAOA \cite{tate2023warm}.

\subsubsection{Quantum approximate optimization algorithm}

The quantum approximate optimization algorithm can be thought of as a discrete and variational generalization of adiabatic evolution \cite{QAOA}.  It is designed to run on quantum computers with discrete gates, as opposed to the continuous-time evolution assumed in annealing.  A QAOA circuit consists of $p$ layers of unitary evolution, where each layer alternates between Hamiltonian evolutions under $B$ and $C$ given by the unitaries
\begin{equation} U_B(\beta_l) = e^{-i \beta_l B}, \ \ \ U_C(\gamma_l) = e^{-i \gamma_l C}, \end{equation}
This produces a state
\begin{equation} \label{QAOA} \ket{\psi_p(\bm \beta, \bm \gamma)} = \prod_{l=1}^p U_B(\beta_l) U_C(\gamma_l) \ket{+}^{\otimes n} \end{equation} 
where $\bm \beta = (\beta_1, \ldots, \beta_p)$ and $\bm \gamma = (\gamma_1, \ldots, \gamma_p)$ are variational ``angle" parameters chosen to minimize $\langle C \rangle$. In the limit $p \to \infty$ the angles can be chosen so that the state evolution mimics a continuous adiabatic schedule (see Appendix \ref{QAOA annealing appendix}), which is guaranteed to converge to the optimal solution \cite{QAOA}. 

One limitation of the QAOA and quantum annealing approaches is that they search through the full $2^N$-dimensional Hilbert space, which may lead to difficulties in identifying feasible solutions in cases where the feasible subspace is small, while the majority of solutions are infeasible. The next approach is designed to address this difficulty.

\subsubsection{Quantum alternating operator ansatz}

The QAOA$^+$ describes a family of quantum algorithms that generalizes the quantum approximate optimization algorithm.  A primary goal of these algorithms is to directly optimize a constrained objective $\mathcal{C}(\bm z)$ by restricting the evolution to the feasible subspace $\mathcal{S}$, building on previous theoretical developments in quantum annealing \cite{hen2016quantum,hen2016driver,leipold2021constructing}. This is accomplished by replacing the transverse field $B$ by a suitably defined $B^+$ that maintains feasibility of an initially feasible state. The QAOA$^+$ evolution is then defined by 
\begin{equation} \label{QAOA+} \ket{\psi_p^+(\bm \beta, \bm \gamma)} = \prod_{l=1}^p U_{B^+}(\beta_l) U_{C}(\gamma_l) \ket{\psi_0^+}. \end{equation} 
A variety of different initial states $\ket{\psi_0^+}$ have been proposed, based on theoretical and practical considerations.  In this subsection we will focus on cases where $\ket{\psi_0^+}$ is taken as the ground state of $B^+$, since this is a necessary condition for the performance guarantee discussed in the next section. From a practical standpoint, it is also possible to choose a simpler state that is easier to prepare \cite{hadfield2019quantum}, though this nullifies the performance guarantee described in the next section and is known to lead to convergence issues \cite{tate2023bridging, cain2022qaoa}. 

As an example, an important body of literature has considered QAOA$^+$ for problems with Hamming weight constraints $\sum_i z_i = b$ \cite{wang2020xy,niroula2022constrained,cook2020quantum,he2023alignment,bartschi2020grover}.  The Hamming weight constraint can be enforced by taking $B^+ = B^{XY}$ with 
\begin{equation} B^{XY} = -\frac{1}{2}\sum_{i < j} X_i X_j + Y_i Y_j = -\sum_{i<j}  \ket{0_i, 1_j}\bra{1_i, 0_j} + \ket{1_i, 0_j}\bra{0_i, 1_j}.\end{equation}  
The Hamming weight of the state is preserved under the application of $B^{XY}$, hence an initial state satisfying a Hamming weight constraint will continue to satisfy the constraint throughout the evolution (\ref{QAOA+}).  The ground state of $B^{XY}$ is the Dicke state with Hamming weight $b$, which can be prepared efficiently using known circuit constructions \cite{aktar2022divide,bartschi2022short,bartschi2019deterministic}. 

The simplest theoretical implementation of QAOA$^+$ uses a unitary operator
\begin{equation} U_{B^{+}} = e^{-i \beta B^{+}},\end{equation}
which we refer to as the ``simultaneous" mixing unitary for $B^{+}$, following the terminology of Ref.~\cite{ConvergenceProof}.  In practice, a simultaneous mixer may involve complicated many-body Hamiltonian dynamics, as exemplified by $B^{XY}$, and it may not be practical to implement this directly in circuit.  In these cases it can be useful to decompose $B^+ = \sum_{j \in J} B^+_{(j)}$, such that each $e^{-i \beta B^+_{(j)}}$ can be implemented exactly in a circuit.  Then instead of $U_{B^+}$ the circuits can employ the more practical ``sequential mixer"
\begin{equation} \label{seq mixer} U^\mathrm{seq}_{B^{+}} =\prod_{j \in J} e^{-i \beta B^{+}_{(j)}}. \end{equation} 
The sequential mixer can be thought of as an approximation to the simultaneous mixer, since in general the $B_{(j)}^+$ do not commute, so $e^{-i \beta B^{+}} \neq \prod_{j \in J} e^{-i \beta B^{+}_{(j)}}$.

\subsection{Convergence guarantee in the adiabatic limit} \label{adiabatic limit}

Binkowski {\it et al.} \cite{ConvergenceProof} have recently derived conditions under which $p \to \infty$ convergence guarantees hold for quantum alternating operator ans\"atze. When these conditions hold, then the state can undergo an adiabatic evolution within the feasible solution subspace, similar to the evolution described in Sec.~\ref{adiabatic section} except with $B^+$ replacing $B$ and with an initial state $\ket{\psi_0^+}$ that is the ground state of $B^+$ \cite{hen2016quantum,hen2016driver,leipold2021constructing}. To obtain this guarantee it is necessary that $B^+$ can be decomposed as $B^+ = \sum_{j\in J} B_j$ where $\mathcal{F} = \{B_j\}$ satisfies properties of a ``mixing family" as defined in Ref.~\cite{ConvergenceProof} and summarized below. Similar properties for $B^+$ were described by Hadfield {\it et al.} in the original introduction of QAOA$^+$. 

\begin{definition}\label{def:mix-fam} Mixing Family (adapted from Ref.~\cite{ConvergenceProof})

Given a constrained optimization problem with a feasible solution space $\mathcal{S}$, a family of Hamiltonians $\mathcal{F} = \{B_j\}_{j \in J}$ is called a mixing family if it satisfies:
\begin{enumerate}[label=(\alph*)]
    \item Feasibility preservation: If $\ket{\bm z} \in \mathcal{S}$ and $\ket{\bm z'} \notin \mathcal{S}$, then $\bra{\bm z} B_j \ket{\bm z'}=0$ for each $B_j$.   \label{condition-a}
    \item Nonpositivity: The restriction of each $B_j$ to the solution subspace $\mathcal{S}$, denoted $B_j|_\mathcal{S}$, is component-wise nonpositive in the computational basis. \label{condition-b}
    \item Full mixing of solutions: The basis-state-transition graph with adjacency matrix $\sum_j B_j |_\mathcal{S}$ is fully connected. \label{condition-c}
\end{enumerate}
\end{definition}
Note that in contrast to Ref.~\cite{ConvergenceProof}, we have used a formulation in which we seek the ground state rather than the highest excited state.  Their equivalent condition (b) requires the $\tilde B_j$ in a mixing operator $\tilde B = \sum_{j \in J}\tilde B_j$ to be nonnegative in the computational basis (Def.~8 in \cite{ConvergenceProof}), such that a state evolving adiabatically under $H(t) = (1-s(t)){B^+}' + s(t)C'$, with $0 \leq s(t) \leq 1$, will transition from the initial highest-excited state of ${B^+}'$ to the final highest-excited state of $C'$. Defining $B^+ = -{ B^+}'$ and $C=-C'$, their condition is equivalent to evolving from the ground state of $B^+$ to the ground state of $C$, when $B^+= \sum_{j \in J}B_j$ with $B_j$ that are component-wise nonpositive as in (b) above. The final condition (c) is equivalent to the condition that $B^+$ is irreducible, as discussed near Definition 8 of Ref.~\cite{ConvergenceProof}. 

Conditions (a)-(c) of Def.~\ref{def:mix-fam} guarantee that any Hamiltonian $H = (1-s(t))B^+ + s(t)C$ with $0 \leq s < 1$, where $C$ is any operator that is diagonal in the computational basis, will have a nonzero energy gap between the ground state and first excited state, $E_1(s) - E_0(s) > 0$, due to the Perron-Frobenius theorem. From $E_1(s) - E_0(s) > 0$ we conclude the adiabatic timescale $T_A(s)$ is finite for $s <1$, hence a finite-time adiabatic path exists from the ground state of $B^+$ to the ground state of $C$, as discussed in Sec.~\ref{adiabatic section}. The QAOA$^+$ evolution can mimic the adiabatic evolution as $p\to \infty$, hence it is capable of preparing the ground state as $p \to \infty$. Detailed mathematical statements and proofs concerning the existence of this path are given in Ref.~\cite{ConvergenceProof}.

%%%%%%%%%%%%%%%%%%%%%%%%%%%%%%%%%%%%%%%%%%%%%%%%%%%%%%%%%%%%%%%%%%
%%%%%%%%%%%%%%%%%%%%%%%%%%%%%%%%%%%%%%%%%%%%%%%%%%%%%%%%%%%%%%%%%%
%%%%%%%%%%%%%%%%%%%%%%%%%%%%%%%%%%%%%%%%%%%%%%%%%%%%%%%%%%%%%%%%%%
\section{Quantum Optimization with a Linear Constraint Containing Sequential Coefficients} \label{linear constraints theory section}

In this section we devise a mixing family to solve linearly constrained problems of the class in Eq.~(\ref{sequential linear BILP single index}).  To do this it will sometimes be use an alternative index notation $``[\kappa,l]"$ that explicitly designates the value of the constraint coefficient $s_i = \kappa$ along with the second index $l$ signifying the $l$th occurrence of the constraint coefficient $\kappa$ up to a total number of occurrences $l_\kappa$,  
\begin{align} \label{sequential-linear-BILP} \bm z_\mathrm{opt} =\ & \underset{\bm z}{\mathrm{arg\ min}}\ \mathcal{C}(\bm z)  \nonumber\\
& \text{s.t.} \ \sum_{\kappa=1}^k \kappa \sum_{l=1}^{l_\kappa} z_{[\kappa,l]} = b \nonumber\\
& z_{[\kappa,l]} \in \{0,1\} \ \forall \kappa,l 
\end{align}
This notation allows us to visualize the constraint variables using a table, as follows:

\begin{equation}\label{eqn:table-notation}
    {\begin{NiceArray}{ccc|c|c|c|}
\cline{4-6}
& & 1 & z_{[1, 1]} & \ldots & z_{[1, l_1]} \\
\cline{4-6}
& & \ldots &  & & \\
\cline{4-6}
 \sum_{\kappa = 1}^{k} \kappa \sum_{l=1}^{l_\kappa} z_{[\kappa, l]}& \leftrightarrow & \kappa \leq k & z_{[\kappa, 1]} & \ldots & z_{[\kappa, l_\kappa]} \\
\cline{4-6}
 &  & \ldots &  && \\
\cline{4-6}
& & k & z_{[k, 1]} & \ldots & z_{[k, l_{k}]} \\
\cline{4-6}
\end{NiceArray}} \ .
\end{equation}

In this notation, each row of the table corresponds to a coefficient value, $1 \leq \kappa \leq k$, and each cell in that row represents an occurrence of a variable with coefficient $\kappa$. Note that each row need not have the same number of cells, as there may be different numbers of variables $l_\kappa$ with each coefficient $\kappa$. Below Eq.~(\ref{sequential linear BILP single index}) we noted requirements on the constraint variables that are necessary for our proofs.  In terms of the present indexing scheme, the requirement that the coefficient one appears at least twice translates to the requirement $l_1 \geq 2$, while the requirement that each unique value of $2 \leq \kappa \leq k$ appears at least once translates to the requirement that $l_\kappa \geq 1$ for $2 \leq \kappa \leq k$. We shall use the two index notations in (\ref{sequential linear BILP single index}) and (\ref{sequential-linear-BILP}) interchangeably in what follows, with the understanding that they are equivalent ways of expressing the constrained problems we consider.

We now present three illustrative example constraints, which will be useful for understanding the mixing operators that we will introduce soon after.

\subsubsection{Base constraint} \label{base constraint}

The minimal constraint satisfying the conditions with $k$ unique coefficient values is $\sum_{\kappa = 1}^k \kappa z_{[\kappa, 1]} + z_{[1,2]}$. Any constraint satisfying Eq.~(\ref{sequential-linear-BILP}) contains the variables used in the base constraint as a subset. The $k = 5$ base constraint is shown here:

\begin{equation}\label{eqn:base-constraint}
    \begin{NiceArray}{ccc|c|c}
\cline{4-5}
& & 1 & z_{[1, 1]} & \multicolumn{1}{c|}{z_{[1, 2]}}  \\
\cline{4-5}
& & 2 & z_{[2,1]}  & \\
\cline{4-4}
  z_{[1,1]} + 2z_{[2,1]} + 3z_{[3,1]} + 4z_{[4,1]} + 5z_{[5,1]} + z_{[1,2]}& \leftrightarrow & 3 & z_{[3, 1]}  \\
\cline{4-4}
 &  & 4 & z_{[4,1]} & \\
\cline{4-4}
& & 5 & z_{[5, 1]} &  \\
\cline{4-4}
\end{NiceArray} \ .
\end{equation}

We can also represent a solution using the table notation.  For example, for $b = 7$, the above constraint is satisfied by (among other solutions) $z_{[1,1]}=z_{[2,1]}=z_{[4,1]}=1$ with the remaining variables set to zero. We represent this via a quantum state or in table notation, 

\begin{equation}\label{eqn:base-constraint}
    \begin{NiceArray}{ccc|c|c}
\cline{4-5}
& & 1 & 1 & \multicolumn{1}{c|}{0}  \\
\cline{4-5}
& & 2 & 1  & \\
\cline{4-4}
  \ket{1_{[1,1]},1_{[2,1]},0_{[3,1]},1_{[4,1]},0_{[5,1]},0_{[1,2]}}& \leftrightarrow & 3 & 0  \\
\cline{4-4}
 &  & 4 & 1 & \\
\cline{4-4}
& & 5 & 0 &  \\
\cline{4-4}
\end{NiceArray} \ .
\end{equation}

\subsubsection{Repeated sequential constraint}

The repeated sequential constraint is the case where there is a fixed $l_{\kappa} = L \geq 2$ for all $\kappa$, representing the scenario where each coefficient appears the same number of times. Visually, this results in a rectangular table where each row contains exactly $L$ entries, as illustrated below for the $k=L=3$ case:

\begin{equation}\label{eqn:repeated-sequential-constraint}
    \begin{NiceArray}{ccc|c|c|c|}
\cline{4-6}
& & 1 & z_{[1, 1]} & {z_{[1, 2]}} & {z_{[1, 3]}}  \\
\cline{4-6}
\sum_{\kappa=1}^3 \sum_{l=1}^3 \kappa z_{[\kappa, l]}&\leftrightarrow & 2 & z_{[2,1]}  & z_{[2,2]} & z_{[2,3]}\\
\cline{4-6}
  & & 3 & z_{[3,1]}  & z_{[3,2]} & z_{[3,3]} \\
\cline{4-6}
\end{NiceArray} \ .
\end{equation}

\subsubsection{Unstructured constraint of type (\ref{sequential-linear-BILP})}

Apart from the containing the base constraint, the constraint variables may not necessarily follow any simple pattern, as in the following example. 

\begin{equation}\label{eqn:unstructured-constraint}
    \begin{NiceArray}{ccc|c|cc}
\cline{4-5}
& & 1 & z_{[1, 1]} & \multicolumn{1}{c|}{z_{[1, 2]}} & \\
\cline{4-5}
& & 2 & z_{[2,1]}  & & \\
\cline{4-6}
  &  & 3 & z_{[3, 1]} & \multicolumn{1}{c|}{z_{[3, 2]}} & \multicolumn{1}{c|}{z_{[3, 3]}}  \\
\cline{4-6}
 z_{[1,1]} + z_{[1,2]} + 2z_{[2,1]} + 3z_{[3,1]} + 3z_{[3,2]} + 3z_{[3,3]} + 4z_{[4,1]} + 5z_{[5,1]} + 6z_{[6,1]} + 6z_{[6,2]} & \leftrightarrow  & 4 & z_{[4,1]} & \\
\cline{4-4}
& & 5 & z_{[5, 1]} &  \\
\cline{4-5}
& & 6 & z_{[6, 1]} & \multicolumn{1}{c|}{z_{[6, 2]}} \\
\cline{4-5}
\end{NiceArray} \ .
\end{equation}

\subsection{Mixing operators for QAOA$^+$}

To develop a mixing family for this problem, we require operators that can transition between all feasible solutions consistent with the constraint in (\ref{sequential-linear-BILP}). For example, a constraint $z_1 + z_2 + 2z_3 = 2$ has solutions $\ket{1_1,1_2,0_3}$ and $\ket{0_1,0_2,1_3}$ and we require an operator to drive transitions between these solutions.  These solutions have different Hamming weights, so the $XY$ mixer cannot be employed. Instead we define ``$m$-to-1 merge" operators $M^{(m)}_{I, i\text{*}}$ to drive these transitions
\begin{equation}\label{merge operator}
M^{(m)}_{I, i\text{*}} 
= \ketbra{0_{i_1}, 0_{i_2}, \ldots 0_{i_{m}}, 1_{i\text{*}}}{1_{i_1}, 1_{i_2}, \ldots 1_{i_{m}}, 0_{i\text{*}}} 
+ \ketbra{1_{i_1}, 1_{i_2}, \ldots 1_{i_{m}}, 0_{i\text{*}}}{0_{i_1}, 0_{i_2}, \ldots 0_{i_{m}}, 1_{i\text{*}}},
\end{equation}
where $I = \{i_1, i_2, \ldots i_m\}$ is a set of qubits and $\sum_{i \in I} s_i = s_{i\text{*}}$ to maintain feasibility.  The ``$XY$"-mixer terms $\frac{1}{2}(X_iX_j + Y_iY_j) = \ket{0_i,1_j}\bra{1_i,0_j}$ correspond to merge operators with $m=1$, while operators with $m \geq 2$ are required for constraints with unequal coefficients $\kappa$ as in (\ref{sequential-linear-BILP}).  The Pauli representations for $M_{I,i*}^{(m)}$ are derived in Appendix \ref{sec:mixer-to-gates} and Ref.~\cite{fuchs2022constraint}, with a few examples in Table \ref{tab:merge}.  The multi-qubit Pauli operators $P_\alpha$ in the expansion of each $M_{I,i\text{*}}^{(m)}$ are mutually commutative, hence $\exp(-i \theta M^{(m)}_{I,i\text{*}})$ can be compiled without error as $\exp(-i \theta M^{(m)}_{I,i\text{*}}) = \prod_\alpha \exp(-i \theta P_\alpha)$ where the product runs over all $P_\alpha$ in the Pauli representation of $M^{(m)}_{I,i\text{*}}.$  More efficient compilations may also be possible.

\begin{table}[]
    \centering
    \begin{tabular}{c|c}
    operator $M_{I,i\text{*}}$ & Pauli representation \\
    \hline
    $M_{a,b}$     & $2^{-1}(X_a X_b + Y_a Y_b)$ \\
    $M_{\{a,b\},c}$ & $2^{-2}(X_aX_bX_c - Y_aY_bX_c + Y_a X_b Y_c + X_a Y_b Y_c)$\\
    $M_{\{a,b,c\},d}$ & $2^{-3}(X_aX_bX_cX_d - Y_aY_bX_cX_d - Y_a X_b Y_c X_d - X_a Y_b Y_c X_d + Y_a X_b X_c Y_d + X_a Y_b X_c Y_d + X_a X_b Y_c Y_d -Y_a Y_b Y_c Y_d)$\\
    \end{tabular}
    \caption{Pauli representations of few-body merge operators.}
    \label{tab:merge}
\end{table}

We now wish to use these merge operators to construct mixing families from Def.~\ref{def:mix-fam} for constrained problems in the class (\ref{sequential-linear-BILP}). 

To do this, we define two sets, suggestively named $\mathcal{F}_{\text{max}}$, the ``maximal $m$-to-1 mixing family", and $\mathcal{F}_{\text{min}}$, the ``minimal $m$-to-1 mixing family", or for brevity the maximal mixing family and minimal mixing family respectively.  We will prove later that these are mixing families. 
In fact, we will show that any set $\mathcal{F}$ such that $\mathcal{F}_{\text{min}} \subseteq \mathcal{F} \subseteq \mathcal{F}_{\text{max}}$ is a mixing family as well. 
As discussed later in Sec.~\ref{Adiabatic limit computations}, it can be advantageous to use a larger mixing family for greater connectivity at the cost of increased circuit depth.
These two distinguished families represent the extreme ends of this tradeoff and are intended to provide a framework in which a user can adjust connectivity as desired. 

The minimal mixing family only uses $m$-merge operators for $m=1$ and $m=2$. Thus, for brevity, we will sometimes denote $M^{(1)}_{\{i\},j}$ as $M_{i,j}$ and $M^{(2)}_{\{i_1,i_2\},i^*}$ as $M_{\{i_1,i_2\},i^*}$ when it is clear from context.

Let the maximal $m$-to-1 mixing family be

    \begin{equation} \label{fmax}
        \mathcal{F}_\text{max} = \{-M^{(m)}_{I, i\text{*}} : i\text{*} \in \{1, \ldots, N\} \ \forall m, I = \{i_1, \ldots i_m\}, \ \text{s.t.} \ i_j \neq i\text{*}, \ \sum_{i \in I}s_{i} = s_{i\text{*}} \},
    \end{equation}
which contains all possible $m$-to-1 merge operators $-M^{(m)}_{I, i\text{*}}$ consistent with a given constraint, $\sum_{i \in I}s_{i} = s_{i\text{*}}$. We also define the ``minimal $m$-to-1 mixing family" 
    \begin{align} \label{fmin}
    \mathcal{F}_\text{min} = &
    \{-M_{[\kappa, l],[\kappa, l + 1]} : \kappa \in \{1, \ldots, k\}, l \in \{1, \ldots l_\kappa - 1\}  \} \nonumber \\
    & \cup \{-M_{\{[1, 1], [\kappa, 1]\}, [\kappa + 1, 1]} : \kappa \in \{2, \ldots, k - 1\} \} \cup \{-M_{\{[1, 1], [1, 2]\}, [2, 1]}\} \ .
    \end{align}

The minimal $m$-to-1 mixing family consists of three types of terms: (i) $M_{[\kappa, l],[\kappa, l + 1]}$, (ii) $M_{\{[1, 1], [\kappa, 1]\},[\kappa + 1, 1]}$ and (iii) $M_{\{[1, 1], [1, 2]\}, [2, 1]}$. It has $\sum_\kappa (l_\kappa - 1) = n - k$ terms of type (i), $k - 2$ terms of type (ii) and a single term of type (iii) for a total size of $|\mathcal{F}_{\text{min}}| = n - 1$. We will show below and in Appendix~\ref{proof appendix} that the properties of our specific constraints and mixing families allow us to drive transitions between all solutions and obtain a systematic performance guarantee.

To develop an intuition for the operators included minimal mixing family, consider a base constraint with $k = 3$
\begin{equation}\label{eqn:simple-constraint}
    z_{[1,1]} + z_{[1,2]} + 2z_{[2,1]} + 3z_{[3,1]}
\end{equation}
whose minimal mixing family consists of one of each term:
\begin{equation}
    \{-M_{[1,1],[1,2]}, -M_{\{[1,1],[2,1]\},[3,1]}, -M_{\{[1,1],[1,2]\},[2,1]}\}.
\end{equation}
In this section, we use an asterisk $(*)$ as a placeholder for variables whose values are not relevant to the transition described. The role of these three types of terms is summarized as follows:

\begin{enumerate}[label={(\roman*)}]
    \item 
    
    Drives transitions between states where adjacent variables with the same coefficient are swapped.
    In terms of our example, this corresponds to $M_{[1,1],[1,2]}$, which swaps the values of $z_{[1,1]}$ and $z_{[1,2]}$ while leaving the total sum unchanged (e.g. $1(1) + 1(0) + 2(*) + 3(*) = 1(0) + 1(1) + 2(*) + 3(*)$)
    Visually, this is represented as swapping two neighboring cells within the same row of the table:

\[
\begin{NiceArray}{c|c|ccc|c|c}[cell-space-limits=3pt]
\cline{2-3}\cline{6-7}
1 & 1 & \multicolumn{1}{c|}{0} & \textcolor{graphorange}{\Block{3-1}{\xleftrightarrow[ \hspace{1cm} ]{M_{[1,1],[1,2]}}}} & 1 & 0 & \multicolumn{1}{c|}{1} \\
\cline{2-3}\cline{6-7}
2 & * & & & 2 & * & \\
\cline{2-2}\cline{6-6}
3 & * & & & 3 & * & \\
\cline{2-2}\cline{6-6}
\end{NiceArray}
\]

    \item Drives transitions between states where variables whose coefficients differ by one are swapped, preserving the constraint value by flipping the value of $z_{[1,1]}$. 
    In terms of our example, this corresponds to $M_{\{[1,1],[2,1]\},[3,1]}$, which sets $z_{[1,1]}$ and $z_{[2,1]}$ to zero in exchange for activating $z_{[3,1]}$ to preserve the sum (e.g. $1(1) + 1(*) + 2(1) + 3(0) = 1(0) + 1(*) + 2(0) + 3(1)$).
    Visually, this is represented shifting a cell with value 1 downwards by one row at the cost of setting the top left cell to zero.

\[
\begin{NiceArray}{c|c|ccc|c|c}[cell-space-limits=3pt]
\cline{2-3}\cline{6-7}
1 & 1 & \multicolumn{1}{c|}{*} & \textcolor{graphgreen}{\Block{3-1}{\xleftrightarrow[ \hspace{1cm} ]{M_{\{[1,1],[2,1]\},[3,1]}}}} & 1 & 0 & \multicolumn{1}{c|}{*} \\
\cline{2-3}\cline{6-7}
2 & 1 & & & 2 & 0 & \\
\cline{2-2}\cline{6-6}
3 & 0 & & & 3 & 1 & \\
\cline{2-2}\cline{6-6}
\end{NiceArray}
\]
    
    \item Drives transitions between states where two coefficient-one variables are exchanged for a coefficient-two variable.
    In terms of our example, this corresponds to $M_{\{[1,1],[1,2]\},[2,1]}$, which sets $z_{[1,1]}$ and $z_{[2,1]}$ to zero in exchange for turning on $z_{[3,1]}$ to preserve the sum (e.g. $1(1) + 1(1) + 2(0) + 3(*) = 1(0) + 1(0) + 2(1) + 3(*)$).
    Visually, this is represented as trading the first two cells in the first row for the leftmost cell in the second row:

    \[
\begin{NiceArray}{c|c|ccc|c|c}[cell-space-limits=3pt]
\cline{2-3}\cline{6-7}
1 & 1 & \multicolumn{1}{c|}{1} & \textcolor{graphblue}{\Block{3-1}{\xleftrightarrow[ \hspace{1cm} ]{M_{\{[1,1],[1,2]\},[2,1]}}}} & 1 & 0 & \multicolumn{1}{c|}{0} \\
\cline{2-3}\cline{6-7}
2 & 0 & & & 2 & 1 & \\
\cline{2-2}\cline{6-6}
3 & * & & & 3 & * & \\
\cline{2-2}\cline{6-6}
\end{NiceArray}
\]
\end{enumerate}

We now provide the reader with a bit of the intuition for why $\mathcal{F}_{\text{min}}$ forms a mixing family.
Key to QAOA$^+$ convergence is a mixing family's ability to connect all feasible states 
via a sequence of transitions, following Def.~\ref{def:mix-fam}\ref{condition-c}.
This is necessary to create a path from all nonoptimal solutions to the optimal ones. In the context of our problem, a mixing family must swap or exchange ones and zeroes within the table in a way that maintains feasibility and connects all feasible states.
Terms of type (i) allow adjacent variables to be swapped within a single row, while terms of type (ii) and (iii) allow for variables to be shifted between adjacent rows in exchange for flipping the value of $z_{[1,1]}$ to maintain feasibility.
Sequences of these interactions allow any solution to be reached from any other, as we show in general in Theorem~\ref{thm:mixer-families}, in examples of basis state transition graphs for a base constraints with all possible $b$ in Fig.~\ref{fig:base-constraint-graph}, and for additional examples in Figs.~\ref{fig:unstructured-graph}-\ref{fig:repeated-sequential-graph}.

We show full connectedness of solutions of an example problem  for all $b$ in Fig.~\ref{fig:base-constraint-graph}. Consider the $b=5$ case in Fig.~\ref{fig:base-constraint-graph}(e). Starting on the rightmost table of the figure, we have a variable $z_{[5,1]}$ with coefficient five, while to reach other feasible solutions we need to populate multiple variables with smaller coefficients in exchange for removing $z_{[5,1]}$.
This is accomplished using sequences of operators of types (i), (ii), and (iii). First, a type (ii) term allows us to replace $z_{[5,1]}$ with variables $z_{[4,1]}$ and $z_{[1,1]}$, maintaining the constraint. A type (i) term then rearranges the topmost row, so a similar process can be repeated with the new variable $z_{[4,1]}$.
Finally, a term of type (iii) allows us to replace $z_{[1,1]}$ and $z_{[1,2]}$ with $z_{[2,1]}$, freeing up space in the top row. Repeating these types of transitions, all solutions can be reached for the constraints we consider.
Critically, this procedure relies on the two conditions we impose on constraints of type Eq.~(\ref{sequential-linear-BILP}); The conditions $l_1 \geq 2$ and $l_k \geq 1$ when $k\geq 2$ ensures that we only need to shift coefficients between adjacent rows using operators of types (ii) and (iii), along with rearranging rows using operators of type (i), to reach all solutions. Similar sequences of transitions are visualized for different example constraints in Figs.~\ref{fig:unstructured-graph}-\ref{fig:repeated-sequential-graph}

For later examples, it will also be useful to consider restrictions of the maximal mixing family that do not include operators acting on more than $\mu$ qubits,

    \begin{equation} \label{fmumax}
        \mathcal{F}_{\mu-\text{max}} = \{-M^{(m)}_{I, i\text{*}} : i\text{*} \in \{1, \ldots, N\} \ \forall m \leq \mu-1, I = \{i_1, \ldots i_m\}, \ \text{s.t.} \ i_j \neq i\text{*}, \ \sum_{i \in I}s_{i} = s_{i\text{*}} \},
    \end{equation}
For circuit implementations the $\mathcal{F}_{\mu-\text{max}}$ can provide a useful intermediate between $\mathcal{F}_{\text{max}}$ and $\mathcal{F}_{\text{min}}$, as we shall consider in more detail later. They can be found by checking all subsets of $\mu$ qubits to see if their values can be changed in a way that satisfies the constraint; for fixed $\mu$, the time to determine $\mathcal{F}_{\mu-\text{max}}$ by checking all subsets scales polynomially in the number of qubits. 

The $\mathcal{F}_{\text{max}}$ and $\mathcal{F}_{\text{min}}$ are useful for proving the following theorem, which provides general guidelines to develop mixing families for our problem. In practice, it shows that circuits based on $\mathcal{F}_{\text{min}}$ obtain a convergence guarantee, while any additional constraint-preserving interactions (which define the set $\mathcal{F}_\text{max}$) can be added as desired while maintaining that guarantee.
\begin{theorem}\label{thm:mixer-families}
    Consider a problem instance of the form (\ref{sequential-linear-BILP}) with a feasible solution subspace $\mathcal{S}^{(b)}$ for an arbitrarily chosen $b$. Let $\mathcal{F}$ be a set of Hamiltonians such that $\mathcal{F}_{\text{min}} \subseteq \mathcal{F} \subseteq \mathcal{F}_\text{max}$. Then, $\mathcal{F}$ satisfies all technical conditions of Def.~\ref{def:mix-fam} and therefor is a mixing family .
\end{theorem}
\begin{proof}
    The full proof that all such $\mathcal{F}$ satisfy the mixing-family conditions from Def.~\ref{def:mix-fam} is given in Appendix \ref{proof appendix}; we present a sketch here. 
    
    The proofs of conditions (a) and (b) of Def. II.1 follow by construction. To prove condition (c), we show that the basis state interaction graph $G^{(N)}_{\mathcal{S}^{(b)}}(\mathcal{F}_{\text{min}}^{(N)})$ corresponding to the minimal $m$-to-1 mixing family $\mathcal{F}_\mathrm{min}^{(N)}$ is connected. We use the superscripts $(N)$ to denote the number of variables $z_i$, which we induct on to prove connectivity (condition (c)) for arbitrary problem instances. Since $\mathcal{F}_\text{max}$ contains $\mathcal{F}_\text{min}$, it follows that any $\mathcal{F}$ such that $\mathcal{F}_{\text{min}} \subseteq \mathcal{F} \subseteq \mathcal{F}_\text{max}$ also satisfies condition (c).

    The base case can be checked by hand. Then, in the inductive case, we assume that for a given instance of a problem (\ref{sequential-linear-BILP}) with $N - 1$ variables, its corresponding basis state interaction graph $G^{(N-1)}_{\mathcal{S}^{(b)}}$ is connected. We then take an arbitrary problem with $N$ variables and consider its basis interaction graph $G^{(N)}_{\mathcal{S}^{(b)}}$. We first partition the vertices into two disjoint subsets based on the value of the $N$th variable $z_N$, then consider the induced subgraphs formed by these vertex sets. We observe that for each of these subgraphs, we can construct an $N - 1$ variable problem  of the form (\ref{sequential-linear-BILP}) whose basis state interaction graph is the same as the subgraph, up to renaming of vertices. By the inductive hypothesis, each of these $N - 1$ variable problems must have connected basis state interaction graphs, which means that $G^{(N)}_{\mathcal{S}^{(b)}}$ contains at most two connected components. We then explicitly find a vertex in each of the two components with certain properties, related to the requirements $l_1 \geq 2$ and $l_\kappa \geq 1$ for $2 \leq \kappa \leq k$ in (\ref{sequential-linear-BILP}), and show that there must be an edge connecting these two vertices. Therefore, $G^{(N)}_{\mathcal{S}^{(b)}}$ is connected and condition (c) is satisfied. 
    
\end{proof}

\begin{figure*}[htbp]
\centering
\setlength{\tabcolsep}{5pt} % 
\subfloat[\centering $b = 1$]{
    \begin{tikzcd}[ampersand replacement=\&]
        \BaseConstraint{1}{0}{0}{0}{0}{0} \&\& \BaseConstraint{0}{1}{0}{0}{0}{0}
        \arrow["{M_{[1,1],[1,2]}}",color={graphorange}, tail reversed, from=1-1, to=1-3]
    \end{tikzcd}
}
\hspace{0.25\textwidth}
\subfloat[\centering $b = 2$]{
    \begin{tikzcd}[ampersand replacement=\&,column sep=large]
        \BaseConstraint{1}{1}{0}{0}{0}{0} \&\& \BaseConstraint{0}{0}{1}{0}{0}{0}
        \arrow["{M_{\{[1,1],[1,2]\},[2,1]}}",color={graphblue}, tail reversed, from=1-1, to=1-3]
    \end{tikzcd}
}

\subfloat[\centering $b = 3$]{
    \begin{tikzcd}[ampersand replacement=\&]
	\BaseConstraint{0}{1}{1}{0}{0}{0} \&\& \BaseConstraint{1}{0}{1}{0}{0}{0} \&\&\& \BaseConstraint{0}{0}{0}{1}{0}{0}
	\arrow["{M_{[1,1],[1,2]}}",color={graphorange}, tail reversed, from=1-1, to=1-3]
	\arrow["{M_{\{[1,1],[2,1]\},[3,1]}}",color={graphgreen}, tail reversed, from=1-3, to=1-6]
\end{tikzcd}
}

\vspace{3mm}

\subfloat[\centering $b = 4$]{
    \begin{tikzcd}[ampersand replacement=\&]
	\BaseConstraint{1}{1}{1}{0}{0}{0} \&\&\& \BaseConstraint{0}{1}{0}{1}{0}{0} \&\& \BaseConstraint{1}{0}{0}{1}{0}{0} \&\&\& \BaseConstraint{0}{0}{0}{0}{1}{0}
	\arrow["{M_{\{[1,1],[2,1]\},[3,1]}}",color={graphgreen}, tail reversed, from=1-1, to=1-4]
	\arrow["{M_{[1,1],[1,2]}}",color={graphorange}, tail reversed, from=1-4, to=1-6]
	\arrow["{M_{\{[1,1],[3,1]\},[4,1]}}",color={graphgreen}, tail reversed, from=1-6, to=1-9]
\end{tikzcd}
}

\vspace{3mm}

\subfloat[\centering $b = 5$]{
    \begin{tikzcd}[ampersand replacement=\&]
    \BaseConstraint{0}{0}{1}{1}{0}{0} \&\&\& \BaseConstraint{1}{1}{0}{1}{0}{0} \&\&\& \BaseConstraint{0}{1}{0}{0}{1}{0} \&\& \BaseConstraint{1}{0}{0}{0}{1}{0} \&\&\&  \BaseConstraint{0}{0}{0}{0}{0}{1}
	\arrow["{M_{\{[1,1],[1,2]\},[2,1]}}",color={graphblue}, tail reversed, from=1-1, to=1-4]
    \arrow["{M_{\{[1,1],[3,1]\},[4,1]}}",color={graphgreen}, tail reversed, from=1-4, to=1-7]
	\arrow["{M_{[1,1],[1,2]}}",color={graphorange}, tail reversed, from=1-7, to=1-9]
	\arrow["{M_{\{[1,1],[4,1]\},[5,1]}}",color={graphgreen}, tail reversed, from=1-9, to=1-12]
\end{tikzcd}
}

\vspace{3mm}

\subfloat[\centering $b = 6$]{
\begin{tikzcd}[ampersand replacement=\&]
	\BaseConstraint{1}{0}{0}{0}{0}{1} \&\&\& \BaseConstraint{0}{1}{0}{0}{0}{1} \\
	\BaseConstraint{0}{1}{1}{1}{0}{0} \&\&\& \BaseConstraint{1}{1}{0}{0}{1}{0} \\
	\BaseConstraint{1}{0}{1}{1}{0}{0} \&\&\& \BaseConstraint{0}{0}{1}{0}{1}{0}
	\arrow["{M_{[1,1],[1,2]}}",color={graphorange}, tail reversed, from=1-1, to=1-4]
	\arrow["{M_{\{[1,1],[4,1]\},[5,1]}}"',color={graphgreen}, tail reversed, from=1-4, to=2-4]
	\arrow["{M_{\{[1,1],[1,2]\},[2,1]}}"',color={graphblue}, tail reversed, from=2-4, to=3-4]
	\arrow["{M_{\{[1,1],[3,1]\},[4,1]}}"',color={graphgreen}, tail reversed, from=3-4, to=3-1]
    \arrow["{M_{[1,1],[1,2]}}",color={graphorange}, tail reversed, from=3-1, to=2-1]
\end{tikzcd}
}\hspace{0.01\textwidth}
\subfloat[\centering $b = 7$]{
\begin{tikzcd}[ampersand replacement=\&]
	\BaseConstraint{1}{1}{1}{1}{0}{0} \&\&\& \BaseConstraint{0}{1}{1}{0}{1}{0} \\
	\BaseConstraint{0}{0}{0}{1}{1}{0} \&\&\& \BaseConstraint{1}{0}{1}{0}{1}{0} \\
	\BaseConstraint{1}{1}{0}{0}{0}{1}\&\&\& \BaseConstraint{0}{0}{1}{0}{0}{1}
	\arrow["{M_{\{[1,1],[3,1]\},[4,1]}}",color={graphgreen}, tail reversed, from=1-1, to=1-4]
	\arrow["{M_{[1,1],[1,2]}}"',color={graphorange}, tail reversed, from=1-4, to=2-4]
	\arrow["{M_{\{[1,1],[2,1]\},[3,1]}}",color={graphgreen}, tail reversed, from=2-4, to=2-1]
	\arrow["{M_{\{[1,1],[4,1]\},[5,1]}}"',color={graphgreen}, tail reversed, from=2-4, to=3-4]
	\arrow["{M_{\{[1,1],[1,2]\},[2,1]}}"',color={graphblue}, tail reversed, from=3-4, to=3-1]
\end{tikzcd}
}\hfill
\subfloat[\centering $b = 8$]{
\begin{tikzcd}[ampersand replacement=\&]
	\BaseConstraint{1}{0}{1}{0}{0}{1} \&\& \BaseConstraint{0}{1}{1}{0}{0}{1} \\
	\BaseConstraint{0}{0}{0}{1}{0}{1} \&\& \BaseConstraint{1}{1}{1}{0}{1}{0} \\
	\BaseConstraint{1}{0}{0}{1}{1}{0} \&\& \BaseConstraint{0}{1}{0}{1}{1}{0}
	\arrow["{M_{[1,1],[1,2]}}",color={graphorange}, tail reversed, from=1-1, to=1-3]
	\arrow["{M_{\{[1,1],[4,1]\},[5,1]}}"',color={graphgreen}, tail reversed, from=1-3, to=2-3]
	\arrow["{M_{\{[1,1],[2,1]\},[3,1]}}"',color={graphgreen}, tail reversed, from=2-3, to=3-3]
	\arrow["{M_{\{[1,1],[2,1]\},[3,1]}}"',color={graphgreen}, tail reversed, from=1-1, to=2-1]
    \arrow["{M_{\{[1,1],[4,1]\},[5,1]}}"',color={graphgreen}, tail reversed, from=2-1, to=3-1]
    \arrow["{M_{[1,1],[1,2]}}",color={graphorange}, tail reversed, from=3-1, to=3-3]
\end{tikzcd}
}
    
\caption{Example basis state graphs of the minimal mixing family, for the $k = 5$ base constraint from Eq.~(\ref{eqn:base-constraint}) with $1 \leq b \leq 8$. In all cases, the basis state graph is connected, as required for the performance guarantees in QAOA$^+$ following Def.~\ref{def:mix-fam}\ref{condition-c}. Due to the symmetry of the constraint, the graph for a given $b$ is isomorphic to the graph for its complement $b' = \sum s_i - b$ by flipping the values of each $z_i$. Since $\sum_i s_i = 16$ in this example, these graphs characterize all values of $b$.}\label{fig:base-constraint-graph}
\end{figure*}
Theorem \ref{thm:mixer-families} provides multiple options for constructing mixing operators for QAOA$^+$ circuits solving linearly constrained problems of the form (\ref{sequential-linear-BILP}). The theorem shows they obtain a $p \to \infty$ performance guarantee from Sec.~\ref{adiabatic limit} when a suitable initial state is chosen.  We address initial state selection in the next subsection.

\subsection{Generalized initial states for QAOA$^+$ ans\"atze} \label{initial states}

Throughout this work we have placed an emphasis on satisfying technical conditions that are necessary for preparing the ground state with QAOA$^+$ in the adiabatic $p\to\infty$ limit.  So far this has focused on the construction of a mixing operator that satisfies the properties of Def.~\ref{def:mix-fam}.  A final necessary component is a suitable initial state.  To conform with the adiabatic limit of Sec.~\ref{adiabatic limit}, the initial state should be the ground state of $B^+$. The structure of this state can be reasoned about in terms of the basis-state-interaction graph of $B^+$, as described in Appendix \ref{initial state appendix}.  We find the initial states for our mixers are more complex than those considered in previous symmetric cases such as QAOA, or QAOA$^+$ with the $XY$-mixer.  We have been unable to identify efficient circuits for preparing these initial states. Instead, we propose an alternative route to the adiabatic limit that makes use of a third Hamiltonian, building on previous ``warm start" ideas that utilize a classically determined initial state while maintaining a performance guarantee \cite{tate2023warm,wurtz2021classically,hen2016quantum,hen2016driver,leipold2021constructing}. The approach is general, and can be applied to any QAOA$^+$ ansatz, which need not be related to the simple linear constraint (\ref{sequential-linear-BILP}). 

Our basic approach is to introduce a third Hamiltonian $A$ with an easy to prepare initial state.  We will choose $A$ such that we can demonstrate the existence of an adiabatic path from the ground state of $A$ to the ground state of $C$, with the operator $B^+$ playing an important role at intermediate evolution times, as described further below.

We would like the initial state of $A$ to be easy to prepare and to satisfy the relevant constraint in (\ref{sequential-linear-BILP}).  To accomplish this, we design $A$ based on the assumption that we know at least one feasible solution $\ket{\bm{z^o}} = \ket{(z_1^o, z_2^o, \ldots, z_N^o)}$. Here we are using a notation with $z^o_i \in \{0,1\}$, and with $Z_i \ket{0_i} = \ket{0_i}$ and $Z_i \ket{1_i} = -\ket{1_i}$. In practice $\ket{\bm z^o}$ could be a classically-determined ``warm-start" solution that approximately solves the problem \cite{tate2023warm,wurtz2021classically}, or it could be determined by any other means.
The general problem of finding a solution to $\sum_{i} s_i z_i = b$ for arbitrary integers $s_i$ and target sum value $b$, often referred to as the subset sum problem, is well-known to be NP-hard \cite{Karp1972}. However, our requirement that each integer 1 to $k$ must appear at least once allows a greedy algorithm to find a solution in linear time, as shown in Appendix~\ref{sec:initial-state-poly-time}. Hence, it is always possible to find a suitable initial state for the constraints we consider.

Following selection of an initial state, we then take
\begin{equation} \label{HA} A = \frac{1}{2}\sum_i \sigma(z_i^o)Z_i \end{equation}
where $\sigma(z) = 2z -1 \in \{1,-1\}$, so that $\ket{\bm{z}^o}$ is the unique ground state $A$. For example, if $z_i^o = 0$, then $\sigma(z_i^o) = 2\times 0 -1 = -1$, so $\ket{z_i^o}$ is the ground state of the corresponding component of $A$, since $\sigma(z_i^o) Z_i \ket{z^o_i} = -\ket{z^o_i}$. The prefactor 1/2 is an arbitrary constant, chosen to give unit energy level spacings in $A$. Including evolution with respect to $A$, we then propose the generalized QAOA$^+$ ansatz
\begin{equation} \label{generalized ansatz} \ket{\psi} = \left(\prod_{l=1}^p e^{-i \gamma_l C}e^{-i \beta_l B} e^{-i \alpha_l A}\right)\ket{\bm z^o} \end{equation}
This ansatz is capable of converging to the optimal solution of $C$ as $p\to \infty$. We will show this first in a case that is relatively easy to understand, then argue it holds for much more general cases after.  

An adiabatic path from the ground state of $A$ to the ground state of $C$ can be constructed from two adiabatic subpaths, as follows.  The first path uses $\gamma_l=0$ for $p_\gamma\to\infty$ layers, with QAOA evolution chosen to reproduce a continuous adiabatic path from the ground state of $A$ to the ground state of $B^+$. An adiabatic path between these states exists because $B^+$ is a mixing operator and $A$ is diagonal in the computational basis, hence the adiabatic Hamiltonian $H(s) =  s(t)B + (1-s(t))A$ has a nonzero energy gap $E_1-E_0>0$ for all $0 \leq s \leq 1$ (a finite gap at $s=0$ is guaranteed since $A$ has a unique ground state by construction), and therefore the adiabatic timescale $T_A$ related to Eq.~(\ref{TA(s)}) is finite, as described in Sec.~\ref{adiabatic limit} and Refs.~\cite{ConvergenceProof,Albash2018RMP}. The second path involves setting $\alpha_l=0$ for $p_\alpha\to\infty$ layers, with QAOA evolution chosen to reproduce a continuous adiabatic path from the ground state of $B^+$ to the ground state of $C$, which exists following Sec.~\ref{adiabatic limit}, similar to the first path.   Combining these paths we have an adiabatic path from the ground state of $A$ to the ground state of $C$, which can be traversed by QAOA$^+$ in the $p_\gamma + p_\alpha = p\to\infty$ limit.

It is actually not necessary to evolve to the ground state of $B^+$ to reach the ground state of $C$ adiabatically from the ground state of $A$. The essential point is that $\alpha(s) A + \gamma(s) C$ is diagonal in the computational basis, hence any Hamiltonian $H = \alpha(s) A + \beta(s) B^+ + \gamma(s) C$ has a nonzero energy gap $E_1 - E_0$ for $\beta > 0$, and a variety of adiabatic paths with varying $\alpha(s),\beta(s),$ and $\gamma(s)$ can be defined for finite-time transitions from the ground state of $A$ arbitrarily close to the ground state of $C$.  We will see examples of such paths in the following section. 

%%%%%%%%%%%%%%%%%%%%%%%%%%%%%%%%%%%%%%%%%%%%%%%%%%%%%%%%%%%%%%%%%%
%%%%%%%%%%%%%%%%%%%%%%%%%%%%%%%%%%%%%%%%%%%%%%%%%%%%%%%%%%%%%%%%%%
%%%%%%%%%%%%%%%%%%%%%%%%%%%%%%%%%%%%%%%%%%%%%%%%%%%%%%%%%%%%%%%%%%
\section{Simulation case study}  \label{Adiabatic limit computations}

In this section we present simulation results that exemplify our QAOA$^+$ implementation (\ref{generalized ansatz}) solving a problem of the type (\ref{sequential-linear-BILP}), its performance guarantee, and its relation to the adiabatic limit. The results have a limited scope that is meant to elucidate central ideas of our theory. Additional analyses of how it performs for varying constraints, choices of mixers, initial states, objective functions, parameter initializations, comparisons against conventional approaches, and other topics are all important for future work, but beyond our scope here.

We consider an adiabatic path directly from the ground state of $A$ to the ground state of $C$ with a time-dependent Hamiltonian
\begin{equation} \label{3 H adiabatic schedule} H(s) = \alpha(s)A + \beta(s)B + \gamma(s)C\end{equation}
where $s = t/T$ is the time, normalized by the total adiabatic time $T$ as discussed in Sec.~\ref{adiabatic section}. The annealing schedule begins with $\alpha(0) =1$ and with $\beta(0) = \gamma(0)=0$, so the initial state is the ground state of $H(0)=A$ from (\ref{HA}). Beyond this, the coefficients change with $s$, and approach $\gamma(1) =1$ and $\alpha(1)=\beta(1)=0$ at the final time $t=T$. Between these limits, we have a continuous schedule of $\alpha(s), \beta(s),$ and $\gamma(s)$, with $\beta(s) >0$ between the initial and final times, as needed to drive transitions between the solutions and maintain an energy gap $E_1-E_0>0$ in the adiabatic limit, following Sec.~\ref{adiabatic limit}.  A simple example schedule of this form is
\begin{equation} \label{annealing schedule} \alpha(s) = \frac{1-s}{1 + ks(1-s)} , \ \ \beta(s)=\frac{k s(1-s)}{1 + ks(1-s)}, \ \  \gamma(s)=\frac{s}{1 + ks(1-s)}, \end{equation} 
where we have normalized the coefficients so that $\alpha(s) + \beta(s) + \gamma(s) = 1$ for all $s.$ We choose $k=4$ in results that follow.

\begin{figure*}
    \centering
    \includegraphics[width=16cm,height=\textheight,keepaspectratio]{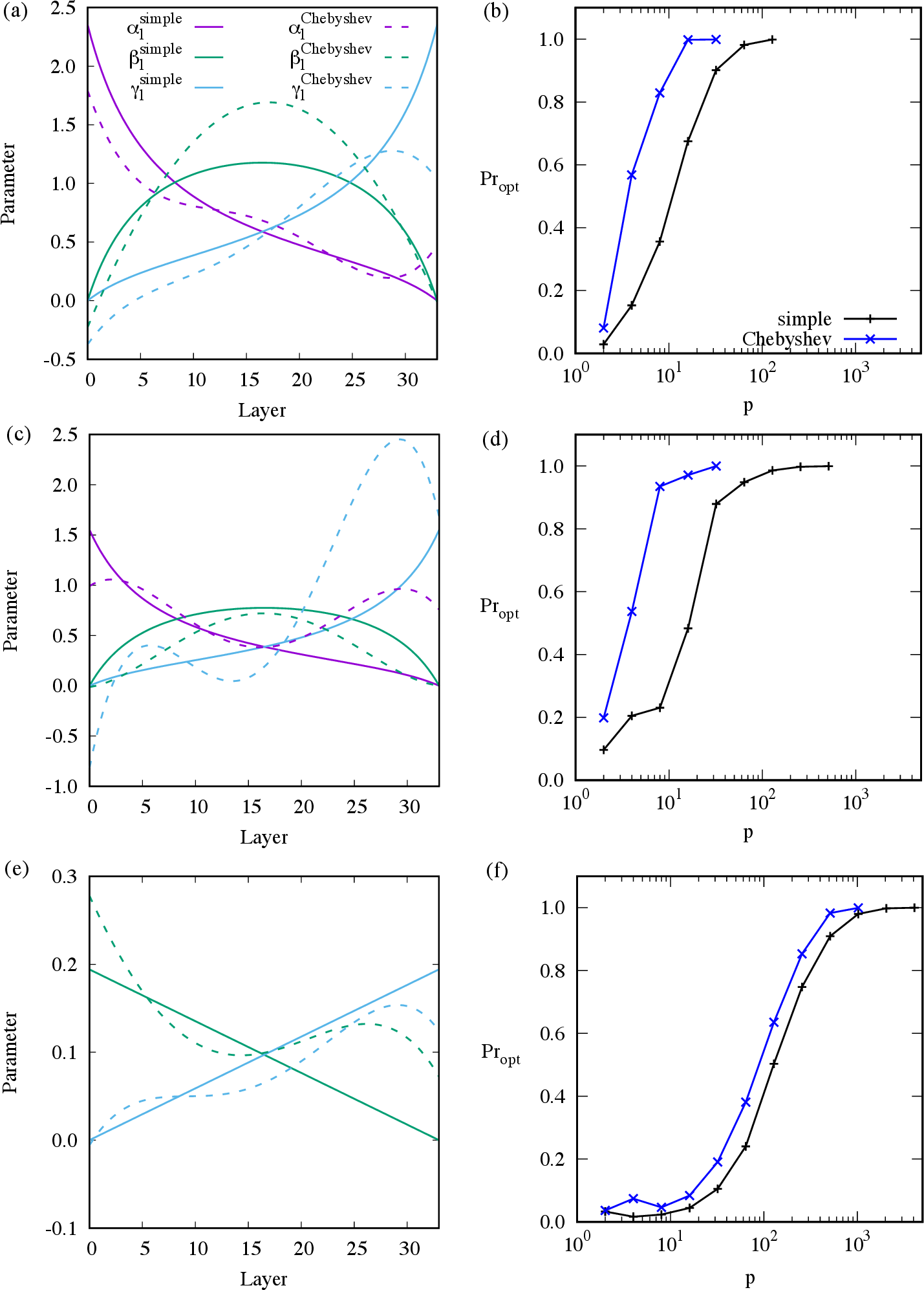}
    \caption{ (a) Parameter schedules at $p=32$ and with the 3-maximal $m$-to-1 mixing family for an example problem, where the $\theta^\mathrm{simple}_l = \theta_l\Delta t$ are parameter schedules from (\ref{annealing schedule}) with $\Delta t$ obtained from a fit to maximize the ground state overlap, and the dashed lines $\theta^\mathrm{Chebyshev}_l$ are from a similar fit to a fifth order Chebyshev polynomial. (b) The optimal solution probability converges to one as the number of QAOA layers $p$ increases; each curves terminates at the first circuit with optimal solution probability $\mathrm{Pr}_\text{opt} \geq 0.999$. (c) shows similar schedules obtained for the minimal $m$-to-1 mixing family and (d) shows its convergence with $p$, while (e,f) shows results for unconstrained optimization of the Lagriangian with QAOA. }
    \label{fig:QAOAperformance}
\end{figure*}

\begin{table}
    \begin{tabular}{|c|c|c|c|}
    \hline
    constraint coefficient $\kappa$ & $c_{[\kappa,1]}$ & $c_{[\kappa,2]}$ & $c_{[\kappa,3]}$ \\
    \hline
    1 & 1.181 & 0.640 & 1.840 \\
    \hline
    2 & 0.643 & 0.015 & 0.352 \\
    \hline
    3 & 2.633 & 0.696 & --- \\
    \hline
    \end{tabular}
    \caption{Variables used in the case study instance of Sec.~\ref{Adiabatic limit computations}, shown to three decimal places.}
    \label{case study variables}
\end{table}

We consider an example problem with $N=8$ qubits, with a linear objective 
\begin{equation} \mathcal{C}(\bm z) = \sum_{\kappa=1}^k\sum_{l=1}^{l_\kappa} c_{[\kappa,l]} z_{[\kappa,l]}, \end{equation}
where the coefficients $c_{[\kappa,l]}$ are generated uniformly at random in $[0,1]$ and normalized so their average is unity, and with constraint variables $\kappa \leq 3$, see Table \ref{case study variables}. The optimal solution state is 
\begin{equation} \ket{\bm z_\mathrm{opt}} = \ket{0_{[1,1]}, 1_{[1,2]}, 0_{[1,3]}, 0_{[2,1]}, 1_{[2,2]}, 1_{[2,3]}, 0_{[3,1]}, 1_{[3,2]}} \end{equation}
and we begin with an arbitrarily chosen initial state
\begin{equation} \ket{\bm z^o} = \ket{1_{[1,1]}, 1_{[1,2]}, 1_{[1,3]}, 0_{[2,1]}, 0_{[2,2]}, 1_{[2,3]}, 1_{[3,1]}, 0_{[3,2]}} \end{equation}

We analyze the approach to the ground state for both QAOA and adiabatic evolution, with two different choices for the mixing operator.  The first is based on the minimal $m$-to-1 mixing family $\mathcal{F}_\text{min}$ from Def.~\ref{fmin}, with a mixing Hamiltonian $B^+ = \sum_{B_j \in \mathcal{F}_\text{min}} B_j$. The second is a $3$-max $m$-to-1 mixing family $\mathcal{F}_{3-\text{max}}$ from (\ref{fmumax}), which is a truncation of the maximal $m$-to-1 mixing family that includes operators acting on at most three qubits at a time. For this mixing family we take $B^+ = N^{-1}\sum_{B_j \in \mathcal{F}_{3-\text{max}}} B_j$, where the prefactor  $N^{-1}$ is chosen heuristically to scale the energies in $B^+$, which produces better results in simulations that follow.   For adiabatic evolution we consider unitary dynamics under the time-dependent Hamiltonian (\ref{annealing schedule}) while for QAOA circuits we utilize sequential mixers (\ref{seq mixer}) based on the component operators $\sim B_j$. The circuits depend on the order in which the noncommuting sequential-mixer components are implemented; here we use the ordering that appears in a $3$-SWAP network \cite{o2019generalized} adapted from Ref.~\cite{swapnetwork}, which in principle can efficiently implement any set of 3-qubit gates on a linear quantum computing architecture.  

First we consider the simple annealing schedule of Eq.~(\ref{annealing schedule}).  We define QAOA angles corresponding to this evolution as $\alpha_l = \alpha(s_l) \Delta t$ where $s_l = l/(p+1)$ and $\Delta t$ is a Trotterized timestep as described in Appendix \ref{QAOA annealing appendix}, with similar expressions for $\beta_l$ and $\gamma_l$. All of these angles can be computed from the single parameter $\Delta t$, which we optimize to obtain the best performance.  We use initial seeds $\Delta t = 0.25, 0.5, \ldots, 10$ in local searches with the Scipy implementation of the BFGS algorithm, optimize each seed to identify a local optimum, then take the best-found optimum in our final results.  For the minimal $m$-to-1 mixing family we find these optimized $\Delta t$ are in the interval $1.1 \leq \Delta t \leq 1.9$ while for the 3-maximal $m$-to-1 mixing family $1.5 \leq \Delta t \leq 5.9$.  The angle parameters themselves take values $0 \lesssim \alpha_l \lesssim 2$, and similarly for $\beta_l$ and $\gamma_l$.

Solid lines in Fig.~\ref{fig:QAOAperformance}(a) show example angles we obtained from this procedure, with $B^+$ defined in terms of the 3-maximal $m$-to-1 mixing family.  The parameters are shown as a continuous curve; the exact values used in (\ref{generalized ansatz}) are points on the curve when the layer $l=1,2,\ldots,32$. 

The performance with schedules of this type is shown in Fig.~\ref{fig:QAOAperformance}(b) in black.  We obtain numerical convergence towards the ground state at $p=256$, with $\mathrm{Pr}_\text{opt} \geq 0.999$.  This demonstrates that a suitably defined adiabatic schedule can be used to parameterize a QAOA$^+$ implementation that converges to the ground state at large $p$, as expected.  

While the previous approach was successful in preparing a high-quality approximation to the ground state, it nonetheless required a large number of circuit layers, which is likely to be impractical on near-term quantum computing hardware.  One way to improve the performance is to allow the $\alpha_l, \beta_l,$ and $\gamma_l$ to vary in ways that are more general than what is allowed by (\ref{annealing schedule}). Rather than using a brute force search approach to identify optimized parameters, we focus on smooth schedules, which have a clear relation to an adiabatic limit and have been broadly successful in previous literature on QAOA \cite{zhou2020quantum,farhi2022quantum,lotshaw2023approximate,wurtz2022counterdiabaticity}.

We consider a parameterization of the angles in terms of Chebyshev polynomials, which are commonly used for function approximations in numerical analysis \cite{numericalrecipes}, with additional details in Appendix \ref{Chebyshev expansion}.    We use fifth-order Chebyshev polynomials to parameterize each of $\alpha(s), \beta(s), \gamma(s)$.  We initialize these parameters to approximate our previous optimized schedules based on (\ref{annealing schedule}), then take these initial points as input to a BFGS optimization run. From this optimization we obtain the parameters shown as dashed lines in Fig.~\ref{fig:QAOAperformance}(a).  These resemble the previous parameters from (\ref{annealing schedule}), but with larger values of $\beta_l$ at intermediate times, and with some oscillations in the other parameters.  The schedules also follow our expectations for an adiabatic schedule, with $\alpha(0) \gg \beta(0),\gamma(0)$ as expected for validity of the adiabatic initial state $\ket{\bm z^o}$, with $\beta(s)\neq 0$ at intermediate times as needed to drive transitions between solutions, and with $\gamma(1) \gg \alpha(1),\beta(1)$ as needed for the final adiabatic state to be the desired ground state $\ket{\bm z_\text{opt}}$ of $C$.  With this parameter choice we obtain numerical convergence to the ground state at $p=32$ as shown in blue in Fig.~\ref{fig:QAOAperformance}(b).  This demonstrates that a variety of smooth parameter schedules are capable of preparing the ground state with high probability, as expected from our analysis.  It also demonstrates that significant improvements in the convergence rate can be obtained through better parameterizations, such as those obtained in a Chebyshev polynomial expansion. 

Figures \ref{fig:QAOAperformance}(c),(d) show results analogous to Figs.~\ref{fig:QAOAperformance}(a),(b) but with $B^+ = \sum_{B_j \in F_\text{min}} B_j$ taken from the minimal $m$-to-1 mixing family (\ref{fmin}). This choice of $B^+$ generates fewer basis state transitions in each layer of QAOA, which leads to slower convergence when parameters are chosen from the schedule (\ref{annealing schedule}) (black curve Fig.~\ref{fig:QAOAperformance}(d)).  With the parameters taken as Chebyshev polynomials we find good numerical convergence to the ground state at $p=32$, with optimized parameters that vary more significantly from the simple schedule (\ref{annealing schedule}). These results confirm that a mixing operator derived from the minimial mixing family is capable of preparing the ground state in our approach, as expected. 

Figures \ref{fig:QAOAperformance}(e),(f) show analogous results with a standard Lagrangian formulation of QAOA. We choose a penalty parameter $\lambda = 1.5 \max(c_{[\kappa,l]})$, which ensures that flipping a bit in a feasible solution cannot yield an infeasible solution with a lower cost. We begin by considering a linear parameter schedule $\beta_l = \Delta t(1-l/(p+1))$ and $\gamma_l = \Delta tl/(p+1)$ as described in Appendix \ref{QAOA annealing appendix}. We optimize $\Delta t$ at each $p$, then use the optimized linear schedules as seeds for Chebyshev polynomial parameterizations, analogous to the QAOA$^+$ examples. The Chebyshev angles achieve better performance than the linear ones, and in each case the optimal solution probabilities converge to $\mathrm{Pr}_\text{opt} > 0.999$ for sufficiently many layers, as expected. In comparison, the QAOA$^+$ implementations achieve higher optimal solution probabilities for each number of layers $p$.

\begin{figure}
    \centering
\includegraphics[width=10cm,height=8cm,keepaspectratio]{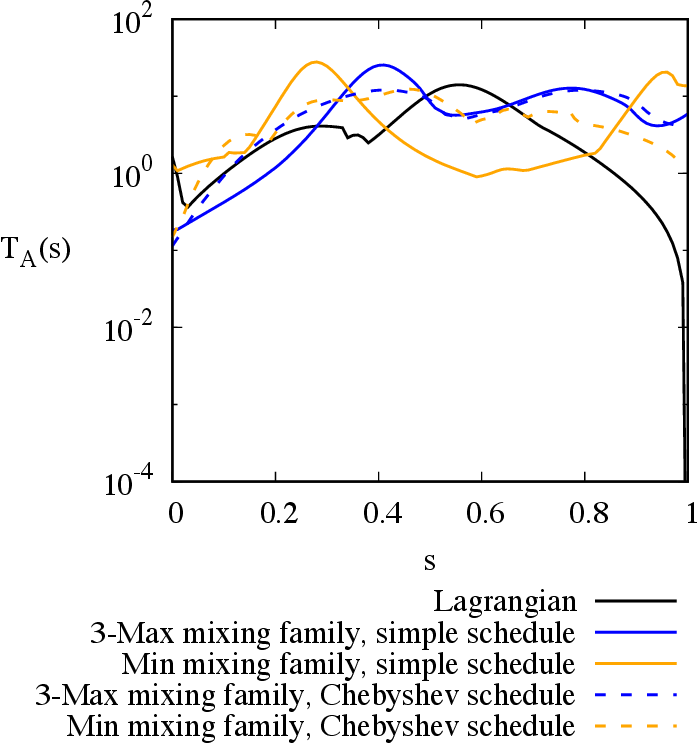}
    \caption{Instantaneous adiabatic timescales $T_A(s)$ from Eq.~(\ref{TA(s)})} for simple schedules (\ref{annealing schedule}), schedules derived from the $p=32$ Chebyshev polynomial fits in Fig.~\ref{fig:QAOAperformance}, and for QAOA with a simple linear schedule as in Sec.~\ref{adiabatic section}.  The $T_A(s)$ do not diverge, confirming the existence of an adiabatic path to the optimal solution, as expected by our theory. The Chebyshev angles suppress the largest values of $T_A(s)$, providing a partial explanation for their improved performance in Figs.~\ref{fig:QAOAperformance}(b),(d) relative to the simple schedule (\ref{annealing schedule}).
    \label{fig:adiabaticparam}
\end{figure}

It is interesting to consider the adiabatic evolution corresponding to the QAOA$^+$ implementations from Fig.~\ref{fig:QAOAperformance}. Recall the adiabatic evolution time depends on the adiabatic timescale $T_A$, which is the maximum over all the instantaneous timescales $T_A(s)$ from Eq.~(\ref{TA(s)}), as discussed in Sec.~\ref{adiabatic section}. To assess this, we computed time-dependent energy level spectra $\{E(s)\}$ for the lowest 20 energy eigenvalues during the adiabatic evolution (\ref{3 H adiabatic schedule}), using the Lanczos algorithm implementation from Ref.~\cite{pylanczos}. Using the spectra we first compute $T_A(s)$ from Eq.~(\ref{TA(s)}) for adiabatic evolution under parameter schedules (\ref{annealing schedule}) as solid lines in Fig.~\ref{fig:adiabaticparam}, with blue curves depicting annealing with $B^+$ derived from the 3-maximal $m$-to-1 mixing family (corresponding to Fig.~\ref{fig:QAOAperformance}(a),(b)) and with orange curves for the minimal $m$-to-1 mixing family (corresponding to Fig.~\ref{fig:QAOAperformance}(c),(d)).  We observe finite $T_A(s)$, corresponding to a nondegenerate ground state energy $E_1(s) - E_{0}(s) > 0$ in Eq.~(\ref{TA(s)}), as expected.  The dashed blue and orange lines show $T_A(s)$ with schedules proportional to the Chebyshev polynomials in Figs.~\ref{fig:QAOAperformance}(a) and \ref{fig:QAOAperformance}(c), respectively, normalized so $|\alpha(s)| +|\beta(s)| + |\gamma(s)|=1$; we do not include the optimized Chebyshev parameters from the Lagrangian formulation in Fig.~\ref{fig:QAOAperformance}(e) because the Chebyshev angles for the number of layers $p=32$ pictured there did not converge to the optimal solution during adiabatic evolution. The Chebyshev schedules for the constraint-preserving mixers tend to yield smaller maxima for the $T_A(s)$, corresponding to a shorter adiabatic runtime to prepare the optimal solution. This is consistent with the observed performance improvements from QAOA$^+$ in Figs.~\ref{fig:QAOAperformance}(b) and \ref{fig:QAOAperformance}(d) under the Chebyshev polynomial parameterization.  

In addition, the Chebyshev-polynomial-based schedules were not constrained to have the desired initial $\ket{\bm z^o}$ and final states $\ket{\bm z_\text{opt}}$ (ground states of $A$ and $C$ respectively), but we find the optimized parameters from QAOA$^+$ do indeed yield initial $\ket{\psi_0}$ and final $\ket{\psi_f}$ ground states of $H(s)$ with overlaps $|\bra{\psi_0} \bm z^o\rangle|, |\bra{\psi_f} \bm z_\text{opt}\rangle| > 0.999$, further supporting the consistency between QAOA$^+$ parameter schedules and adiabatic evolution.

\begin{figure*}
\includegraphics[width=\textwidth,height=12cm,keepaspectratio]{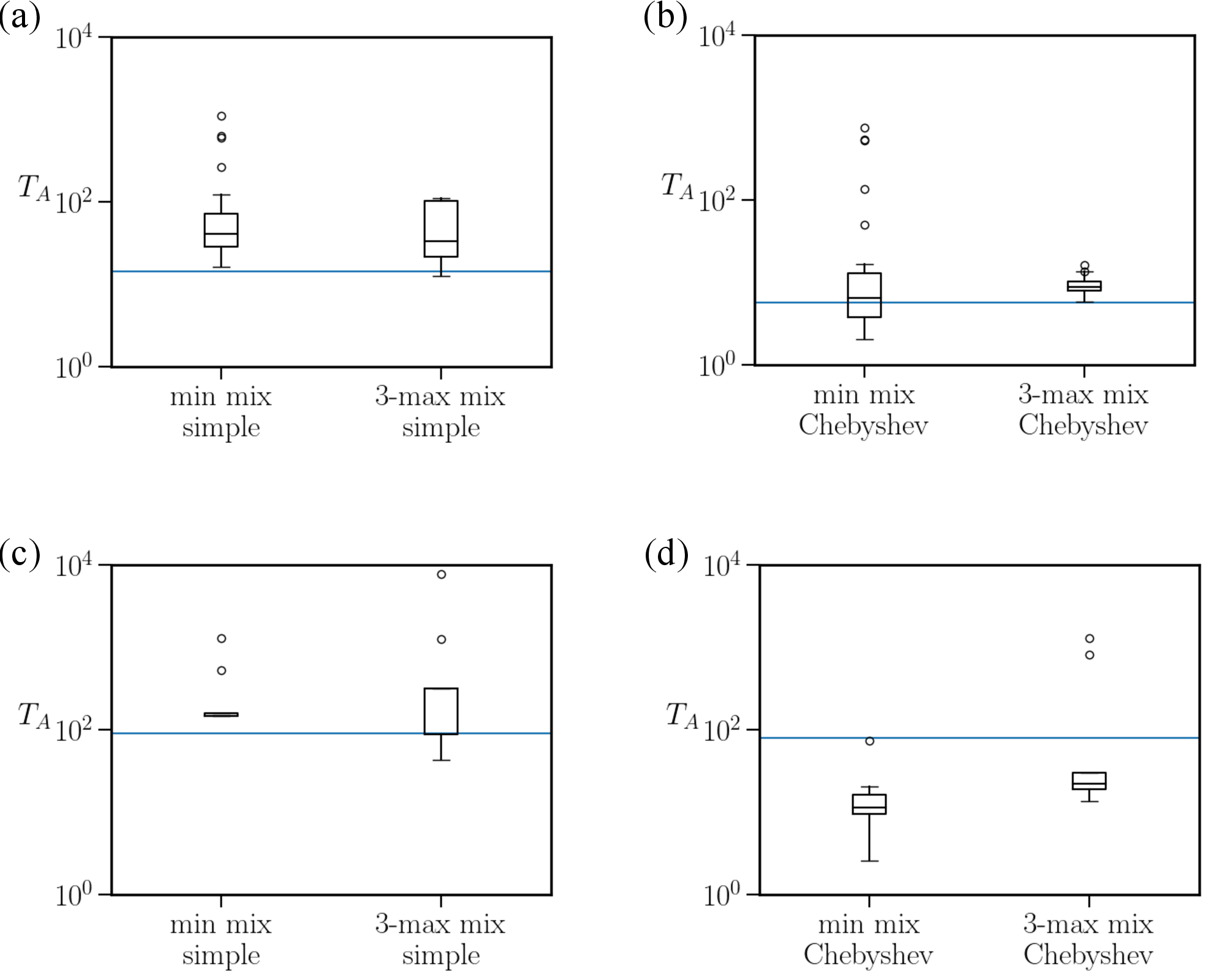}

\caption{Distribution of adiabatic timescales $T_A$ from from Eq.~(\ref{TA(s)})) over all feasible initial states apart from the optimal solution, for the instance in Table \ref{case study variables} with (a) the simple parameter schedule (\ref{annealing schedule}) and (b) optimized Chebyshev parameters.  Panels (c) and (d) show similar results for the instance in Table \ref{case study variables 2}. The line in the middle of each box shows the median $T_A$, the edges of the box show the interquartile rangles, the whiskers extending from the box show the farthest data point within a factor of 1.5 of the interquartile range from the box, and points indicate results outside of this range. Solid lines show $T_A$ from the Lagrangian formulation with a linear schedule or Chebyshev optimized parameters. In all cases the observed timescales are finite, consistent with the analytic theory.}
    \label{fig:timescale dists}
\end{figure*}

Finally, we consider the adiabatic timescales for different initial states and constraints. We demonstrate the timescale is finite in each case we consider, which indicates QAOA$^+$ obtains a performance guarantee, following reasoning from previous sections.  To do this we consider all feasible initial states for the problem summarized in Table \ref{case study variables}, as well as a different problem in Table \ref{case study variables 2} corresponding to a base constraint as in Sec.~\ref{base constraint}.  Figures \ref{fig:timescale dists}(a) and (c) show the adiabatic timescales for these problems, using the simple schedules from Eq.~(\ref{annealing schedule}).   The timescales are all finite, as expected.  We have checked the outliers correspond to avoided level crossings, which produce large $T_A$ due to a small but nonzero gap between the ground and excited state energies in Eq.~(\ref{TA(s)}). The very small range of $T_A$ for the minimal mixing family in Fig.~\ref{fig:timescale dists}(c) occurs because the maximum timescale happens at the end of the schedule during the final approach to the ground state, which yields a similarly large $T_A$ across different initial states. Overall the $T_A$ vary widely for different initial states and for different choices of the mixing families, and they do not always compare favorably to the timescale obtained in the Lagrangian formulation (solid line).  To better understand the behavior of the adiabatic timescales, we next optimized Chebyshev polynomial schedules, starting from angles that approximate the simple schedules, similar to before.  We minimize the objective $-F -F_0 + T_A/(10T_A^\text{simple})$, which minimizes the adiabatic timescale $T_A$ relative to its initial value $T_A^\text{simple}$ while maintaining approximately unit fidelities of the final $F = |\langle \psi(1)\ket{\bm z_\text{opt}}|^2$ and initial $F_0 = |\langle \psi(0)\ket{\varphi_0}|^2$ states, where $\ket{\varphi_0} = \ket{\bm z^o}$ for the constraint-preserving mixers and $\ket{\varphi_0} = \ket{+}^{\otimes n}$ for the Lagrangian formulation. The resulting schedules obtain lower $T_A$ in Fig.~\ref{fig:timescale dists}(b) and (d), which also compare more favorably between the constraint-preserving and Lagrangian formulations.  The results highlight the impact that different mixers and initial states can have on the results, which are topics worthy of further study, but beyond our scope here.  The main conclusion from Fig.~\ref{fig:timescale dists} is that the timescales are finite for different example initial states, constraints, and problems, which indicates that QAOA$^+$ must converge to the optimal solution for a sufficiently large number of layers, in agreement with our general analytic theory from Sec.~\ref{linear constraints theory section} as expected.

To summarize, in this section we considered a simulated case study of QAOA$^+$ solving a specific example instance.  This included demonstrations that QAOA$^+$ circuits (\ref{generalized ansatz}) with various parameters (Fig.~\ref{fig:QAOAperformance}(a)) can prepare the optimal solution with high fidelity as $p\to \infty$ (Fig.~\ref{fig:QAOAperformance}(b)), that a variety of mixing operators can be chosen for the circuits as expected from Theorem \ref{thm:mixer-families} (Figs.~\ref{fig:QAOAperformance}(c),(d)), that it is possible to translate back and forth between QAOA$^+$ circuits and adiabatic evolution (Figs.~\ref{fig:QAOAperformance}-\ref{fig:adiabaticparam}), and that varying initial states and constraints can be chosen while maintaining a performance guarantee (Fig.~\ref{fig:timescale dists}).  Hence we have obtained unified examples of the general physical picture and theoretical conclusions from Secs.~\ref{quantum opt section}-\ref{linear constraints theory section}, which also applies to cases beyond these examples.

\begin{table}
    \begin{tabular}{|c|c|c|c|}
    \hline
    constraint coefficient $\kappa$ & $c_{[\kappa,1]}$ & $c_{[\kappa,2]}$ \\
    \hline
    1 & 0.786 & 0.951  \\
    \hline
    2 & 1.093 & ---  \\
    \hline
    3 & 1.713 &  --- \\
    \hline
    4 & 1.251 &  --- \\
    \hline
    5 & 1.313 & ---  \\
    \hline
    6 & 0.652 & --- \\
    \hline
    7 & 0.241 &  ---  \\
    \hline
    \end{tabular}
    \caption{Problem instance considered in Fig.~\ref{fig:timescale dists}(b).}
    \label{case study variables 2}
\end{table}

\section{Conclusions}

We considered QAOA$^+$ implementations solving constrained optimization problems with a class of linear constraints containing sequential integer coefficients. (\ref{sequential-linear-BILP}).  Unlike QAOA implementations which utilize Lagrangian formulations for unconstrained optimization over both feasible and infeasible solutions, our QAOA$^+$ approach is designed to only populate feasible solutions, thereby limiting the search to the computationally-relevant subspace \cite{hadfield2019quantum}.  By considering QAOA$^+$ mixing operators in terms of graphs, and proving connectivity properties of the graphs, we were able to demonstrate that technical convergence criteria from Ref.~\cite{ConvergenceProof} were satisfied.  This led to families of QAOA$^+$ circuits solving nonsymmetric linearly constrained problems with provable performance guarantees as the number of circuit layers $p \to\infty$. The circuits generalize previous approaches for unconstrained optimization \cite{QAOA} and for constrained optimization with symmetric linear constraints \cite{wang2020xy}, while also maintaining enough structure in the constraint to avoid NP-hard optimization barriers that prevent analytic circuit construction for generic linearly constrained problems \cite{leipold2021constructing}. Our results provide explicit circuits with polynomialy many gates per layer that allow QAOA$^+$ to address more general linearly constrained problems while maintaining a performance guarantee. 

As a practical matter, if a linear constraint does not conform exactly to the structure we imposed, then it is possible to add dummy variables to reformulate the constraint in terms our structure, while adding penalty terms to the objective to penalize nonzero values for those variables.  This allows a broader set of linear constraints to be addressed, in an intermediate approach that combines aspects of the Lagrangian and constraint-preserving approaches.

Furthermore, for our circuits we used an optimization ansatz that begins from an arbitrary initial solution state, which is defined as the ground state of a nonstandard third Hamiltonian which is included in the ansatz.  This approach solves a real problem in devising a suitable initial state for solving nonsymmetric constrained problems, where circuits to prepare the ideal ``adiabatic initial states" of a mixing operator for QAOA$^+$ are unknown or impractical. Unlike these difficult to prepare states, our initial state is a simple computational basis state, which could represent an approximate solution to the problem as in previous ``warm start" approaches \cite{tate2023warm,wurtz2021classically,saleem2023approaches,maciejewski2024improving}, an algorithm we presented, or could be devised by any other means. Our approach to the initial state is general and could be applied to future QAOA$^+$ implementations solving a wide variety of constrained optimization problems.

With these circuits we demonstrated our performance guarantee is obtained for an example case-study problem, first employing a simple adiabatic-inspired schedule, then employing a significantly improved schedule based on Chebyshev polynomials, which required a significantly reduced depth for optimal solution probabilities $\mathrm{Pr}_\text{opt}>0.999$. This provided a concrete example of the theory, and confirmed the performance guarantee is obtained in an example, as expected. The Chebyshev parameterization has not been explored previously to our knowledge, and may also be useful in devising efficient and performant parameterizations for a variety of QAOA$^+$ implementations in the future. 

There are several interesting future directions that could follow from this work.  We have not performed a systematic analysis of the performance or resource costs of our ansatz, which is an obvious next step to characterizing its expected utility.  In a similar vein, it would be useful to obtain a systematic understanding of how the choice of initial state effects the performance of the ansatz, and to better understand the utility of the Chebyshev polynomial parameterization of the angle parameters. Perhaps most promisingly, we expect significant performance improvements can be obtained by developing counterdiabatic- \cite{chandarana2022digitized} and imaginary-time-evolution-based \cite{morris2024performant} extensions of our approach. 

A natural extension of our algorithm would be to perform multiple iterations with initial states $\ket{\bm z^o}$ that are taken as the best solution determined from all previously measured samples, in a similar spirit to Refs.~\cite{saleem2023approaches,maciejewski2024improving}. These updates are naturally incorporated into our framework while maintaining a performance guarantee, and may allow for higher-quality solutions to be obtained from low depth circuits. 

Finally, developing QAOA$^+$ ans\"atze with performance guarantees for problems with more general constraints, such as multiple linear constraints of the type we consider here, or constraints with more general structures, is an important open problem. 

\section{Acknowledgements}

We thank Rebekah Herrman, Ryan Bennink, and Titus Morris for feedback on the manuscript, and thank Bryan O'Gorman for technical assistance with implementing the generalized SWAP network.  We also thank Sarah Powers, Jim Ostrowski, George Siopsis, Rizwanal Alum, Nora Bauer, Sarah Chehade, Aidan Hennessey, Luke Choi, Pranav Singamsetty, and Heon Lee for helpful discussions. P.C.L was supported by DARPA ONISQ program under award W911NF-20-2-0051.  Manuscript revisions were completed with support by the U.S. Department of Energy, Office of Science, Office of Advanced Scientific Computing Research under field work proposal ERKJ445, 
 Award Number~89243024SSC000129.  This work was supported in part by the U.S. Department of Energy, Office of Science, Office of Workforce Development for Teachers and Scientists (WDTS) under the Science Undergraduate Laboratory Internships program.

\bibliographystyle{unsrt}
\bibliography{bibliography}

@misc{ConvergenceProof,
      title={Elementary Proof of QAOA Convergence}, 
      author={Lennart Binkowski and Gereon Koßmann and Timo Ziegler and René Schwonnek},
      year={2023},
      eprint={2302.04968},
      archivePrefix={arXiv},
      primaryClass={quant-ph}
}

@article{lotshaw2023approximate,
  title={Approximate Boltzmann distributions in quantum approximate optimization},
  author={Lotshaw, Phillip C and Siopsis, George and Ostrowski, James and Herrman, Rebekah and Alam, Rizwanul and Powers, Sarah and Humble, Travis S},
  journal={Physical Review A},
  volume={108},
  number={4},
  pages={042411},
  year={2023},
  publisher={APS}
}

@article{shaydulin2024evidence,
  title={Evidence of scaling advantage for the quantum approximate optimization algorithm on a classically intractable problem},
  author={Shaydulin, Ruslan and Li, Changhao and Chakrabarti, Shouvanik and DeCross, Matthew and Herman, Dylan and Kumar, Niraj and Larson, Jeffrey and Lykov, Danylo and Minssen, Pierre and Sun, Yue and others},
  journal={Science Advances},
  volume={10},
  number={22},
  pages={eadm6761},
  year={2024},
  publisher={American Association for the Advancement of Science}
}

@article{wurtz2022counterdiabaticity,
  title={Counterdiabaticity and the quantum approximate optimization algorithm},
  author={Wurtz, Jonathan and Love, Peter J},
  journal={Quantum},
  volume={6},
  pages={635},
  year={2022},
  publisher={Verein zur F{\"o}rderung des Open Access Publizierens in den Quantenwissenschaften}
}

@article{he2023alignment,
  title={Alignment between initial state and mixer improves QAOA performance for constrained optimization},
  author={He, Zichang and Shaydulin, Ruslan and Chakrabarti, Shouvanik and Herman, Dylan and Li, Changhao and Sun, Yue and Pistoia, Marco},
  journal={npj Quantum Information},
  volume={9},
  number={1},
  pages={121},
  year={2023},
  publisher={Nature Publishing Group UK London}
}

@inproceedings{cook2020quantum,
  title={The quantum alternating operator ansatz on maximum k-vertex cover},
  author={Cook, Jeremy and Eidenbenz, Stephan and B{\"a}rtschi, Andreas},
  booktitle={2020 IEEE International Conference on Quantum Computing and Engineering (QCE)},
  pages={83--92},
  year={2020},
  organization={IEEE}
}

@inproceedings{shaydulin2023qaoawith,
  title={QAOA with NP > 200},
  author={Shaydulin, Ruslan and Pistoia, Marco},
  booktitle={2023 IEEE International Conference on Quantum Computing and Engineering (QCE)},
  volume={1},
  pages={1074--1077},
  year={2023},
  organization={IEEE}
}

@article{pelofske2024short,
  title={Short-depth QAOA circuits and quantum annealing on higher-order ising models},
  author={Pelofske, Elijah and B{\"a}rtschi, Andreas and Eidenbenz, Stephan},
  journal={npj Quantum Information},
  volume={10},
  number={1},
  pages={30},
  year={2024},
  publisher={Nature Publishing Group UK London}
}

@inproceedings{pelofske2023quantum,
  title={Quantum annealing vs. QAOA: 127 qubit higher-order ising problems on NISQ computers},
  author={Pelofske, Elijah and B{\"a}rtschi, Andreas and Eidenbenz, Stephan},
  booktitle={International Conference on High Performance Computing},
  pages={240--258},
  year={2023},
  organization={Springer}
}

@article{ebadi2022quantum,
  title={Quantum optimization of maximum independent set using Rydberg atom arrays},
  author={Ebadi, Sepehr and Keesling, Alexander and Cain, Madelyn and Wang, Tout T and Levine, Harry and Bluvstein, Dolev and Semeghini, Giulia and Omran, Ahmed and Liu, J-G and Samajdar, Rhine and others},
  journal={Science},
  volume={376},
  number={6598},
  pages={1209--1215},
  year={2022},
  publisher={American Association for the Advancement of Science}
}

@article{harrigan2021quantum,
  title={Quantum approximate optimization of non-planar graph problems on a planar superconducting processor},
  author={Harrigan, Matthew P and Sung, Kevin J and Neeley, Matthew and Satzinger, Kevin J and Arute, Frank and Arya, Kunal and Atalaya, Juan and Bardin, Joseph C and Barends, Rami and Boixo, Sergio and others},
  journal={Nature Physics},
  volume={17},
  number={3},
  pages={332--336},
  year={2021},
  publisher={Nature Publishing Group UK London}
}

@article{brady2023iterative,
  title={Iterative quantum algorithms for maximum independent set: a tale of low-depth quantum algorithms},
  author={Brady, Lucas T and Hadfield, Stuart},
  journal={arXiv preprint arXiv:2309.13110},
  year={2023}
}

@article{brady2021behavior,
  title={Behavior of analog quantum algorithms},
  author={Brady, Lucas T and Kocia, Lucas and Bienias, Przemyslaw and Bapat, Aniruddha and Kharkov, Yaroslav and Gorshkov, Alexey V},
  journal={arXiv preprint arXiv:2107.01218},
  year={2021}
}

@article{bravyi2020obstacles,
  title={Obstacles to variational quantum optimization from symmetry protection},
  author={Bravyi, Sergey and Kliesch, Alexander and Koenig, Robert and Tang, Eugene},
  journal={Physical review letters},
  volume={125},
  number={26},
  pages={260505},
  year={2020},
  publisher={APS}
}

@article{crooks2018performance,
  title={Performance of the quantum approximate optimization algorithm on the maximum cut problem},
  author={Crooks, Gavin E},
  journal={arXiv preprint arXiv:1811.08419},
  year={2018}
}

@article{shaydulin2021classical,
  title={Classical symmetries and the quantum approximate optimization algorithm},
  author={Shaydulin, Ruslan and Hadfield, Stuart and Hogg, Tad and Safro, Ilya},
  journal={Quantum Information Processing},
  volume={20},
  pages={1--28},
  year={2021},
  publisher={Springer}
}

@article{farhi2020quantum,
  title={The quantum approximate optimization algorithm needs to see the whole graph: A typical case},
  author={Farhi, Edward and Gamarnik, David and Gutmann, Sam},
  journal={arXiv preprint arXiv:2004.09002},
  year={2020}
}

@article{zhu2022adaptive,
  title={Adaptive quantum approximate optimization algorithm for solving combinatorial problems on a quantum computer},
  author={Zhu, Linghua and Tang, Ho Lun and Barron, George S and Calderon-Vargas, FA and Mayhall, Nicholas J and Barnes, Edwin and Economou, Sophia E},
  journal={Physical Review Research},
  volume={4},
  number={3},
  pages={033029},
  year={2022},
  publisher={APS}
}

@article{moondra2024promise,
  title={Promise of Graph Sparsification and Decomposition for Noise Reduction in QAOA: Analysis for Trapped-Ion Compilations},
  author={Moondra, Jai and Lotshaw, Philip C and Mohler, Greg and Gupta, Swati},
  journal={arXiv preprint arXiv:2406.14330},
  year={2024}
}

@article{herrman2021globally,
  title={Globally optimizing QAOA circuit depth for constrained optimization problems},
  author={Herrman, Rebekah and Treffert, Lorna and Ostrowski, James and Lotshaw, Phillip C and Humble, Travis S and Siopsis, George},
  journal={Algorithms},
  volume={14},
  number={10},
  pages={294},
  year={2021},
  publisher={MDPI}
}

@article{ponce2023graph,
  title={Graph decomposition techniques for solving combinatorial optimization problems with variational quantum algorithms},
  author={Ponce, Moises and Herrman, Rebekah and Lotshaw, Phillip C and Powers, Sarah and Siopsis, George and Humble, Travis and Ostrowski, James},
  journal={arXiv:2306.00494},
  year={2023}
}

@article{lotshaw2023simulations,
  title={Simulations of frustrated Ising Hamiltonians using quantum approximate optimization},
  author={Lotshaw, Phillip C and Xu, Hanjing and Khalid, Bilal and Buchs, Gilles and Humble, Travis S and Banerjee, Arnab},
  journal={Philosophical Transactions of the Royal Society A},
  volume={381},
  number={2241},
  pages={20210414},
  year={2023},
  publisher={The Royal Society}
}

@article{shaydulin2023parameter,
  title={Parameter transfer for quantum approximate optimization of weighted maxcut},
  author={Shaydulin, Ruslan and Lotshaw, Phillip C and Larson, Jeffrey and Ostrowski, James and Humble, Travis S},
  journal={ACM Transactions on Quantum Computing},
  volume={4},
  number={3},
  pages={1--15},
  year={2023},
  publisher={ACM New York, NY}
}

@article{herrman2022multi,
  title={Multi-angle quantum approximate optimization algorithm},
  author={Herrman, Rebekah and Lotshaw, Phillip C and Ostrowski, James and Humble, Travis S and Siopsis, George},
  journal={Scientific Reports},
  volume={12},
  number={1},
  pages={6781},
  year={2022},
  publisher={Nature Publishing Group UK London}
}

@article{lotshaw2021empirical,
  title={Empirical performance bounds for quantum approximate optimization},
  author={Lotshaw, Phillip C and Humble, Travis S and Herrman, Rebekah and Ostrowski, James and Siopsis, George},
  journal={Quantum Information Processing},
  volume={20},
  number={12},
  pages={403},
  year={2021},
  publisher={Springer}
}

@article{akshay2020reachability,
  title={Reachability deficits in quantum approximate optimization},
  author={Akshay, Vishwanathan and Philathong, Hariphan and Morales, Mauro ES and Biamonte, Jacob D},
  journal={Physical review letters},
  volume={124},
  number={9},
  pages={090504},
  year={2020},
  publisher={APS}
}

@article{akshay2022circuit,
  title={Circuit depth scaling for quantum approximate optimization},
  author={Akshay, Vishwanathan and Philathong, H and Campos, E and Rabinovich, Daniil and Zacharov, Igor and Zhang, Xiao-Ming and Biamonte, Jacob D},
  journal={Physical Review A},
  volume={106},
  number={4},
  pages={042438},
  year={2022},
  publisher={APS}
}

@article{lotshaw2022scaling,
  title={Scaling quantum approximate optimization on near-term hardware},
  author={Lotshaw, Phillip C and Nguyen, Thien and Santana, Anthony and McCaskey, Alexander and Herrman, Rebekah and Ostrowski, James and Siopsis, George and Humble, Travis S},
  journal={Scientific Reports},
  volume={12},
  number={1},
  pages={12388},
  year={2022},
  publisher={Nature Publishing Group UK London}
}

@article{cerezo2021variational,
  title={Variational quantum algorithms},
  author={Cerezo, Marco and Arrasmith, Andrew and Babbush, Ryan and Benjamin, Simon C and Endo, Suguru and Fujii, Keisuke and McClean, Jarrod R and Mitarai, Kosuke and Yuan, Xiao and Cincio, Lukasz and others},
  journal={Nature Reviews Physics},
  volume={3},
  number={9},
  pages={625--644},
  year={2021},
  publisher={Nature Publishing Group UK London}
}

@article{niroula2022constrained,
  title={Constrained quantum optimization for extractive summarization on a trapped-ion quantum computer},
  author={Niroula, Pradeep and Shaydulin, Ruslan and Yalovetzky, Romina and Minssen, Pierre and Herman, Dylan and Hu, Shaohan and Pistoia, Marco},
  journal={Scientific Reports},
  volume={12},
  number={1},
  pages={17171},
  year={2022},
  publisher={Nature Publishing Group UK London}
}

@article{wang2020xy,
  title={XY mixers: Analytical and numerical results for the quantum alternating operator ansatz},
  author={Wang, Zhihui and Rubin, Nicholas C and Dominy, Jason M and Rieffel, Eleanor G},
  journal={Physical Review A},
  volume={101},
  number={1},
  pages={012320},
  year={2020},
  publisher={APS}
}

@article{basso2021quantum,
  title={The quantum approximate optimization algorithm at high depth for maxcut on large-girth regular graphs and the sherrington-kirkpatrick model},
  author={Basso, Joao and Farhi, Edward and Marwaha, Kunal and Villalonga, Benjamin and Zhou, Leo},
  journal={arXiv preprint arXiv:2110.14206},
  year={2021}
}

@article{hastings2021classical,
  title={A Classical Algorithm Which Also Beats $\frac{1}{2} + \frac{2}{\pi} \frac{1}{\sqrt{D}}$ For High Girth MAX-CUT},
  author={Hastings, Matthew B},
  journal={arXiv preprint arXiv:2111.12641},
  year={2021}
}

@article{farhi2022quantum,
  title={The quantum approximate optimization algorithm and the Sherrington-Kirkpatrick model at infinite size},
  author={Farhi, Edward and Goldstone, Jeffrey and Gutmann, Sam and Zhou, Leo},
  journal={Quantum},
  volume={6},
  pages={759},
  year={2022},
  publisher={Verein zur F{\"o}rderung des Open Access Publizierens in den Quantenwissenschaften}
}

@article{zhou2020quantum,
  title={Quantum approximate optimization algorithm: Performance, mechanism, and implementation on near-term devices},
  author={Zhou, Leo and Wang, Sheng-Tao and Choi, Soonwon and Pichler, Hannes and Lukin, Mikhail D},
  journal={Physical Review X},
  volume={10},
  number={2},
  pages={021067},
  year={2020},
  publisher={APS}
}

@article{brady2021optimal,
  title={Optimal protocols in quantum annealing and quantum approximate optimization algorithm problems},
  author={Brady, Lucas T and Baldwin, Christopher L and Bapat, Aniruddha and Kharkov, Yaroslav and Gorshkov, Alexey V},
  journal={Physical Review Letters},
  volume={126},
  number={7},
  pages={070505},
  year={2021},
  publisher={APS}
}

@article{lucas2014ising,
  title={Ising formulations of many NP problems},
  author={Lucas, Andrew},
  journal={Frontiers in physics},
  volume={2},
  pages={74887},
  year={2014},
  publisher={Frontiers}
}

@inproceedings{bartschi2022short,
  title={Short-depth circuits for Dicke state preparation},
  author={B{\"a}rtschi, Andreas and Eidenbenz, Stephan},
  booktitle={2022 IEEE International Conference on Quantum Computing and Engineering (QCE)},
  pages={87--96},
  year={2022},
  organization={IEEE}
}

@article{aktar2022divide,
  title={A divide-and-conquer approach to Dicke state preparation},
  author={Aktar, Shamminuj and B{\"a}rtschi, Andreas and Badawy, Abdel-Hameed A and Eidenbenz, Stephan},
  journal={IEEE Transactions on Quantum Engineering},
  volume={3},
  pages={1--16},
  year={2022},
  publisher={IEEE}
}

@inproceedings{bartschi2019deterministic,
  title={Deterministic preparation of Dicke states},
  author={B{\"a}rtschi, Andreas and Eidenbenz, Stephan},
  booktitle={International Symposium on Fundamentals of Computation Theory},
  pages={126--139},
  year={2019},
  organization={Springer}
}

@article{tate2023bridging,
  title={Bridging classical and quantum with SDP initialized warm-starts for QAOA},
  author={Tate, Reuben and Farhadi, Majid and Herold, Creston and Mohler, Greg and Gupta, Swati},
  journal={ACM Transactions on Quantum Computing},
  volume={4},
  number={2},
  pages={1--39},
  year={2023},
  publisher={ACM New York, NY}
}

@article{cain2022qaoa,
  title={The QAOA gets stuck starting from a good classical string},
  author={Cain, Madelyn and Farhi, Edward and Gutmann, Sam and Ranard, Daniel and Tang, Eugene},
  journal={arXiv preprint arXiv:2207.05089},
  year={2022}
}

@article{tate2023warm,
  title={Warm-Started QAOA with Custom Mixers Provably Converges and Computationally Beats Goemans-Williamson's Max-Cut at Low Circuit Depths},
  author={Tate, Reuben and Moondra, Jai and Gard, Bryan and Mohler, Greg and Gupta, Swati},
  journal={Quantum},
  volume={7},
  pages={1121},
  year={2023},
  publisher={Verein zur F{\"o}rderung des Open Access Publizierens in den Quantenwissenschaften}
}

@article{marsh2019quantum,
  title={A quantum walk-assisted approximate algorithm for bounded NP optimisation problems},
  author={Marsh, Samuel and Wang, JB},
  journal={Quantum Information Processing},
  volume={18},
  number={3},
  pages={61},
  year={2019},
  publisher={Springer}
}

@inproceedings{bartschi2020grover,
  title={Grover mixers for QAOA: Shifting complexity from mixer design to state preparation},
  author={B{\"a}rtschi, Andreas and Eidenbenz, Stephan},
  booktitle={2020 IEEE International Conference on Quantum Computing and Engineering (QCE)},
  pages={72--82},
  year={2020},
  organization={IEEE}
}

@article{perlin2024q,
  title={Q-CHOP: Quantum constrained Hamiltonian optimization},
  author={Perlin, Michael A and Shaydulin, Ruslan and Hall, Benjamin P and Minssen, Pierre and Li, Changhao and Dubey, Kabir and Rines, Rich and Anschuetz, Eric R and Pistoia, Marco and Gokhale, Pranav},
  journal={arXiv preprint arXiv:2403.05653},
  year={2024}
}

@article{herman2023constrained,
  title={Constrained optimization via quantum zeno dynamics},
  author={Herman, Dylan and Shaydulin, Ruslan and Sun, Yue and Chakrabarti, Shouvanik and Hu, Shaohan and Minssen, Pierre and Rattew, Arthur and Yalovetzky, Romina and Pistoia, Marco},
  journal={Communications Physics},
  volume={6},
  number={1},
  pages={219},
  year={2023},
  publisher={Nature Publishing Group UK London}
}

@article{wurtz2021classically,
  title={Classically optimal variational quantum algorithms},
  author={Wurtz, Jonathan and Love, Peter J},
  journal={IEEE Transactions on Quantum Engineering},
  volume={2},
  pages={1--7},
  year={2021},
  publisher={IEEE}
}

@book{numericalrecipes,
title={Numerical Recipes in Fortran 77: The Art of Scientific Computing},
author={William H. Press and Brian P. Flannery and Saul A. Teukolsky},
isbn={0-521-43064-X},
edition={second},
publisher={Cambridge University Press},
year={1996},
url={https://cambridge.org/us/universitypress/subjects/mathematics/numerical-recipes/numerical-recipes-fortran-77-art-scientific-computing-volume-1-2nd-edition?format=HB&isbn=9780521430647}
}

@misc{QAOA,
      title={A Quantum Approximate Optimization Algorithm}, 
      author={Edward Farhi and Jeffrey Goldstone and Sam Gutmann},
      year={2014},
      eprint={1411.4028},
      archivePrefix={arXiv},
      primaryClass={quant-ph}
}

@article{Albash2018RMP,
  title = {Adiabatic quantum computation},
  author = {Albash, Tameem and Lidar, Daniel A.},
  journal = {Rev. Mod. Phys.},
  volume = {90},
  issue = {1},
  pages = {015002},
  numpages = {64},
  year = {2018},
  month = {Jan},
  publisher = {American Physical Society},
  doi = {10.1103/RevModPhys.90.015002},
  url = {https://link.aps.org/doi/10.1103/RevModPhys.90.015002}
}

@misc{pylanczos,
note={https://pypi.org/project/pylanczos/}
}

@misc{swapnetwork,
author={Bryan O'Gorman},
note={https://github.com/quantumlib/Cirq/blob/v1.3.0/cirq-core/cirq/contrib/acquaintance/strategies/complete.py}
}

@article{o2019generalized,
  title={Generalized swap networks for near-term quantum computing},
  author={O'Gorman, Bryan and Huggins, William J and Rieffel, Eleanor G and Whaley, K Birgitta},
  journal={arXiv preprint arXiv:1905.05118},
  year={2019}
}

@article{morris2024performant,
  title={Performant near-term quantum combinatorial optimization},
  author={Morris, Titus D and Lotshaw, Phillip C},
  journal={arXiv preprint arXiv:2404.16135},
  year={2024}
}

@article{chandarana2022digitized,
  title={Digitized-counterdiabatic quantum approximate optimization algorithm},
  author={Chandarana, Pranav and Hegade, Narendra N and Paul, Koushik and Albarr{\'a}n-Arriagada, Francisco and Solano, Enrique and Del Campo, Adolfo and Chen, Xi},
  journal={Physical Review Research},
  volume={4},
  number={1},
  pages={013141},
  year={2022},
  publisher={APS}
}

@article{hadfield2019quantum,
  title={From the quantum approximate optimization algorithm to a quantum alternating operator ansatz},
  author={Hadfield, Stuart and Wang, Zhihui and O’gorman, Bryan and Rieffel, Eleanor G and Venturelli, Davide and Biswas, Rupak},
  journal={Algorithms},
  volume={12},
  number={2},
  pages={34},
  year={2019},
  publisher={MDPI}
}

@article{saleem2023approaches,
  title={Approaches to constrained quantum approximate optimization},
  author={Saleem, Zain H and Tomesh, Teague and Tariq, Bilal and Suchara, Martin},
  journal={SN Computer Science},
  volume={4},
  number={2},
  pages={183},
  year={2023},
  publisher={Springer}
}

@article{maciejewski2024improving,
  title={Improving Quantum Approximate Optimization by Noise-Directed Adaptive Remapping},
  author={Maciejewski, Filip B and Biamonte, Jacob and Hadfield, Stuart and Venturelli, Davide},
  journal={arXiv preprint arXiv:2404.01412},
  year={2024}
}

@article{fuchs2022constraint,
  title={Constraint preserving mixers for the quantum approximate optimization algorithm},
  author={Fuchs, Franz Georg and Lye, Kjetil Olsen and M{\o}ll Nilsen, Halvor and Stasik, Alexander Johannes and Sartor, Giorgio},
  journal={Algorithms},
  volume={15},
  number={6},
  pages={202},
  year={2022},
  publisher={MDPI}
}

@article{hen2016driver,
  title={Driver Hamiltonians for constrained optimization in quantum annealing},
  author={Hen, Itay and Sarandy, Marcelo S},
  journal={Physical Review A},
  volume={93},
  number={6},
  pages={062312},
  year={2016},
  publisher={APS}
}

@article{hen2016quantum,
  title={Quantum annealing for constrained optimization},
  author={Hen, Itay and Spedalieri, Federico M},
  journal={Physical Review Applied},
  volume={5},
  number={3},
  pages={034007},
  year={2016},
  publisher={APS}
}

@article{leipold2021constructing,
  title={Constructing driver Hamiltonians for optimization problems with linear constraints},
  author={Leipold, Hannes and Spedalieri, Federico M},
  journal={Quantum Science and Technology},
  volume={7},
  number={1},
  pages={015013},
  year={2021},
  publisher={IOP Publishing}
}

@article{farhi2000quantum,
  title={Quantum computation by adiabatic evolution},
  author={Farhi, Edward and Goldstone, Jeffrey and Gutmann, Sam and Sipser, Michael},
  journal={arXiv preprint quant-ph/0001106},
  year={2000}
}

@article{farhi2001quantum,
  title={A quantum adiabatic evolution algorithm applied to random instances of an NP-complete problem},
  author={Farhi, Edward and Goldstone, Jeffrey and Gutmann, Sam and Lapan, Joshua and Lundgren, Andrew and Preda, Daniel},
  journal={Science},
  volume={292},
  number={5516},
  pages={472--475},
  year={2001},
  publisher={American Association for the Advancement of Science}
}

@article{albash2018adiabatic,
  title={Adiabatic quantum computation},
  author={Albash, Tameem and Lidar, Daniel A},
  journal={Reviews of Modern Physics},
  volume={90},
  number={1},
  pages={015002},
  year={2018},
  publisher={APS}
}

@article{saleem2020max,
  title={Max-independent set and the quantum alternating operator ansatz},
  author={Saleem, Zain Hamid},
  journal={International Journal of Quantum Information},
  volume={18},
  number={04},
  pages={2050011},
  year={2020},
  publisher={World Scientific}
}

@Inbook{Karp1972,
author="Karp, Richard M.",
editor="Miller, Raymond E.
and Thatcher, James W.
and Bohlinger, Jean D.",
title="Reducibility among Combinatorial Problems",
bookTitle="Complexity of Computer Computations: Proceedings of a symposium on the Complexity of Computer Computations, held March 20--22, 1972, at the IBM Thomas J. Watson Research Center, Yorktown Heights, New York, and sponsored by the Office of Naval Research, Mathematics Program, IBM World Trade Corporation, and the IBM Research Mathematical Sciences Department",
year="1972",
publisher="Springer US",
address="Boston, MA",
pages="85--103",
isbn="978-1-4684-2001-2",
doi="10.1007/978-1-4684-2001-2_9",
url="https://doi.org/10.1007/978-1-4684-2001-2_9"
}
\appendix

\section{Relating QAOA and Annealing schedules} \label{QAOA annealing appendix}

Continuous adiabatic evolution is defined in terms of a time-ordered integral
\begin{equation} U = \mathcal{T} e^{-i\int_0^T dt H(t)}  = \mathcal{T} e^{-iT\int_0^1 ds H(s)} \end{equation}
where $\mathcal{T}$ is the time ordering operator, $H(t) =(1-s(t))B + s(t)C$ following (\ref{Ht}), $T$ is the total adiabatic evolution time, and where the final equality parameterizes the integral in terms of $dt=Tds$. Taking $s = l/N_\text{steps}$ we approximate the time-ordered integral as
\begin{equation} U \approx \mathcal{T} \prod_{l=1}^{N_\text{steps}} e^{-i H(l/N_\text{steps}) \Delta t} \end{equation}
where $\Delta t = T/N_\text{steps}$. Each step can be further Trotterized to yield separate terms for each component Hamiltonian
\begin{equation} \label{U Trotter QAOA} U \approx \mathcal{T} \prod_{l=1}^{N_\text{steps}} e^{-i \beta(s) \Delta t B} e^{-i \gamma(s) \Delta t C}. \end{equation}
The error is negligible in each approximation above when $\Delta t$ is sufficiently small. A QAOA evolution (\ref{QAOA}) can be defined directly from (\ref{U Trotter QAOA}) after relating $N_\text{steps}$ to the number of QAOA layers $p$. First note that for quantum annealing schedules the final Hamiltonian at $s=1$ is just the target Hamiltonian $C$, with $\beta(1)=0$ and $\gamma(1)=1$. In the Trotterized evolution (\ref{U Trotter QAOA}) this produces a term $e^{-i \beta(1) \Delta t B} e^{-i \gamma(1) \Delta t C}=e^{-i \Delta t C}$ when $l=N_\text{steps}$, and this term does not effect the computational basis populations because $C$ is diagonal in the computational basis.  Hence we can obtain the same computational results if we neglect the final term and take the sum out to $N_\text{steps}-1$ terms.  Each of these terms is then attributed to a single layer of QAOA, hence we take $p = N_\text{steps}-1$.  This produces
\begin{equation} \label{U Trotter QAOA final} U \approx \mathcal{T} \prod_{l=1}^{p} e^{-i \beta_l B} e^{-i \gamma_l C} \end{equation}
where $\beta_l = \beta(l/(p+1))\Delta t$, with $\Delta t = T/(p+1)$, and with a similar expression for $\gamma_l$. Similar reasoning also applies to QAOA$^+$ and adiabatic evolution with three Hamiltonians, for example in (\ref{generalized ansatz}).

\section{Pauli representations of merge operators}\label{sec:mixer-to-gates}

Here we consider the mixers $M_{I,i\text{*}}$ from Eq.~(\ref{merge operator}) expressed in the Pauli operator basis, as well as the corresponding circuits for the exponentiated operators $\exp(-i \theta M_{I,i\text{*}})$.  We consider $M_{I,i\text{*}}$ acting on the space of $|I|+1$ qubits $i \in I \cup \{i\text{*}\}$ with Hilbert space dimension $2^{|I|+1}$.  Corresponding expressions in the full $N$-qubit Hilbert space includes a tensor product with identity operators on the remaining qubit Hilbert spaces. 

Let $\sigma^{\alpha}_i$ denote a Pauli operator acting on qubit $i$ with superscript $\alpha \in \{X, Y, Z, \mathbb{1}\}$ denoting the specific Pauli operator. Let $\sigma^{\bm \alpha} = \bigotimes_{i=1}^N \sigma_i^{\alpha_i}$ denote a product Pauli operator that acts on all qubits. The set $\{\sigma^{\bm \alpha} \}$ of all product Pauli operators forms a basis of the $4^{|I|+1}$ dimensional Liouville space of operators that act on the $2^{|I|+1}$ dimensional Hilbert space.  Our $m$-to-1 merge operator can be expressed in terms of this basis as 
\begin{equation} M_{I,i\text{*}} = \frac{1}{2^{|I|+1}}\sum_{\bm \alpha} \mathrm{Tr}(\sigma^{\bm \alpha} M_{I,i\text{*}}) \sigma^{\bm \alpha} \end{equation}
Defining the shorthand notations $\ket{\bm z_{I,i\text{*}}} = \ket{0_{i_1}, 0_{i_2}, \ldots 0_{i_{m}}, 1_{i\text{*}}}$ and $\ket{\overline{\bm z_{I,i\text{*}}}}= \ket{1_{i_1}, 1_{i_2}, \ldots 1_{i_{m}}, 0_{i\text{*}}}$, the merge operator (\ref{merge operator}) is $M_{I,i\text{*}} = \ket{\bm z_{I,i\text{*}}} \bra{\overline{\bm z_{I,i\text{*}}}} + \ket{\overline{\bm z_{I,i\text{*}}}} \bra{\bm z_{I,i\text{*}}}$ and in the Pauli basis this gives

\begin{equation} \label{M Pauli 2}
M_{I,i\text{*}}=\frac{1}{2^{|I|}} \sum_{\bm \alpha} \text{Re}\left(\prod_{j=1}^{m}\bra{z_{i_j}}\sigma^{\alpha_j}_{i_j}\ket{\overline{z_{i_j}}}\right) \sigma^{\bm \alpha}
\end{equation}

Any term $\sigma^{\bm \alpha}$ containing an $\mathbb{1}_j$ or $Z_j$ operator will vanish in (\ref{M Pauli 2}) since $\bra{z_j}\mathbb{1}_{j}\ket{\overline{z_j}} = \bra{z_j}Z_{j}\ket{\overline{z_j}} = 0$. Similarly, terms with an odd number of $Y$ components will vanish when taking the real part.  This leaves terms containing $X_j$ and an even number of $Y_j$. These terms come in two types. The first type has $Y_i$ operators acting only on qubits $i \in I$. For $m$ Pauli $Y$ terms  these will have a coefficient $\prod_{j=1}^m \bra{0_j} Y_j \ket{1_j} = (-1)^{m/2}$. The second type has a term $Y_{i\text{*}}$ with a coefficient $\bra{1_{i\text{*}}} Y_j \ket{0_{i\text{*}}}\times \prod_{j=1}^{m-1} \bra{0_j} Y_j \ket{1_j} = -(-1)^{m/2}$. In both cases the coefficient is not affected by the Pauli $X$ part of the operator since $\bra{0_j}X_j\ket{1_j} = \bra{1_j}X_j\ket{0_j} = 1$.

Following the previous paragraph we can now obtain a simple description of $M_{I,i\text{*}}$ in terms of operators $\sigma^{\bm \alpha(m)}_I = \bigotimes_{i \in I} \sigma_i^{\alpha}$ with $m$ terms $\sigma_i^Y = Y_i$ and $|I|-m$ terms $\sigma_i^X = X_i$ in the product.    In terms of these operators $M_{I,i\text{*}}$ is
\begin{equation} \label{Merge Pauli} M_{I,i\text{*}} = \frac{1}{2^{|I|}} \sum_{m\ \mathrm{even}}^{|I|} (-1)^{m/2} \left( \sum_{\bm \alpha(m)} \sigma_I^{\bm \alpha(m)} X_{i\text{*}} - \sum_{\bm \alpha(m-1)} \sigma_I^{\bm \alpha(m-1)} Y_{i\text{*}} \right)\end{equation}
where $\sum_{\bm \alpha(m)}$ denotes the sum over all choices of $ \sigma_I^{\bm \alpha(m)}$ with $Y_i$ operators on different sets of $m$ qubits, and similarly for $\sum_{\bm \alpha(m-1)}$, with the understanding that $\sum_{\bm \alpha(m-1)}=0$ when $m=0$. The number of Pauli terms grows exponentially with $|I|$. A few examples are given in Table \ref{tab:merge}.  Example circuits that perform the multiple $X$ and $Y$ rotations are given in Figs.~\ref{fig:RXX-circ} and \ref{fig:RXY-circ}. 

The Pauli terms within the Pauli expansion of a given $m$ are mutually commutative, which allows the merge operators to be compiled from their Pauli expansions without error, as noted in the main text.  The source of the commutativity is that each term in the sum (\ref{Merge Pauli}) is a product of an even number of Pauli $Y$ operators along with a set of Pauli $X$ operators, as noted above. While the individual operators $X_i$ and $Y_i$ anticommute with one another, $X_iY_i = -Y_i X_i$, for pairs they will commute since $X_iX_jY_iY_j = (-1)^2 Y_i Y_j X_i X_j = Y_i Y_j X_i X_j$.  Each pair of operators in the expansion of each $M_{I,i\text{*}} $ will have an even number of these anticommutations, resulting in strings that commute, as follows.  First consider the base string containing only $X$.  It commutes with all other strings containing an even number of $Y$, by the reasoning above.  The simplest next case to consider is two different strings with two $Y$ operators each.  If the $Y$ operators are at four distinct locations, with $Y_iY_j$ on the first string and $Y_kY_l$ on the second string (with $i\neq j \neq k \neq l)$, then each $Y$ will anticommute with a corresponding $X$ on the other strings, resulting in an even number (four) of anticommutations. Hence, in this case the strings commute.  The other option is that the two strings each have $Y_i$ at a given location $i$, but that their other $Y$ operators are at separate locations $j\neq k$.  In this case the $Y$ at the same location $i$ in each string commute with each other because they are the same operator, while the $Y$ at different locations at each anticommute with $X$ on the other string, again resulting in an even number of anticommutations.  Hence, in the strings again commute, and combining with the previous case, we see that all strings with two $Y$ commute with each other.  Similarly reasoning applies when consider strings with greater numbers of $Y_i$ operators, with the result that only even numbers of anticommutations are possible, resulting in mutually commutative strings.

\begin{figure}

\[
\begin{array}{cc}
    \Qcircuit @C=1em @R=.7em {
\lstick{\ket{z_1}} & \ctrl{5} & \ctrl{3} & \ctrl{1} & \gate{R_X(\theta)} & \ctrl{1} & \ctrl{3} & \ctrl{5} & \qw & \\
\lstick{\ket{z_2}} & \qw  & \qw    & \targ &  \qw                & \targ    & \qw  & \qw    & \qw & \\
& & & & & & & & & \rstick{R_{X_1 X_2 \ldots X_m}(\theta)\ket{z_1 z_1 \ldots z_N}}\\
    \lstick{\ket{z_3}} & \qw & \targ    & \qw   &  \qw & \qw      & \targ      & \qw &  \qw & \\
   \cdots &&&& \cdots &&&&&& \\
 \lstick{\ket{z_m}}  &\targ    \qw & \qw   &    \qw   & \qw  & \qw & \qw & \targ & \qw & {\gategroup{1}{1}{6}{9}{1em}{\}}}
}
\end{array}
\]

\caption{Quantum circuit diagram for an $m$-qubit simultaneous rotation gate about Pauli-$X$.}
\label{fig:RXX-circ}
\end{figure}

\begin{figure}
    \centering
    \[
    \begin{array}{cc}
        \Qcircuit @C=1em @R=.7em {
    \lstick{\ket{z_1}} & \qw & \multigate{6}{R_{X_1 X_2 \ldots X_N}(\theta)} & \qw & \qw \\
    \lstick{\ket{z_2}} & \qw & \ghost{R_{X_1 X_2 \ldots X_N}(\theta)} & \qw & \qw \\
    &  \cdots& & \cdots & & \\
    & & & & & \rstick{R_{X_1 X_2 \ldots Y_i \ldots X_N}(\theta)\ket{z_1 z_2 \ldots z_N}}\\
    \lstick{\ket{z_i}} & \gate{R_Z(-\frac{\pi}{2})} & \ghost{R_{X_1 X_2 \ldots X_N}(\theta)} & \gate{R_Z(\frac{\pi}{2})} & \qw \\
    & \cdots & & \cdots &  &\\
    \lstick{\ket{z_N}} & \qw & \ghost{R_{X_1 X_2 \ldots X_N}(\theta)} & \qw & \qw {\gategroup{1}{1}{7}{5}{1em}{\}}} \\
}
    \end{array}
    \]

\caption{Quantum circuit diagram for an $(m-1)$-qubit simultaneous rotation gate about Pauli $X$-axis and a $1$-qubit rotation about Pauli $Y$-axis for qubit $i$. Applying $R_Z$ gates along the boundaries, for different qubits, allows arbitrary combinations of $X$ and $Y$ rotations. }
    \label{fig:RXY-circ}
\end{figure}

\section{Proof of Theorem \ref{thm:mixer-families}} \label{proof appendix}

Here we prove that the technical conditions of a mixing family from Def.~\ref{def:mix-fam} are satisfied by each $\mathcal{F}$ with $\mathcal{F}_\mathrm{min} \subseteq \mathcal{F} \subseteq \mathcal{F}_\mathrm{max}$, where $\mathcal{F}_\mathrm{min}$ and $\mathcal{F}_\mathrm{max}$ are defined in (\ref{fmin}) and (\ref{fmax}), respectively. The proof follows from lemmas demonstrating each condition (a)-(c), with the vast majority of complexity appearing in the proof of (c), as follows. 

\subsection{Proof that condition \ref{condition-a} of Def. \ref{def:mix-fam} is satisfied}

\begin{lemma}\label{lemma:no-edges-between}	

    Let $\mathcal{F}$ be a mixing family for a linearly constrained problem (\ref{sequential-linear-BILP}) with feasible subspace $\mathcal{S}$, such that $\mathcal{F} \subseteq \mathcal{F}_\mathrm{max}$. Then for $\ket{\bm z} \in \mathcal{S}$ and $\ket{\bm z'} \notin \mathcal{S}$, $\bra{\bm z} B_j \ket{\bm z'} = 0$.
\end{lemma}	

\begin{proof} 

    A matrix element of $B_j=M_{I,i*}$ is given by (\ref{merge operator}) as
    \begin{align} \bra{\bm z} M_{I,i*} \ket{\bm z'} = & \bra{\bm z} \left(\ketbra{0_{i_1}, 0_{i_2}, \ldots 0_{i_{m}}, 1_{i\text{*}}}{1_{i_1}, 1_{i_2}, \ldots 1_{i_{m}}, 0_{i\text{*}}}\otimes \mathbb{1}_{j \notin I,i*}  \right)\ket{\bm z'} \nonumber\\
    & + \bra{\bm z} \left(\ketbra{1_{i_1}, 1_{i_2}, \ldots 1_{i_{m}}, 0_{i\text{*}}}
    {0_{i_1}, 0_{i_2}, \ldots 0_{i_{m}}, 1_{i\text{*}}}\otimes \mathbb{1}_{j \notin I,i*}  \right)\ket{\bm z'},\end{align}
    where we have explicitly included the identity operator $\mathbb{1}_{j \notin I,i*}$ on the subspace of qubits that are unaffected by $M_{I,i*}$.  The matrix element $\bra{\bm z} M_{I,i*}\ket{\bm z'}$ is nonzero only when either $z_{i} = 0 \ \forall i \in I$ with $z_{i*}=1$ or $z_{i} = 1 \ \forall i \in I$ with $z_{i*}=0$, and $z_j = z_j'$ for $j \notin I \cup \{i\text{*}\}$.  Hence $z_i' = (1-z_i) \ \forall i \in I \cup \{i\text{*}\}$. Furthermore, for each $B_j = -M_{I,i*} \in \mathcal{F} \subseteq \mathcal{F}_\text{max}$, the $I$ and $i\text{*}$ are such that $\sum_{i\in I} s_i = s_{i^*}$ from the definition (\ref{fmax}) of $\mathcal{F}_\text{max}$. From these conditions it follows that $b = \sum_i s_i z_i = \sum_i s_i z_i'$ for each nonzero matrix element of each $B_j \in \mathcal{F}_\text{max}$. Hence $\ket{\bm z}$ and $\ket{\bm z'}$ satisfy the same constraint, so they are in the same feasible subspace $\ket{\bm z},\ket{\bm z'} \in \mathcal{S}$ when $\bra{\bm z} B_j \ket{\bm z'}$ is nonzero with $B_j \in \mathcal{F}$. Since the matrix element $\bra{\bm z} M_{I,i*}\ket{\bm z'}$ is nonzero only when $\ket{\bm z}$ and $\ket{\bm z'}$ are in the same feasible solution space, it follows that if $\ket{\bm z}$ and $\ket{\bm z'}$ are not in the same feasible solution subspace $\mathcal{S}$, then $\bra{\bm z} B_j \ket{\bm z'}=0$ for each $B_j \in \mathcal{F}$. 
\end{proof}	

\subsection{Proof that condition \ref{condition-b} of Def.~\ref{def:mix-fam} is satisfied}

\begin{lemma}\label{lemma:non-negative}
Let $\mathcal{F}$ be a generic mixing family, following Def.~\ref{def:mix-fam}, for a linear constraint (\ref{sequential-linear-BILP}) with feasible subspace $\mathcal{S}$. Then $\forall B_j \in \mathcal{F},  B_j|_{\mathcal{S}}$ is component-wise nonpositive in the computational basis.
\end{lemma}

\begin{proof}
 Each operator $B_j = -M^{(m)}_{I, i\text{*}} \in \mathcal{F}$ has nonpositive matrix elements (of zero and negative one) in the computational basis, by the definition Eq.~(\ref{merge operator}).
\end{proof}

\subsection{Proof that condition \ref{condition-c} of Def.~\ref{def:mix-fam} is satisfied}

\begin{lemma}\label{lemma:connected}	
    Let $G_{\mathcal{S}^{(b)}}(\mathcal{F}) = (V_{\mathcal{S}^{(b)}}, E_{\mathcal{S}^{(b)}})$ be a basis-state-interaction graph associated with a mixing family $\mathcal{F}$ and a linearly constrained problem (\ref{sequential-linear-BILP}) with feasible solution subspace $\mathcal{S}^{(b)}$. The vertex set $V_{\mathcal{S}^{(b)}}$ contains vertices for each feasible solution $\ket{\bm z} \in \mathcal{S}^{(b)}$ of a linearly constrained problem (\ref{sequential-linear-BILP}), while the edge set $E_{\mathcal{S}^{(b)}}$ contains edges $(\ket{\bm z},\ket{\bm z'})$ between vertices $\ket{\bm z},\ket{\bm z'} \in \mathcal{S}^{(b)}$ whenever there exists a $B_j \in \mathcal{F}$ such that $\bra{\bm z} B_j \ket{\bm z'}\neq 0$. Let $\mathcal{F}_{min} \subseteq \mathcal{F} \subseteq \mathcal{F}_{max}$, with $\mathcal{F}_\text{min}$ and $\mathcal{F}_\text{max}$ the minimal and maximal mixing families from (\ref{fmin}) and (\ref{fmax}), respectively. Then, $G_{\mathcal{S}^{(b)}}(\mathcal{F})$ is connected $\forall b \in \mathbb{N}$.
\end{lemma}	

\begin{proof}

We will prove the lemma by induction on the number $N$ of variables $z_i$ in the constraint. We consider the basis state interaction graph $G^{(N)}_{\mathcal{S}^{(b)}}(\mathcal{F}_{\text{min}}^{(N)}) = (V^{(N,b)}, E^{(N,b)}_{\text{min}})$ associated with the minimal $m$-to-1 mixing family $\mathcal{F}_\mathrm{min}^{(N)}$, since connectivity of this graph implies connectivity of the graph for any other $\mathcal{F}^{(N)}$ with $\mathcal{F}_\mathrm{min}^{(N)} \subseteq \mathcal{F}^{(N)} \subseteq \mathcal{F}_\mathrm{max}^{(N)}$. This is because these other graphs simply add edges to the graph from $\mathcal{F}_\text{min}^{(N)}$. For clarity, throughout this proof we carry superscripts $(N)$ to explicitly denote the number of variables, and we use shorthand to denote the $N$-variable problem graph with a constraint equating to $b$ as $G^{(N,b)} \equiv G^{(N)}_{\mathcal{S}^{(b)}}(\mathcal{F}_\text{min}^{(N)}).$

We begin with the base case, with the constraint $z_{[1,1]} + z_{[1,2]} = b$. This constraint contains the fewest variables possible, since variables with coefficient $s_i=1$ must occur at least twice (see text immediately below Eq.~(\ref{sequential-linear-BILP})). The mixing family is $\mathcal{F}_\mathrm{min}^{(2)} = \{M_{[1,1],[1,2]}\}$. For $b \notin \{0,1,2\}$ there are no solutions and the graph is trivially connected.  For $b \in \{0,1,2\}$ we have:
\begin{itemize}
    \item $b=0$: $V^{(0)} = \{\ket{00}\}$, so $G^{(2,b)}$ is trivially connected.
    \item $b=1$: $V^{(1)} = \{\ket{01}, \ket{10}\}$ with $M_{[1,1],[1,2]}\ket{10} = \ket{01} \implies \big( \ket{10}, \ket{01} \big) \in E^{(1)}_{\mathrm{min}}$ so $G^{(2,b)}$ is connected.
    \item $b=2$: $V^{(2)} = \{\ket{11}\}$, so $G^{(2,b)}$ is trivially connected.
\end{itemize} 
This proves the base case.

For the inductive step, suppose that all graphs $G^{(N-1,b)}$ associated with an $(N - 1)$-variable linearly constrained problem as in (\ref{sequential-linear-BILP}) are connected. Now consider an arbitrary $N$-variable linearly constrained problem as defined in (\ref{sequential-linear-BILP}) together with a mixing family  
$\mathcal{F}_\mathrm{min}^{(N)}$ which is associated with a graph $G^{(N,b)}$. 

Consider a copy of this problem which is identical except that the $N$th variable is removed. Without loss of generality, assume $z_N = z_{[k, l_k]}$, meaning that this variable has coefficient $k$. Here, we again utilize the index notation $z_{[\kappa,l]}$ from (\ref{sequential-linear-BILP}). With this notation, a bitstring can be visualized as a table of bits as seen in Eq. \ref{eqn:table-notation}, where bits with the same coefficient $\kappa$ are shown in the same row, while the columns run over all $l_\kappa$ bits with that coefficient. The array is shown as rectangular for simplicity, but it is important to note that different rows can have different numbers of columns, depending on the $l_\kappa$.

Such a problem has an associated mixing family $\mathcal{F}_\mathrm{min}^{(N - 1)} = \mathcal{F}_\mathrm{min}^{(N)} \setminus \{ M_k \}$, where $M_k$ is the operator that acts on several qubits in such a way that the state of the $N$th qubit changes as $\ket{0_N} \leftrightarrow \ket{1_N}$ while the other qubits change to preserve the value of the constraint, following the definition of the minimal $m$-to-1 mixing family in (\ref{fmin}). Such a copy will then have the associated graph $G^{(N - 1, b')}$. 

To prove connectivity of $G^{(N,b)}$, we will first show that the graph $G^{(N,b)}$ is the union of two disjoint subgraphs $G_0$ and $G_1$ along with at least one edge between vertices in $G_0$ and $G_1$; $G_0$ and $G_1$ are isomorphic to graphs $G^{(N-1,b)}$ and $G^{(N-1,b-k)}$ associated with the $(N - 1)$-variable problem described in the previous paragraph for $b' = b$ and $b' = b - k$, respectively. Importantly, we assumed that the $N$th variable has coefficient $k$, so removing it preserves the sequential coefficient requirement, ensuring the $(N - 1)$-variable problem still satisfies the conditions of~(\ref{sequential-linear-BILP}). Hence they are each connected by the inductive hypothesis. To show the whole graph $G^{(N,b)}$ is connected, we then find an edge between vertices in the two induced subgraphs $G_0$ and $G_1$.

First, partition the total vertex set $V^{(N,b)}$ into parts $V_0$ and $V_1$ depending on the value of $z_N \in \{0,1\}$, such that 
\begin{equation}
    V^{(N,b)} = V_0 \bigsqcup V_1, \ \text{where} \ V_{x} = \{\ket{\bm z} \in V^{(N,b)} : z_N = x\}
\end{equation}
 and $\bigsqcup$ denotes the disjoint union. These vertex sets give us corresponding induced subgraphs $G_x=(V_x,E_x)$ where $E_x$ is the set of all edges $e = (u,v) \ \in E^{(N,b)}_{\mathrm{min}}$ with $u,v \in V_x$.  We will show that $G_0$ is isomorphic to $G^{(N-1,b)}$, while $G_1$ is isomorphic to $G^{(N-1,b-k)}$. To show the graphs are isomorphic, we need to show there is a bijective mapping of their vertices that is edge-preserving. 

By definition, each vertex $\ket{\bm z} \in V^{(b - kx)}$ represents a solution to the constraint $\sum_{i=1}^{N-1} s_i z_i = b-k x$. We can relate these to vertices in $V_x$ using the bijective mapping $f_x$ that appends a bit $x \in \{0,1\}$ to a bitstring,
\begin{equation}
    f_{x} : (z_1, \ldots, z_{N-1}) \mapsto (z_1, \ldots, z_{N-1}, x),
\end{equation}
with inverse $f_x^{-1} : (z_1, \ldots, z_{N-1}, x) \mapsto (z_1, \ldots, z_{N-1})$. Let $\ket{\bm z'} = \ket{f_{x}(\bm z)}$. Then,
\begin{equation} \label{isomorphic vertices}
    \sum_{i=1}^{N} s_i z'_{i} = \sum_{i=1}^{N-1} s_i z_{i} + k x = b 
\end{equation}
so the bitstring $\bm z'$ is a solution to the $N$-variable constraint with $z'_N = x$ and therefore $\ket{\bm z'} \in V_x$.

The final step in showing the graphs are isomorphic is to show $f_x$ is edge-preserving.  By definition, each edge $(\ket{\bm z}, \ket{\bm w}) \in E^{(N-1,b - k x)}_{\mathrm{min}}$ corresponds to an operator $B_j \in \mathcal{F}_{\mathrm{min}}^{(N-1)}$ for which $\bra{\bm z} B_j \ket{\bm w} \neq 0$.  The operator $B_j$ is also present in the mixing family $\mathcal{F}_\text{min}^{(N)}$, with the understanding that in the $N$-variable problem the operator acts as the identity $\mathbb{1}_N$ on the $N$th qubit subspace; for rigor and clarity, we can denote these as $B_j^{(N-1)} \in \mathcal{F}_\text{min}^{(N-1)}$ and $B_j^{(N)} \in \mathcal{F}_\text{min}^{(N)}$ with $B_j^{(N)}= B_j^{(N-1)}\otimes \mathbb{1}_N$. Therefore, $\bra{\bm z} B_j^{(N-1)} \ket{\bm w} = \bra{\bm z} B_j^{(N-1)}\ket{\bm w} \bra{x} \mathbb{1} \ket{x} = \bra{f_x(\bm z)} B_j^{(N)} \ket{f_x(\bm w)} \neq 0$.  So for each edge $(\ket{\bm z}, \ket{\bm w}) \in E^{(N-1,b - k x)}_{\mathrm{min}}$, we have $(\ket{f_x(\bm z)}, \ket{f_x(\bm w)}) \in E_x$. 

We can also show that for each edge $e' = (\ket{\bm z'}, \ket{\bm w'}) \in E_{x}$, there exists an edge $(\ket{f^{-1}_x(\bm z')}, \ket{f^{-1}_x(\bm w')}) \in E^{(N-1,b - k x)}_{\mathrm{min}}$, as follows.  The mixing families for the $N$-variable and $(N-1)$-variable problems are related as $\mathcal{F}_\mathrm{min}^{(N)} = \mathcal{F}_{\mathrm{min}}^{(N-1)} \cup \{M_k\}$, as noted previously.  Since $M_k$ changes the state $\ket{0_N} \leftrightarrow \ket{1_N}$, while $\ket{z_N} = \ket{x}$ is fixed in each $G_x$, it follows that each edge $e \in E_x$ is not related to the operator $M_k$.  Therefore, edges $e \in G_x$ can only arise from the operators $B_j \neq M_k$, and these operators are also contained in the mixing family $\mathcal{F}_\mathrm{min}^{(N-1)}$. Then, similarly to the previous case, for each edge $e' = (\ket{\bm z'}, \ket{\bm w'}) \in E_x$ there exists a $B_j \in \mathcal{F}_\mathrm{min}^{(N)}$ such that $\bra{\bm z'}B_j\ket{\bm w'} \neq 0$, and for any such $B_j$ we also have that $B_j \in \mathcal{F}_\mathrm{min}^{(N-1)}$, $\bra{f_x^{-1}(\bm z')}B_j\ket{f_x^{-1}(\bm w')} \neq 0$, and therefore the edge $(\ket{f^{-1}_x(\bm z')}, \ket{f^{-1}_x(\bm w')}) \in E^{(b - k x)}$.

In total, we have shown there is a bijective mapping $f_x$ between the vertices of $G^{(N-1,b - k x)}$ and $G_x$ such that vertices $\ket{\bm z}$ and $\ket{\bm w}$ are adjacent in $G^{(N-1,b - k x)}$ if and only if $\ket{f_x(\bm z)}$ and $\ket{f_x(\bm w)}$ are adjacent in $G_x$. Therefore, these graphs are isomorphic. Applying the inductive hypothesis, we conclude that $G_0$ and $G_1$ are each connected. 

If $b < k$, then it follows that $G_1$ is empty and $G^{(N,b)} = G_0$, which is connected. Otherwise, to complete the proof, we show there exists an edge connecting the subgraphs $G_0$ and $G_1$ within the total graph $G^{(N,b)}$. Since $G_0$ and $G_1$ are connected, finding an edge between them shows $G^{(N,b)}$ is connected.

Our method of showing connectivity will depend on the number of variables $l_k$ with coefficient $k$ in the $N$-variable problem. Let $k^{(N-1)}$ denote the value of $k$ in the $(N-1)$-variable problem. If $l_k > 1$, at least one other variable in the $N$-variable problem has coefficient $k$ in addition to $z_N$. Then, removing $z_N$ does not change $k^{(N-1)} = k$. Otherwise, $l_k = 1$ and $z_N$ is the only variable with coefficient $k$. Removing $z_N$ then gives us $k^{(N-1)} = k - 1$ following the sequential constraint requirement. In terms of the table representation, these two cases correspond to whether or not removing $z_N$ changes the number of rows in the table, see Fig.~\ref{fig:inductive-step}. Lemmas \ref{lemma:prepare-for-swap} and \ref{lemma:prepare-for-merge} give us what we need to prove that an edge exists between the two subgraphs in each of these two cases, as we demonstrate below.

    If $l_k > 1$, then a coefficient $s_i = k$ already exists in the $(N-1)$-variable problem, as visualized by incorporating $z_N = z_{[k, l_{k}]}$ into an existing row of the table in Fig.~\ref{fig:inductive-step}. By definition (\ref{fmin}), the mixing family is $\mathcal{F}_\mathrm{min}^{(N)} = \mathcal{F}_\mathrm{min}^{(N-1)} \cup \{ M_{[k, l_k - 1], [k, l_k]} \}$. 
    We know that $b \geq k$, so by Lemma \ref{lemma:prepare-for-swap} $\exists \ket{\bm z} \in V^{(N-1,b)}$ such that $z_{[k, l_k - 1]} = 1$. Let $\ket{\bm z'} = \ket{f_0(\bm z)} \in V_0$ and let $\ket{\bm w'} = M_{[k, l_k-1], [k, l_k]}\ket{\bm z'}.$ Then $\ket{\bm w'} \in V_1$, by the definitions of $M_{[k, l_k - 1], [k, l_k]}$ and $\ket{\bm z'} = \ket{f_0(\bm z)}$. 
    Therefore, there is an edge $(\ket{\bm w'}, \ket{\bm z'})$ connecting vertices in $V_0$ and $V_1$ within the total graph $G^{(N,b)}$.

    Otherwise, $l_k = 1$, as visualized in the rightmost table of Fig.~\ref{fig:inductive-step}.
    By the definition (\ref{fmin}) of $\mathcal{F}_\text{min}$, $\mathcal{F}_\mathrm{min}^{(N)} = \mathcal{F}_\mathrm{min}^{(N-1)} \cup \{ M^{(2)}_{\{[1, 1], [k - 1, 1]\}, [k, 1]} \}$. Applying Lemma \ref{lemma:prepare-for-merge} to the $(N-1)$-variable problem where $b \geq k = k^{(N-1)} + 1$, $\exists \ket{\bm z} \in V^{(N-1,b)}$ such that $z_{[1, 1]} = 1$ and $z_{[k^{(N-1)} = k - 1, 1]}= 1$. Let $\ket{\bm z'} = \ket{f_0(\bm z)} \in V_0$ and $\ket{\bm w'} = M^{(2)}_{\{[1, 1], [k - 1, 1]\}, [k, 1]} \ket{\bm z'}$. We have $\ket{\bm w'} \in V_1$ by the definitions of $M^{(2)}_{\{[1, 1], [k - 1, 1]\}, [k, 1]}$ and $\ket{\bm z'} = \ket{f_0(\bm z)}$.  Therefore, there is an edge $(\ket{\bm w'}, \ket{\bm z'})$ connecting vertices in $V_0$ and $V_1$ within the total graph $G^{(N,b)}$.
    
We conclude that $G^{(N,b)}$ is connected. 

\end{proof}

The following two lemmas are used to show that an edge exists between two subgraphs defined in the above proof. Informally, Lemma~\ref{lemma:prepare-for-swap} says that for large enough $b$ and any fixed cell $z_i$, there is always a solution to Eq.~(\ref{sequential-linear-BILP}) such that $z_i = 1$. Similarly, Lemma~\ref{lemma:prepare-for-merge} says that for large enough $b$, there is always a solution to Eq.~(\ref{sequential-linear-BILP}) such that the top-left and bottom-left cells are both set to 1.

\begin{figure}
    \centering
    \[
    {\begin{NiceArray}{r|c|c|c|cr|c|c|c|ccr|c|c|c|c}
\cline{2-4}\cline{7-9}\cline{13-15}
1 & z_{[1, 1]} & \ldots & z_{[1, l_1]} && 1 & z_{[1, 1]} & \ldots & z_{[1, l_1]} &&& 1 & z_{[1, 1]} & \ldots & z_{[1, l_1]}\\
\cline{2-4}\cline{7-9}\cline{13-15}
\ldots &&&& \xrightarrow{+ z_N} & \ldots &&&&& \ \text{or} \ & \ldots &&&&\\
\cline{2-4}\cline{7-10}\cline{13-15}
k^{(N-1)} & z_{[k^{(N-1)}, 1]} & \ldots & && k^{(N-1)} = k & z_{[k, 1]} & \ldots &&\multicolumn{1}{c|}{\textcolor{blue}{z_{[k, l_k]}}} &&  k^{(N-1)} = k - 1& z_{[k - 1, 1]} & \ldots & \\
\cline{2-4}\cline{7-10}\cline{13-15}
\multicolumn{11}{c}{} & k & \textcolor{blue}{z_{[k, 1]} }& \multicolumn{3}{c}{}\\
\cline{13-13}
\end{NiceArray}} 
\]
\caption{Tables illustrating the inductive step and two possible cases in the proof of Lemma~\ref{lemma:connected}, with the $N$th variable labeled in blue.}
\label{fig:inductive-step}
\end{figure}

\begin{lemma}\label{lemma:prepare-for-swap}
    Suppose we have a linearly constrained problem specified by (\ref{sequential-linear-BILP}). Suppose there are $l_\kappa$ variables with coefficient $\kappa$, and denote the $l$th one of these variables as $z_{[\kappa,l]}$.  Consider a specific variable $z_{[\kappa\text{*},l\text{*}]}$. Then, $\forall b$ with $\kappa\text{*} \leq b \leq \sum_{\kappa=1}^{k} \kappa l_\kappa$,  $\exists \ket{\bm z} \in V^{(b)}$ such that $z_{[\kappa\text{*}, l\text{*}]} = 1$. 
\end{lemma} 

\begin{proof}
    
    We will prove the lemma by induction on $b$. In the base case, $b = \kappa\text{*}$ so define $\ket{z}$ by $z_{[\kappa\text{*}, l\text{*}]} = 1$ and $z_{[\kappa, l]} = 0$ otherwise. Then, 
    \begin{equation}
        \sum_{\kappa=1}^{k} \sum_{l=1}^{l_\kappa} \kappa z_{[\kappa, l]} = \kappa\text{*}
    \end{equation}
    as desired. This proves the base case.

    For the inductive step on $b$, suppose the lemma holds for $b - 1$. Then, there exists a solution $\ket{\bm z^{(b-1)}}$ such that $\sum_{\kappa=1}^{k}\sum_{l=1}^{l_\kappa}\kappa z_{[\kappa,l]}^{(b-1)}  = b - 1$ with $z^{(b-1)}_{[\kappa\text{*},l\text{*}]}=1$. From this we will show there exists a state $\ket{\bm z^{(b)}}$ such that $\sum_{\kappa=1}^{k}\sum_{l=1}^{l_\kappa}\kappa z_{[\kappa,l]}^{(b)} = b$ with $z^{(b)}_{[\kappa\text{*},l\text{*}]}=1$.

    Since $\sum_{\kappa=1}^k\sum_{l=1}^{l_\kappa} \kappa z^{(b-1)}_{[\kappa,l]} = b-1 < b \leq \sum_{\kappa=1}^k\sum_{l=1}^{l_\kappa} \kappa$, there must be at least one $z_{[\kappa,l]}^{(b-1)} = 0$. Let $\kappa'$ be the smallest $\kappa$ such that $\exists l'$ such that $z_{[\kappa', l']}^{(b-1)} = 0$. We will define $\ket{\bm z^{(b)}}$ depending on one of three possible cases of the values of $\kappa'$. The three cases are 1) $\kappa'=1$, 2) $\kappa'=\kappa\text{*}+1$, and 3) $1 < \kappa' \leq k$ and $\kappa' \neq \kappa\text{*} + 1$.
    \begin{enumerate}
        \item If $\kappa' = 1$, then let $\ket{\bm z^{(b)}}$ be defined as $z^{(b)}_{[\kappa', l']} = 1$ and $z^{(b)}_{[\kappa, l]} = z^{(b-1)}_{[\kappa, l]}$ otherwise. This gives us the desired $z^{(b)}_{[\kappa\text{*},l\text{*}]} = z^{(b-1)}_{[\kappa\text{*},l\text{*}]} = 1$. We also know $z^{(b - 1)}_{[\kappa', l']} = 0$ by definition so we have
    
        \begin{equation}
            \sum_{\kappa=1}^{k}  \sum_{l=1}^{l_\kappa} \kappa z^{(b)}_{[\kappa, l]} = (b - 1) + 1 = b \implies \ket{\bm z ^{(b)}} \in V^{(b)}.
        \end{equation}

\item If $\kappa' = \kappa\text{*} + 1$, we will define $\ket{\bm z^{(b)}}$ as follows: if $\kappa\text{*} > 1$, let $z^{(b)}_{[\kappa\text{*} - 1, 1]} = 0$. Set the remaining values $z^{(b)}_{[1, 2]} = 0$, $z^{(b)}_{[\kappa', l']} = 1$, and $z^{(b)}_{[\kappa, l]} = z^{(b-1)}_{[\kappa, l]}$ otherwise. We know $z^{(b - 1)}_{[\kappa', l']} = 0$ by definition and that $z^{(b-1)}_{[1, 2]} = 0$ by minimality of $\kappa'$ with $\kappa' > 1$. Therefore,

        \begin{equation}
            \sum_{\kappa=1}^{k}  \sum_{l=1}^{l_\kappa} \kappa z^{(b)}_{[\kappa, l]} = (b - 1) - (\kappa\text{*} - 1) - 1 + \kappa' = (b - 1) - (\kappa' - 2) - 1 + \kappa'  = b  \implies \ket{\bm z^{(b)}} \in V^{(b)}.
        \end{equation}

        \item Otherwise, we have that $\kappa' > 1$ and $\kappa' \neq \kappa\text{*}+1$. We will define $\ket{\bm z^{(b)}}$ as follows: $z^{(b)}_{[\kappa' - 1,1]} = 0$, $z^{(b)}_{[\kappa', l']} = 1$, and $z^{(b)}_{[\kappa, l]} = z^{(b-1)}_{[\kappa, l]}$ otherwise. We know $z^{(b - 1)}_{[\kappa', l']} = 0$ by definition and that $z^{(b-1)}_{[\kappa' - 1, 1]} = 0$ by minimality of $\kappa'$ with $\kappa' > 1$ and $\kappa' - 1 \neq \kappa\text{*}$. Therefore,
        \begin{equation}
            \sum_{\kappa=1}^{k} \sum_{l=1}^{l_l} \kappa z^{(b)}_{[\kappa, l]} =
            (b - 1) + \kappa' - (\kappa' - 1) = b \implies \ket{\bm z^{(b)}} \in V^{(b)}.
        \end{equation}
        \end{enumerate}

    So in all cases such a $\ket{\bm z^{(b)}}$ exists with the desired $z^{(b)}_{[1, 1]} = z^{(b - 1)}_{[1, 1]} = 1$, $z^{(b)}_{[k, 1]} = z^{(b - 1)}_{[k, 1]} = 1$, completing the induction. Solutions of this type for an example problem are shown in Fig.~\ref{fig:tables-swap-lemma}. 
\end{proof}

\begin{figure}
   \centering
    \begin{equation}
    \begin{array}{c|c|c|cc|c|c|cc|c|c|cccc|c|c|c}
        \multicolumn{4}{c}{b=s_{[\kappa\text{*}, l\text{*}]} = 1} & \multicolumn{3}{c}{b=2} & & \multicolumn{3}{c}{b=3} & & & 
 & \multicolumn{4}{c}{b=\sum_{i}s_i=12}\\
 \cline{2-3}\cline{6-7}\cline{10-11}\cline{16-17}
        1 & 1 & 0 & & 1 & 1 & 1 & & 1 & 1 & 0 & & & & 1 & 1 & 1\\
 \cline{2-3}\cline{6-7}\cline{10-11}\cline{16-17}
        2 & 0 & 0 & \xrightarrow{+1} & 2 & 0 & 0 & \xrightarrow{+1} & 2 & 1 & 0 & \xrightarrow{+1} & \ldots & \xrightarrow{+1} & 2 & 1 & 1\\
 \cline{2-3}\cline{6-7}\cline{10-11}\cline{16-17}
        3 & 0 & 0 & & 3 & 0 & 0 & & 3 & 0 & 0 & & &  & 3 & 1 & 1\\
 \cline{2-3}\cline{6-7}\cline{10-11}\cline{16-17}  
    \end{array}
\end{equation}
    \caption{Illustration of induction in Lemma~\ref{lemma:prepare-for-swap} for $\kappa\text{*} = 1, l\text{*} = 1$.  The leftmost table shows an example bitstring with the desired property when $b=1$, while subsequent tables show solutions for increasing $b$.} 
    \label{fig:tables-swap-lemma}
\end{figure}

\begin{lemma}\label{lemma:prepare-for-merge}
     Suppose we have a linearly constrained problem specified by (\ref{sequential-linear-BILP}). Then, $\forall b$ with $k + 1 \leq b \leq \sum_{\kappa=1}^{k} \kappa l_{\kappa} $ $\exists \ket{\bm z} \in V^{(b)}$ \text{s.t.} $z_{[1, 1]} = 1$, $z_{[k, 1]} = 1.$
\end{lemma}

\begin{proof}
    
    We will prove the lemma by induction on b. In the base case, $b = k + 1$ so define $\ket{\bm z}$ by $z_{[1, 1]} = 1$, $z_{[k, 1]} = 1$ and $z_{[\kappa, l]} = 0$ otherwise. Then, 
    \begin{equation}
        \sum_{\kappa=1}^{k} \sum_{l=1}^{l_\kappa} \kappa z_{[\kappa, l]} = 1 + k 
    \end{equation}
    as desired. This proves the base case.

    For the inductive step on $b$, suppose the lemma holds for $b - 1$. Then, there exists a $\ket{\bm z^{(b-1)}}$ such that $\sum_{\kappa=1}^{k}\sum_{l=1}^{l_\kappa}\kappa z_{[\kappa,l]}^{(b-1)} = b - 1$ with $z_{[1,1]}^{(b-1)} = 1$, $z_{[k,1]}^{(b-1)} = 1$. 

    Since $\sum_{\kappa=1}^{k}\sum_{l=1}^{l_\kappa}\kappa z_{[\kappa,l]}^{(b-1)} = b - 1 < b \leq \sum_{\kappa=1}^{k} \kappa l_\kappa$, there must be at least one $z_{[\kappa,l]}^{(b-1)} = 0$. Let $\kappa'$ be the smallest $\kappa$ such that $\exists l'$ such that $z_{[\kappa', l']}^{(b-1)} = 0$.

    We will define $\ket{\bm z^{(b)}}$ depending on one of three possible cases of the values of $\kappa'$. The three cases are 1) $\kappa'=1$, 2) $\kappa'=2$, and 3) $2 < \kappa' \leq k$.
    \begin{itemize}
        \item If $\kappa' = 1$, then let $\ket{\bm z^{(b)}}$ be defined as $z^{(b)}_{[\kappa' = 1, l']} = 1$ and $z^{(b)}_{[\kappa, l]} = z^{(b-1)}_{[\kappa, l]}$ otherwise. We know $z^{(b-1)}_{[\kappa' = 1, l']} = 0$ by definition. Therefore,
        \begin{equation}
            \sum_{\kappa=1}^{k}  \sum_{l=1}^{l_\kappa} \kappa z^{(b)}_{[\kappa, l]} = (b - 1) + 1 = b \implies \ket{\bm z^{(b)}} \in V^{(b)}
        \end{equation}.

        \item If $\kappa' = 2$, let $\ket{\bm z^{(b)}}$ be defined as $z^{(b)}_{[1, 2]} = 0$, $z^{(b)}_{[\kappa' = 2, l']} = 1$ and $z^{(b)}_{[\kappa, l]} = z^{(b-1)}_{[\kappa, l]}$ otherwise. We know $z^{(b-1)}_{[\kappa' = 2, l']} = 0$ by definition and $z^{(b-1)}_{[1, 2]} = 1$ by minimality of $\kappa'$ with $\kappa = 2$. Therefore,
        
        \begin{equation}
            \sum_{\kappa=1}^{k}  \sum_{l=1}^{l_\kappa} \kappa z^{(b)}_{[\kappa, l]} = (b - 1) - 1 + 2 = b \implies \ket{\bm z^{(b)}} \in V^{(b)}.
        \end{equation}
        \item Otherwise $2 < \kappa' \leq k$ Let $\ket{\bm z^{(b)}}$ be defined as $z^{(b)}_{[\kappa', l']} = 1$, $z^{(b)}_{[\kappa' - 1, 1]} = 0$ and $z^{(b)}_{[\kappa, l]} = z^{(b-1)}_{[\kappa, l]}$ otherwise. We know $z^{(b-1)}_{[\kappa', l']} = 0$ by definition and $z^{(b-1)}_{[\kappa'-1, 1]} = 1$ by minimality of $\kappa$ with $\kappa > 2$. Therefore, 
        \begin{equation}
            \sum_{\kappa=1}^{k} \sum_{l=1}^{l_\kappa} s_{[\kappa, l]} z^{(b)}_{[\kappa, l]} =
            (b - 1) + \kappa' - (\kappa' - 1) = b \implies \ket{\bm z^{(b)}} \in V^{(b)}
        \end{equation}
        \end{itemize}

    So in all cases such a $\ket{\bm z^{(b)}}$ exists with  the desired $z^{(b)}_{[1, 1]} = z^{(b - 1)}_{[1, 1]} = 1$, $z^{(b)}_{[k, 1]} = z^{(b - 1)}_{[k, 1]} = 1$, completing the induction. Solutions of this type for an example problem are shown in Fig.~\ref{fig:tables-merge-lemma}. 
\end{proof}

\begin{figure}
    \centering
    \begin{equation}
    \begin{array}{c|c|c|cc|c|c|cc|c|c|cccc|c|c|c}
        \multicolumn{4}{c}{b=k+1 = 4} & \multicolumn{3}{c}{b=5} & & \multicolumn{3}{c}{b=6} & & 
 & \multicolumn{5}{c}{b=\sum_{i}s_i=12}\\
 \cline{2-3}\cline{6-7}\cline{10-11}\cline{16-17}
        1 & 1 & 0 & & 1 & 1 & 1 & & 1 & 1 & 0 & & & & 1 & 1 & 1\\
 \cline{2-3}\cline{6-7}\cline{10-11}\cline{16-17}
        2 & 0 & 0 & \xrightarrow{+1} & 2 & 0 & 0 & \xrightarrow{+1} & 2 & 1 & 0 & \xrightarrow{+1} & \ldots & \xrightarrow{+1} & 2 & 1 & 1\\
 \cline{2-3}\cline{6-7}\cline{10-11}\cline{16-17}
        3 & 1 & 0 & & 3 & 1 & 0 & & 3 & 1 & 0 & & &  & 3 & 1 & 1\\
 \cline{2-3}\cline{6-7}\cline{10-11}\cline{16-17}  
    \end{array}
\end{equation}
    \caption{Illustration of induction in Lemma \ref{lemma:prepare-for-merge}, similar to Fig.~\ref{fig:tables-swap-lemma}.}
    \label{fig:tables-merge-lemma}
\end{figure}

\section{Graph symmetry analysis of adiabatic initial states} \label{initial state appendix}

Here we analyze adiabatic initial states for various mixers, including unconstrained mixers used in penalty-based approaches to QAOA, as well as symmetric and nonsymmetric constraint-preserving mixers.  Our analysis is based on the structure of basis-state-interaction graphs corresponding to a specified mixer, and reveals interesting features relating the symmetry of the graph to the structure in the initial state.  Our main finding is that previous approaches to unconstrained QAOA and to QAOA$^+$ with symmetric constraints require symmetric initial states, while for generic mixers and constraints, such as we consider, the initial states are in general nonsymmetric. 

\subsection{Initial states with symmetric mixing families}

 It will be instructive to first consider symmetric cases from previous QAOA and QAOA$^+$ approaches.  We will relate symmetry in the transverse field mixer $B = -\sum_i X_i$ and the $XY$ mixer $B^{XY}= -(1/2) \sum_{i<j} X_i X_j + Y_i Y_j$ to the structure of their corresponding adiabatic initial states. This will be useful for our extension to generic linear constraints in the next subsection. 

\begin{figure}
    \centering
    \includegraphics[width=\textwidth,height=15cm,keepaspectratio]{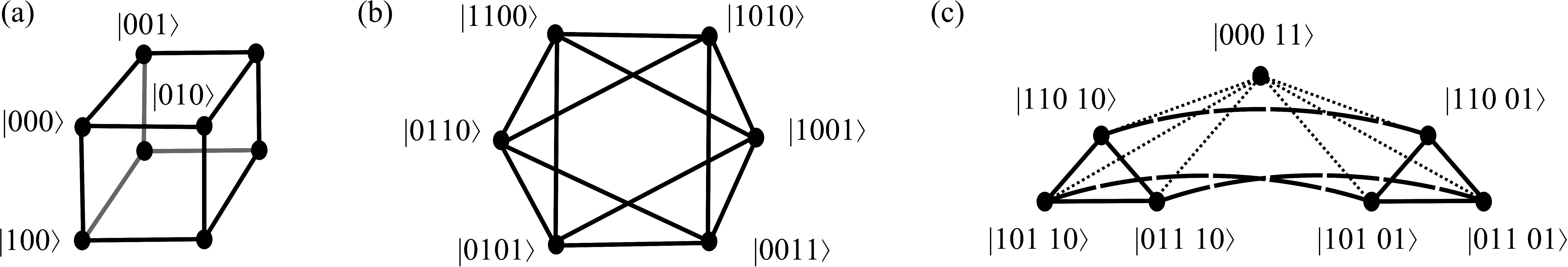}
    \caption{Example basis state interaction graphs for (a) unconstrained optimization with the transverse field mixer $B$ on three qubits, (b) an $XY$ mixer $B_{XY}$ for a Hamming weight conserving constraint $z_1 + z_2 + z_3 + z_4 = 2$, and (c) a maximal $m$-to-1 mixing family for a constraint $z_1 + z_2 + z_3 + 2z_4 + 2z_5 = 4$.  In (c), the transitions driven by three-qubit operators $M_{\{j,k\},i\text{*}}$ are shown with dotted lines, while two-qubit operator $M_{j,i*}$ transitions are shown by solid and dashed black lines.  The basis state transition graphs in (a) and (b) are regular, which entails an adiabatic initial state that is a symmetric superposition over all pictured basis states, while in (c) the nonregular graph entails an asymmetric adiabatic initial state. }
    \label{fig:interaction graph initial state}
\end{figure}

 First consider unconstrained optimization problems with the standard transverse field mixer $B$.  Its basis state interaction graph is a hypercube, see Fig.~\ref{fig:interaction graph initial state}(a) or the previous Fig.~\ref{fig:transition basis graphs}(b).  Each vertex is adjacent to $N$ other vertices, which are related through an edge generated by an operator $X_j$ for $j = 1,\ldots, N$.  The maximal eigenstate of a regular graph adjacency matrix is given by a symmetric vector of all basis components.  For $-B = \sum_i X_i$ this gives a maximal eigenstate (ground state of $B$)
 \begin{equation} \ket{\psi^{(B)}_0} = \frac{1}{2^{n/2}}\sum_{\bm z} \ket{\bm z}. \end{equation} 
 This is the adiabatic initial state for the transverse field mixer. 
 
 Now let us consider the more complicated XY mixer, which was developed to address optimization problems with a constraint on the Hamming weight $H$ of the solutions.  An example of the basis state interaction graph for this mixer is shown in Fig.~\ref{fig:interaction graph initial state}(b). The graph is regular with degree $H(N-H)$, since the components of $B_{XY}$ drive transitions such as $\ket{0_i,1_j} \leftrightarrow \ket{1_i,0_j}$, and for any basis state $\ket{\bm z} \in \mathcal{S}$ there are $H$ ways to choose a label $1_j$ and $N-H$ ways to choose a $0_i$ for these transitions. Since the graph is regular, the adiabatic initial state is symmetric 
 \begin{equation} \ket{\psi^{(B^{XY})}_0} = \frac{1}{\mathrm{dim}(\mathcal{S})^{1/2}}\sum_{\bm z \in \mathcal{S}} \ket{\bm z}. ,\end{equation} 
 These states are known as a Dicke states and efficient circuits have been devised to prepare them  \cite{bartschi2019deterministic,bartschi2022short}.   It is worth noting that the alternative ``ring XY-mixer" $\sim \sum_{i} X_{i}X_{i+1} + Y_iY_{i+1}$ does not yield a regular interaction graph, hence, requires an alternative initial state to conform to the adiabatic limit. 

\subsection{Initial states for generic linear-constraint mixing families}

As a variety of different mixing families are possible in Def.~\ref{def:mix-fam}, it is natural to wonder whether a suitable mixing family can be chosen to yield a regular interaction graph in our approach, and hence, a simple uniform initial state analogous to previous symmetric cases.  However, we have constructed a simple constraint $z_1 + z_2 + z_3 + 2z_4 + 2z_5 = 4$ for which no symmetric graph can be constructed from the merge operators $M_{I,i*}$. We show the basis state interaction graph for the maximal $m$-to-1 mixing family for this problem in Fig.~\ref{fig:interaction graph initial state}(c). There are three types of interactions, corresponding to transitions $\ket{0_i,1_j} \leftrightarrow \ket{1_i,0_j}$ between the variables $z_i, z_j$ with coefficients $s_i = s_j = 1$ in the constraint (solid edges), transitions between variables with coefficients $s_{i'}=s_{j'}=2$ (dashed edges), and transitions that merge variables with different coefficients (dotted edges). We attempted to prune operators from the maximum mixing family to generate a regular basis state interaction graph for this problem but found we were unable to accomplish this goal. We conclude that, in general, asymmetry in the basis state interaction graph is an unavoidable aspect of using mixing families composed from the merge operators $M_{I,i*}$ for nonsymmetric constraints of the form (\ref{sequential-linear-BILP}).
 
 One way to yield a symmetric interaction graph is to generalize the mixing family to include transition operators $\ket{\bm z}\bra{\bm z'}$ for each basis state $\ket{\bm z},\ket{\bm z'} \in \mathcal{S}$, rather than requring the family be composed from the $m$-to-1 merge operators $M_{I,i*}$, which are limited by the use of a single target variable $z_{i*}$. If operators that drive transitions $\ket{\bm z}\bra{\bm z'}$ for each basis state $\ket{\bm z},\ket{\bm z'} \in \mathcal{S}$ are included, then the basis state interaction graph will be the complete graph, which is regular. However, circuits implementing such operators appear to be complicated in general cases, while we expect the $M_{I,i*}$ to be much more practical.

 \section{Finding feasible initial states classically in linear time}\label{sec:initial-state-poly-time}

To show that a feasible initial state can be found classically in polynomial time, we present a following greedy algorithm. The algorithm begins with the all-zeros state, and in each iteration, it finds the variable with the largest available coefficient. If this coefficient is less than the remaining target sum, the variable is set to one, and the target is decremented accordingly. Theorem~\ref{thm:alg-1-correct} shows that this algorithm always returns a feasible state and terminates in $\mathcal{O}(n) \in \text{poly}(n)$ time. Note that we assume the ordering specified in the input to Algorithm~\ref{algorithm:find-initial-state} without loss of generality, as any coefficient set specified in Eq.~(\ref{sequential linear BILP single index}) can be pre-processed in linear time using the reordering procedure specified by Algorithm~\ref{algorithm:reordering-step}, as shown in Lemma~\ref{lemma:reordering-step}. 

\begin{algorithm}[H]
\SetAlgoLined

\caption{Find a feasible state for Eq. \ref{sequential-linear-BILP}}\label{algorithm:find-initial-state}

\KwIn{$\mathbf{s} = (s_1, \ldots,s_n)$, with $s_i = i$ for $1 \leq i \leq k$ and $1 \leq s_{k+1} \leq \ldots \leq s_n \leq k$.}
\KwIn{$b \in \mathbb{Z}$ with $0 \le b \le \sum_i s_i$}
\KwOut{$\mathbf{z} = (z_1, \ldots z_n) \in \{0, 1\}^n$ such that $\sum_i s_i z_i = b$.}
$\mathbf{z} \gets (0, \ldots, 0)$ of length $n$\;
\If{$b = 0$}{
    \Return{$\mathbf{z}$}\;
}
$b' \gets b$\;
\For{$i = n, \ldots, 1$}{
    \eIf{$s_i < b'$}{
        $z_i \gets 1$\;
        $b' \gets b' - s_i$\;
    }{
        $z_{b'} \gets 1$\tcp*{In the proof we show that $1 \leq b' \leq n$ here, so $z_{b'}$ is well-defined.}
        \Return{$\mathbf{z}$}\;
    }
}
\end{algorithm}

\begin{theorem}\label{thm:alg-1-correct}
    Algorithm \ref{algorithm:find-initial-state} always halts in $\mathcal{O}(n)$ time and returns $\mathbf{z} = (z_1, \ldots z_n) \in \{0, 1\}^n$ such that $\sum_i s_i z_i = b$.
\end{theorem}

\begin{proof}

We first prove runtime. Initializing $\mathbf{z}$ with $n$ elements can be done in linear time and the for loop iterates through at most $n$ values, performing a constant number of operations per iteration. Thus, the algorithm has an overall time complexity of $\mathcal{O}(n)$.

Next, we prove correctness. Observe that by the ordering requirement on the input, we first process the duplicate coefficients $s_n, \ldots, s_{k+1} $ in decreasing order before proceeding to $s_k, \ldots, s_1$, which are precisely the values $k, \ldots, 1$ respectively.

At the beginning of each iteration of the loop, we maintain the invariants: (i) $b' = b -\sum_j s_jz_j$ and (ii) $b' > 0$. 
Invariant (i) holds because $b'$ begins as $b$ and then in each iteration where we do not return, $z_i$ is set to 1, and $b'$ is decreased by the corresponding coefficient $s_i$, maintaining the invariant.
Invariant (ii) holds because because $b'$ only decreases from within the \textit{if} block where $s_i < b'$, meaning that the updated value, $b' - s_i$, is strictly positive.

We now use this to show that the \textit{else} block (the case where $s_i \geq b'$) is always triggered before the loop ends. Suppose this were not the case. Then, it must have always been the case that $s_i < b'$ in every iteration and all $z_i$ get set to one. By invariants (i) and (ii) and the condition that  $b \leq \sum_j s_j$, we have $0 < b' = b - \sum_j s_j \leq 0$, a contradiction. 

Therefore, the \textit{else} block must be triggered during some iteration, $i$, of the loop. During that iteration, we have that $b' \leq s_i \leq k$. By invariant (ii) and the fact that $k \leq n$, we can see that $1 \leq b' \leq n$, ensuring that $b'$ is a valid index and thus $z_{b'}$ is well-defined.
Additionally, because the loop processes entries in reverse order, and the first $k$ coefficients are $s_i = i$, we are guaranteed that $s_1, \ldots, s_{b'}$ remain unused (i.e. $z_1 =  \ldots = z_{b'} = 0$). Setting $z_{b'} \gets 1$, the sum becomes $\sum_j s_j z_j = (b - b') + s_{b'} = b$ and we correctly return $\mathbf{z}$.

\end{proof}

\begin{algorithm}[H]
\SetAlgoLined
\caption{Reordering Step}\label{algorithm:reordering-step}

\KwIn{$\mathbf{s} = (s_1, \ldots, s_n)$, a list containing each integer from $1$ to $k \leq n$ potentially multiple times}
\KwOut{$\mathbf{s}'$, which is a copy of $\mathbf{s}$, now rearranged such that $s'_i = i$ for $1 \leq i \leq k$ and that the remaining elements are in increasing order.}

Count occurrences $c_1, \ldots, c_k$ of each value in $\mathbf{s}$\;
$\mathbf{s}' \gets (1, \ldots, k, 0, \ldots, 0)$ with $n - k$ zeros at the end\;
$i \gets k + 1$\;
\For{$j = 1, \ldots, k$}{
    \If{$c_j > 1$}{
        Place the remaining $c_j - 1$ copies of $j$ into positions $i, i+1, \ldots, i + c_j - 2$\;
        $i \gets i + (c_j - 1)$\;
    }
}
\Return{$\mathbf{s}'$}\;
\end{algorithm}

\begin{lemma}\label{lemma:reordering-step}

Algorithm~\ref{algorithm:reordering-step} is correct and runs in $\mathcal{O}(n)$ time.

\end{lemma}
\begin{proof}

Algorithm~\ref{algorithm:reordering-step} is analogous to \textit{counting sort} with the exception that it reserves the first $k$ slots for the required copy of the integers $1, \ldots,  k$. Specifically, the algorithm creates a second array, $(c_1, \ldots, c_k)$, where $c_i$ represents the number of times the value $i$ occurs in $\mathbf{s}$. Then, $\mathbf{s}$ is overwritten: $1, \ldots, k$ are placed at the beginning and the remaining values are filled in with zeros. Then,  the remaining values are inserted in order based on their count minus one (for the existing copy). This ensures the output $\mathbf{s}$ meets the desired  ordering requirement.

Counting the values in $\mathbf{s}$ can be done through a single pass of the $n$ elements. Allocating and filling $\mathbf{s}$ requires $\mathcal{O}(n)$ time. The reorganization for loop requires placing a total of $\sum_{j=1}^{k} (c_j - 1) = n - k \leq n$ elements and thus requires $\mathcal{O}(n)$ time. Therefore, the total runtime is $\mathcal{O}(n)$.

\end{proof}

\section{Chebyshev polynomials and the relation to adiabatic evolution} \label{Chebyshev expansion}

Chebyshev polynomials provide low-degree approximations to continuous functions with a high-accuracy compared to other approximate polynomial expansions.  Here we use these polynomials to approximate optimized QAOA$^+$ angles and continuous annealing paths.  We summarize the main steps below; further details of the method are given in Ref.~\cite{numericalrecipes}. 

A function $f(x)$ can be expressed exactly as a sum over an infinite number of Chebyshev polynomials
\begin{equation} T_n(x) = \cos (n \arccos (x) ). \end{equation}
For a numerical approximation a finite sum of $N$ terms is used to approximate the function as 
\begin{equation} \label{f Chebyshev approx} f(x) \approx \sum_{k=1}^N c_k T_{k-1}(x) - \frac{1}{2} c_1 \end{equation} 
where the coefficients are
\begin{equation} \label{cj} c_j = \frac{2}{N} \sum_{k=1}^N f\left[\cos\left(\frac{\pi(k-\frac{1}{2})}{N}\right)\right]\cos\left(\frac{\pi(j-1)(k-\frac{1}{2})}{N}\right)\end{equation}

We now describe how to parameterize annealing schedules such as (\ref{3 H adiabatic schedule}) using Chebyshev polynomials.  First note the Chebyshev polynomials are defined for $x \in [-1,1]$ while $s \in [0,1]$, hence we take $x = -1 + 2s$. Then any schedule, such as the $\alpha(s), \beta(s),$ or $\gamma(s)$ defined in (\ref{annealing schedule}), can be expanded to any chosen order $N$ using (\ref{f Chebyshev approx}) with coefficients computed from (\ref{cj}).

We now consider parameters $\alpha_l, \beta_l, \gamma_l$ from the $l$th layer of a $p$-depth QAOA$^+$ circuit.  To relate these to the Chebyshev polynomials we first take $s=s(l) = l/(p+1)$ (similar to Appendix \ref{QAOA annealing appendix}) then take $x_l = 2s(l)-1$ to give $x_l$ that are distributed inside the interval $[-1,1].$ We then define $\alpha_l = \alpha(x_l)$ where $\alpha(x_l)$ is given by (\ref{f Chebyshev approx}), with similar expressions for $\beta_l$ and $\gamma_l$.  For optimized Chebyshev angles, we fit the parameters $c_k$ in an optimization routine.  This can be related directly to a continuous annealing schedule after taking the $x_l$ as continuous variables, following similar reasoning to Appendix \ref{QAOA annealing appendix}.

\begin{figure}
    \centering

\[\begin{tikzcd}[ampersand replacement=\&,column sep=3em]
	\&\&\&\&\&\&\&\& \UnstructuredConstraintOne{1}{1}{1}{1}{0}{0}{1}{1}{1} \\
	\&\& \UnstructuredConstraintOne{0}{0}{1}{1}{1}{1}{0}{1}{1}\&\& \UnstructuredConstraintOne{1}{1}{0}{1}{1}{1}{0}{1}{1} \&\& \UnstructuredConstraintOne{1}{1}{1}{0}{0}{1}{1}{1}{1} \&\& \UnstructuredConstraintOne{1}{1}{1}{0}{1}{0}{1}{1}{1} \\
	\UnstructuredConstraintOne{0}{1}{1}{1}{1}{1}{1}{0}{1} \&\& \UnstructuredConstraintOne{1}{0}{1}{1}{1}{1}{1}{0}{1} \&\& \UnstructuredConstraintOne{0}{1}{0}{0}{1}{1}{1}{1}{1} \&\& \UnstructuredConstraintOne{0}{1}{0}{1}{0}{1}{1}{1}{1} \&\& \UnstructuredConstraintOne{0}{1}{0}{1}{1}{0}{1}{1}{1} \\
	\UnstructuredConstraintOne{1}{1}{1}{1}{1}{1}{1}{1}{0} \&\& \UnstructuredConstraintZero{1}{1}{1}{1}{1}{1}{1}{1}{1} \&\& \UnstructuredConstraintOne{1}{0}{0}{0}{1}{1}{1}{1}{1} \&\& \UnstructuredConstraintOne{1}{0}{0}{1}{0}{1}{1}{1}{1} \&\& \UnstructuredConstraintOne{1}{0}{0}{1}{1}{0}{1}{1}{1}
	\arrow["{M_{\{[1,1],[1,2]\},[2,1]}}",color={graphblue}, tail reversed, from=2-3, to=2-5]
	\arrow["{M_{\{[1,1],[4,1]\},[5,1]}}"',color={graphgreen}, tail reversed, from=2-3, to=3-3]
	\arrow["{M_{\{[1,1],[3,1]\},[4,1]}}"',color={graphgreen}, tail reversed, from=2-5, to=3-5]
	\arrow["{M_{[3,2],[3,3]}}",color={graphorange}, tail reversed, from=2-7, to=2-9]
	\arrow["{M_{\{[1,1],[2,1]\},[3,1]}}"',color={graphgreen}, tail reversed, from=2-7, to=3-7]
	\arrow["{M_{[3,1],[3,2]}}",color={graphorange}, tail reversed, from=2-9, to=1-9]
	\arrow["{M_{\{[1,1],[2,1]\},[3,1]}}"',color={graphgreen}, tail reversed, from=2-9, to=3-9]
	\arrow["{M_{\{[1,1],[5,1]\},[6,1]}}",color={graphgreen}, tail reversed, from=3-1, to=4-1]
	\arrow["{M_{[1,1],[1,2]}}",color={graphorange}, tail reversed, from=3-3, to=3-1]
	\arrow["{M_{[1,1],[1,2]}}"',color={graphorange}, tail reversed, from=3-5, to=4-5]
	\arrow["{M_{[3,1],[3,2]}}"',color={graphorange}, tail reversed, from=3-7, to=3-5]
	\arrow["{M_{[3,2],[3,3]}}"',color={graphorange}, tail reversed, from=3-9, to=3-7]
	\arrow["{M_{[6,1],[6,2]}}",color={graphorange}, tail reversed, from=4-1, to=4-3]
	\arrow["{M_{[3,1],[3,2]}}",color={graphorange}, tail reversed, from=4-5, to=4-7]
	\arrow["{M_{[1,1],[1,2]}}",color={graphorange}, tail reversed, from=4-7, to=3-7]
	\arrow["{M_{[3,2],[3,3]}}",color={graphorange}, tail reversed, from=4-7, to=4-9]
	\arrow["{M_{[1,1],[1,2]}}",color={graphorange}, tail reversed, from=4-9, to=3-9]
\end{tikzcd}\]
    \caption{Basis state graph for the example unstructured constraunt in Eq.~(\ref{eqn:unstructured-constraint}) with $b = 28$.} 
    \label{fig:unstructured-graph}
\end{figure}

\begin{figure}
    \centering
    \[\begin{tikzcd}[ampersand replacement=\&,row sep=small,column sep=small]
	\&\& \SequentialConstraint{0}{0}{0}{1}{1}{0}{1}{1}{1} \&\& \SequentialConstraint{1}{1}{0}{1}{0}{0}{1}{1}{1} \\
	\& \SequentialConstraint{0}{0}{0}{1}{0}{1}{1}{1}{1} \&\& \SequentialConstraint{1}{1}{0}{0}{1}{0}{1}{1}{1} \&\& \SequentialConstraint{1}{0}{1}{1}{0}{0}{1}{1}{1} \\
	\SequentialConstraint{0}{0}{0}{0}{1}{1}{1}{1}{1} \&\& \SequentialConstraint{1}{1}{0}{0}{0}{1}{1}{1}{1} \&\& \SequentialConstraint{1}{0}{1}{0}{1}{0}{1}{1}{1} \&\& \SequentialConstraint{0}{1}{1}{1}{0}{0}{1}{1}{1} \\
	\&\&\& \SequentialConstraint{1}{0}{1}{0}{0}{1}{1}{1}{1} \&\& \SequentialConstraint{0}{1}{1}{0}{1}{0}{1}{1}{1} \\
	\&\&\&\& \SequentialConstraint{0}{1}{1}{0}{0}{1}{1}{1}{1} \& \SequentialConstraint{1}{1}{1}{1}{1}{0}{0}{1}{1} \\
	\&\&\&\& \SequentialConstraint{1}{1}{1}{1}{0}{1}{0}{1}{1} \&\& \SequentialConstraint{1}{1}{1}{1}{1}{0}{1}{0}{1} \\
	\&\&\& \SequentialConstraint{1}{1}{1}{0}{1}{1}{0}{1}{1} \&\& \SequentialConstraint{1}{1}{1}{1}{0}{1}{1}{0}{1} \&\& \SequentialConstraint{1}{1}{1}{1}{1}{0}{1}{1}{0} \\
	\&\& \SequentialConstraint{0}{0}{1}{1}{1}{1}{0}{1}{1} \&\& \SequentialConstraint{1}{1}{1}{0}{1}{1}{1}{0}{1} \&\& \SequentialConstraint{1}{1}{1}{1}{0}{1}{1}{1}{0} \\
	\& \SequentialConstraint{0}{1}{0}{1}{1}{1}{0}{1}{1} \&\& \SequentialConstraint{0}{0}{1}{1}{1}{1}{1}{0}{1} \&\& \SequentialConstraint{1}{1}{1}{0}{1}{1}{1}{1}{0} \\
	\SequentialConstraint{1}{0}{0}{1}{1}{1}{0}{1}{1} \&\& \SequentialConstraint{0}{1}{0}{1}{1}{1}{1}{0}{1} \&\& \SequentialConstraint{0}{0}{1}{1}{1}{1}{1}{1}{0} \\
	\& \SequentialConstraint{1}{0}{0}{1}{1}{1}{1}{0}{1} \&\& \SequentialConstraint{0}{1}{0}{1}{1}{1}{1}{1}{0} \\
	\&\& \SequentialConstraint{1}{0}{0}{1}{1}{1}{1}{1}{0}
	\arrow["{M_{[2,2],[2,3]}}"',color={graphorange}, tail reversed, from=1-3, to=2-2]
	\arrow["{M_{\{[1,1],[1,2]\},[2,1]}}",color={graphblue}, tail reversed, from=1-3, to=2-4]
	\arrow["{M_{[2,1],[2,2]}}",color={graphorange}, tail reversed, from=1-5, to=2-4]
	\arrow["{M_{[2,1],[2,2]}}"',color={graphorange}, tail reversed, from=2-2, to=3-1]
	\arrow["{M_{\{[1,1],[1,2]\},[2,1]}}",color={graphblue}, tail reversed, from=2-2, to=3-3]
	\arrow["{M_{[2,2],[2,3]}}",color={graphorange}, tail reversed, from=2-4, to=3-3]
	\arrow["{M_{[1,2],[1,3]}}",color={graphorange}, tail reversed, from=2-4, to=3-5]
	\arrow["{M_{[1,2],[1,3]}}"',color={graphorange}, tail reversed, from=2-6, to=1-5]
	\arrow["{M_{[2,1],[2,2]}}",color={graphorange}, tail reversed, from=2-6, to=3-5]
	\arrow["{M_{\{[1,1],[2,1]\},[3,1]}}",color={graphgreen}, tail reversed, from=3-1, to=10-1]
	\arrow["{M_{[1,2],[1,3]}}",color={graphorange}, tail reversed, from=3-3, to=4-4]
	\arrow["{M_{[2,2],[2,3]}}",color={graphorange}, tail reversed, from=3-5, to=4-4]
	\arrow["{M_{[1,1],[1,2]}}"',color={graphorange}, tail reversed, from=3-7, to=2-6]
	\arrow["{M_{[1,1],[1,2]}}"',color={graphorange}, tail reversed, from=4-6, to=3-5]
	\arrow["{M_{[2,1],[2,2]}}"',color={graphorange}, tail reversed, from=4-6, to=3-7]
	\arrow["{M_{\{[1,1],[2,1]\},[3,1]}}",color={graphgreen}, tail reversed, from=4-6, to=5-6]
	\arrow["{M_{[1,1],[1,2]}}",color={graphorange}, tail reversed, from=5-5, to=4-4]
	\arrow["{M_{[2,2],[2,3]}}",color={graphorange}, tail reversed, from=5-5, to=4-6]
	\arrow["{M_{\{[1,1],[2,1]\},[3,1]}}"',color={graphgreen}, tail reversed, from=5-5, to=6-5]
	\arrow["{M_{[3,1],[3,2]}}",color={graphorange}, tail reversed, from=5-6, to=6-7]
	\arrow["{M_{[2,2],[2,3]}}"',color={graphorange}, tail reversed, from=6-5, to=5-6]
	\arrow["{M_{[3,1],[3,2]}}",color={graphorange}, tail reversed, from=6-5, to=7-6]
	\arrow["{M_{[2,2],[2,3]}}",color={graphorange}, tail reversed, from=6-7, to=7-6]
	\arrow["{M_{[3,2],[3,3]}}",color={graphorange}, tail reversed, from=6-7, to=7-8]
	\arrow["{M_{[2,1],[2,2]}}"',color={graphorange}, tail reversed, from=7-4, to=6-5]
	\arrow["{M_{[3,1],[3,2]}}",color={graphorange}, tail reversed, from=7-4, to=8-5]
	\arrow["{M_{[2,1],[2,2]}}",color={graphorange}, tail reversed, from=7-6, to=8-5]
	\arrow["{M_{[3,2],[3,3]}}",color={graphorange}, tail reversed, from=7-6, to=8-7]
	\arrow["{M_{[2,2],[2,3]}}",color={graphorange}, tail reversed, from=7-8, to=8-7]
	\arrow["{M_{\{[1,1],[1,2]\},[2,1]}}",color={graphblue}, tail reversed, from=8-3, to=7-4]
	\arrow["{M_{[3,1],[3,2]}}"',color={graphorange}, tail reversed, from=8-3, to=9-4]
	\arrow["{M_{\{[1,1],[1,2]\},[2,1]}}",color={graphblue}, tail reversed, from=8-5, to=9-4]
	\arrow["{M_{[3,2],[3,3]}}",color={graphorange}, tail reversed, from=8-5, to=9-6]
	\arrow["{M_{[2,1],[2,2]}}",color={graphorange}, tail reversed, from=8-7, to=9-6]
	\arrow["{M_{[1,2],[1,3]}}",color={graphorange}, tail reversed, from=9-2, to=8-3]
	\arrow["{M_{[3,1],[3,2]}}",color={graphorange}, tail reversed, from=9-2, to=10-3]
	\arrow["{M_{[3,2],[3,3]}}",color={graphorange}, tail reversed, from=9-4, to=10-5]
	\arrow["{M_{\{[1,1],[1,2]\},[2,1]}}",color={graphblue}, tail reversed, from=9-6, to=10-5]
	\arrow["{M_{[1,1],[1,2]}}"',color={graphorange}, tail reversed, from=10-1, to=9-2]
	\arrow["{M_{[3,1],[3,2]}}",color={graphorange}, tail reversed, from=10-1, to=11-2]
	\arrow["{M_{[1,2],[1,3]}}"',color={graphorange}, tail reversed, from=10-3, to=9-4]
	\arrow["{M_{[3,2],[3,3]}}",color={graphorange}, tail reversed, from=10-3, to=11-4]
	\arrow["{M_{[1,2],[1,3]}}",color={graphorange}, tail reversed, from=10-5, to=11-4]
	\arrow["{M_{[1,1],[1,2]}}"',color={graphorange}, tail reversed, from=11-2, to=10-3]
	\arrow["{M_{[3,2],[3,3]}}",color={graphorange}, tail reversed, from=11-2, to=12-3]
	\arrow["{M_{[1,1],[1,2]}}",color={graphorange}, tail reversed, from=11-4, to=12-3]
\end{tikzcd}\]
    \caption{Basis state graph for the $k = 3, L = 3$ repeated sequential constraint from Eq.~(\ref{eqn:repeated-sequential-constraint}) with $b = 13$.} 
    \label{fig:repeated-sequential-graph}
\end{figure}

\end{document}